\documentclass[10pt, 3p, fleqn]{elsarticle}

\usepackage[british]{babel}
\usepackage[utf8]{inputenc}
\usepackage{graphics}
\usepackage{graphicx}
\graphicspath{{./}{ManuscriptFigures/}{ManuscriptFigures/concentration_results/}}
\usepackage{amsmath,amssymb,esint}
\usepackage{lineno}
\usepackage{multirow}
\usepackage{subfig}
\usepackage{natbib}
\usepackage{eucal}
\usepackage{mathtools}
\usepackage{placeins}
\usepackage[normalem]{ulem}
\usepackage{enumerate}
\usepackage[x11names]{xcolor}
\usepackage[colorlinks]{hyperref}
\usepackage{lmodern}
\hypersetup{
  pdftitle={Mass-conservative and phase-consistent transport of phase-confined scalars in front tracking},
  pdfauthor={Ayman Mazloum, Gabriele Gennari, Fabian Denner, Berend van Wachem}
}
\usepackage{longtable}
\usepackage{relsize}
\usepackage{setspace}
\usepackage{MnSymbol}
\usepackage[british]{babel}
\usepackage{booktabs}
\usepackage{siunitx}
\usepackage{tikz}
\usetikzlibrary{calc}
\usepackage{pgfplots}
\pgfplotsset{compat=1.18}
\usepackage{pgfplotstable}
\usetikzlibrary{shapes.geometric, arrows, intersections, through}
\usetikzlibrary{arrows.meta}
\usetikzlibrary{positioning}
\tikzset{>={Latex[width=1mm,length=1mm]}}
\usepackage{comment}

\usepackage[ruled,vlined]{algorithm2e}

\newproof{proof}{Proof}
\usepackage{varwidth}
\usepackage{adjustbox}

\usetikzlibrary{calc}
\tikzset{math3d/.style={z={(-0.65cm,-0.30cm)},y={(0cm,1cm)},x={(0.9cm,-0.15cm)}}}
\usetikzlibrary{shapes.geometric, arrows, intersections, through}
\usetikzlibrary{decorations.text, decorations.shapes, backgrounds}
\usetikzlibrary{arrows.meta}
\usetikzlibrary{colorbrewer}
\usetikzlibrary{patterns}
\usepgfplotslibrary{colorbrewer}
\usepgfplotslibrary{colormaps}

\tikzstyle{process} = [rectangle, minimum width=4.5cm, minimum height=1cm, text centered, text width = 4.5cm, draw=black]
\tikzstyle{decision} = [diamond, minimum width=4.5cm, minimum height=1cm, aspect=2,inner sep=-0.5ex,text centered, text width = 4.5cm, draw=black]
\tikzstyle{arrow} = [thick,->,>=stealth]
\tikzstyle{line} = [draw, -latex']

\biboptions{sort&compress}

\journal{}

\makeatletter
\renewcommand\elsparagraph{\@startsection{paragraph}{4}{0pt}%
  {10pt plus 6pt minus 3pt}{6pt}{\normalfont\itshape}}
\makeatother

\begin{document}

\begin{frontmatter}

\title{Mass-conservative and phase-consistent transport of phase-confined scalars in front tracking}

\author[affil2]{Ayman Mazloum}
\author[affil2]{Gabriele Gennari}
\author[affil3]{Fabian Denner}
\author[affil2]{Berend van Wachem\corref{cor1}}
\ead{berend.van.wachem@multiflow.org}

\address[affil2]{Chair of Mechanical Process Engineering, Otto-von-Guericke-Universit\"{a}t Magdeburg, Universit\"atsplatz 2, 39106 Magdeburg, Germany}
\address[affil3]{Department of Mechanical Engineering, Polytechnique Montréal, Montréal, H3T 1J4, Québec, Canada}

\cortext[cor1]{Corresponding author: }

\begin{abstract}
In front-tracking multiphase-flow simulations, the Lagrangian front defines the interface geometry and phase regions, while phase-confined scalars are transported independently on a fixed Eulerian mesh. This separation can make the scalar inconsistent with the tracked interface: even a conservative high-order scheme smears the scalar across the interface, leaks it into the other phase and distorts near-interface concentrations despite conserving total mass. We present a transport method that maintains consistency between the scalar and the tracked front while preserving global mass conservation. Instead of the cell-averaged concentration, the method transports the phase-averaged concentration, which is smooth across the interface. This concentration is obtained by dividing the cell-averaged concentration by the phase fraction, extended linearly across the interface and advected. The cell-averaged concentration is rebuilt by multiplying the transported concentration by the front's volume-preserving phase fraction. Because this rebuild can alter the total mass, a uniform shift of the phase-averaged concentration restores it. The method is compared with direct transport of the cell-averaged concentration by the same scheme in five three-dimensional cases: three drops (in translation, in a reversible vortex and with a localized source) and a rising bubble with and without dissolved surfactant. In the four cases without surfactant it reduces the error against the exact solution by more than a factor of ten at the finest resolution and converges faster for the drops. In the surfactant-laden bubble, direct transport leaks $5.5$--$24\%$ of the surfactant into the wake, depending on resolution, whereas the proposed method keeps it in the bubble.
\end{abstract}
\begin{keyword}
scalar transport \sep front tracking \sep phase-averaged concentration \sep mass conservation \sep scalar leakage \sep surfactant
\end{keyword}
\end{frontmatter}

\section{Introduction}

Scalar transport in interfacial flows is encountered in a wide range of multiphase-flow simulations. In gas--liquid and liquid--liquid flows, transported scalar quantities may either be defined in both phases or be physically confined to one phase or to the interface itself. Examples include soluble species released from a rising bubble by mass transfer across a fluid interface and transported in the surrounding liquid \citep{Marschall2012,Bothe2013}, temperature fields used to model boiling and evaporation phenomena \citep{RuedaVillegas2016,Irfan2017}, vapour mass fractions transported in the gas phase during droplet evaporation \citep{Irfan2017,Palmore2019}, and bulk surfactant concentrations that adsorb and desorb at deformable interfaces, coupled with interfacial surfactant concentrations transported along moving surfaces \citep{Muradoglu2008,Shin2018}. Passive scalar fields are also used to study scalar mixing in bubble-induced turbulence \citep{Hidman2023}.

These applications share a common numerical feature: the scalar field is advanced by the flow while the phase boundary moves and deforms. This is especially important for phase-confined scalars, which are physically meaningful only in one phase. For such quantities, global mass conservation is not sufficient: a numerical scalar field may preserve the total scalar mass while progressively losing consistency with the phase region defined by the moving interface. This inconsistency may appear as wrong-phase scalar content, numerical tails, or non-physical extrema near the interface.

In volume-of-fluid (VOF) formulations, a scalar can likewise lose its consistency with the phase that carries it in the transport of phase-specific scalars. The scalar field must be advected consistently with the volume-fraction field that defines the phase in which the scalar exists; otherwise, conservative scalar transport may still generate artificial interfacial transfer and wrong-phase leakage. This requirement has been recognized in VOF-based mass-transfer formulations, where phase-specific scalar transport is formulated consistently with the volume-fraction representation in order to maintain localization of the transported quantity relative to the phase geometry \citep{Alke2009,Haroun2010a,Bothe2013,Deising2016}. In such single-field formulations, the numerical diffusion of the phase-averaged concentration at high P\'eclet numbers has been reduced by an additional compressive term in the species equation \citep{Maes2018}. More recently, Tourbier et al.~\citep{Tourbier2024} addressed this issue for geometric VOF by developing a consistent scalar-transport methodology coupled to the geometric VOF advection scheme isoAdvector \citep{Roenby2016}, which represents the interface in each cell as an isosurface of the volume fraction and advects it geometrically.

Other interface-capturing approaches address the same requirement in different ways. In conservative diffuse-interface methods, \citet{Jain2022} constructed a scalar-transport model whose transport is consistent with that of the volume-fraction field, which keeps the scalar in its phase and preserves its positivity. In level-set methods with the ghost-fluid treatment of interfacial jump conditions \citep{Fedkiw1999}, quantities defined on one side of the interface are extended across it by extrapolation \citep{Aslam2004}, so that, for example, the vapour mass fraction and temperature of an evaporating droplet can be advanced near the interface without numerical diffusion across it \citep{Tanguy2007}.

In front-tracking methods \citep{Unverdi1992,Tryggvason2001,Tryggvason2011a}, a phase-confined scalar can lose its consistency with the phase that carries it, because the interface and the Eulerian scalar fields are represented on different numerical grids. The interface is tracked explicitly by Lagrangian markers or surface elements, which is attractive because interfacial quantities such as concentration, curvature, surface gradients, and Marangoni stresses can be evaluated directly on the moving front. However, bulk quantities such as passive scalars, dissolved species, vapour mass fractions, temperature fields, and surfactant concentrations are typically stored and transported on the fixed Eulerian grid. These fields must therefore remain consistent with the phase region defined by the independently tracked front, or with the Eulerian indicator field reconstructed from it. For example, front-tracking models of evaporation solve Eulerian temperature and vapour-mass-fraction transport equations coupled to interfacial phase-change conditions \citep{Irfan2017}, while soluble-surfactant models solve coupled bulk and interfacial transport equations with adsorption--desorption kinetics, surfactant-dependent surface tension, and Eulerian advection schemes for the bulk concentration \citep{Muradoglu2008,Muradoglu2014}. Comparisons reported for soluble-surfactant transport show that even high-resolution conservative schemes may produce overshoots, undershoots, numerical tails, and mismatch with the indicator function near the moving front \citep{Muradoglu2014}. This holds for schemes such as the quadratic upstream interpolation scheme (QUICK), essentially non-oscillatory and weighted essentially non-oscillatory schemes (ENO/WENO) and monotonic upstream-centred schemes for conservation laws with total-variation-diminishing limiters (MUSCL--TVD). Thus, the key requirement is consistency between the Eulerian transported scalar and the phase geometry defined by the Lagrangian front.

We refer to this loss of consistency between the transported scalar and the phase region defined by the front as scalar--front inconsistency: the scalar may remain globally mass conservative while part of it is carried outside its phase. To address the scalar--front inconsistency, we propose a transport method for phase-confined scalars in front-tracking simulations. Instead of the cell-averaged concentration, the scalar mass per unit volume of the cell, the method transports the phase-averaged concentration, the scalar mass per unit volume of the phase that carries the scalar. The two differ by the phase support, the fraction of each cell occupied by the carrying phase: it is one inside the phase, zero outside it, and between zero and one in the cells cut by the interface. Across the interface, the cell-averaged concentration therefore falls to zero together with the phase support, and a high-order reconstruction of this steep drop broadens the transported profile relative to the front, whereas the phase-averaged concentration has no such drop. In every time step, the phase-averaged concentration is recovered in the cells that the carrying phase fills sufficiently, extended across the interface by linear continuation along the grid lines and advected with a high-order scheme; the extension serves the purpose of the extrapolation used in ghost-fluid methods \citep{Aslam2004,Tanguy2007}. The scalar is then rebuilt as the product of the advected concentration and a phase support, obtained from the front by a bounded, volume-preserving projection of the reconstructed indicator. Finally, a single shift of the concentration over the carrying phase restores the scalar mass and keeps the scalar non-negative.

The contribution of this work is threefold. First, it formulates the transport of a phase-confined scalar in front tracking in terms of its phase-averaged concentration. The scalar is rebuilt on a volume-preserving phase support and corrected by a mass projection, so that by construction it stays in its phase, conserves its total mass, remains non-negative and keeps a uniform concentration exactly. Second, it extends the phase-averaged concentration from the cells inside the phase into the cells beyond the interface by an explicit rule, without solving an equation: the cells are filled layer by layer, and each cell receives the value on the straight line through the two known values behind it along a grid direction, averaged over the directions in which such values exist. Unlike the extrapolation of ghost-fluid methods \citep{Aslam2004,Tanguy2007}, which solves partial differential equations along the interface normal by iteration in pseudo-time, this rule needs neither a normal field nor iterations, and on a uniform grid it continues a linear concentration exactly wherever two known cells lie on the grid line. Third, it assesses the method with error measures that separate three aspects: the error of the phase-averaged concentration inside the phase, the agreement of the scalar with the support derived from the front, and the leakage into the other phase. These are evaluated in five three-dimensional cases: a uniform concentration translated along one axis; a concentration gradient in a reversible vortex, carried in turn by the drop and by the phase outside it, so that the method is tested for a scalar in either phase, one phase at a time; a moving volumetric source; a rising bubble; and a soluble surfactant in the rising bubble, whose adsorption and desorption are first validated on a sphere at rest against a converged reference solution. In every case the method is compared with the same scheme transporting the cell-averaged concentration.

Section~\ref{sec:motivation} illustrates the motivation of this paper. Section~\ref{sec:governing} introduces the governing equations, while Section~\ref{sec:numerics} presents the numerical framework, including front tracking, the indicator projection, the phase-concentration transport, the flow discretization and the solution procedure. Section~\ref{sec:results} presents the five validation cases, the contribution of each ingredient of the method, the dependence on the advection scheme, the sensitivity to the parameters of the method and its computational cost. The article is concluded in Section~\ref{sec:conclusion}. Implementation details are given in \ref{app:correction_implementation} and \ref{app:scalar_implementation}, the validation of the adsorption and desorption on a sphere at rest in \ref{app:surfactant_sphere}, and the localized source with seven advection schemes in \ref{app:scheme_source}.

{\color{black}
\section{Motivation}
\label{sec:motivation}

An Eulerian scalar field can conserve its volume integral without remaining consistent with the phase distribution reconstructed from a tracked front. To illustrate this behaviour, we consider a front-tracking calculation of the translation of a spherical bubble in a prescribed uniform velocity field directed along the $x$-axis of the Cartesian coordinates $\mathbf{x}=(x,y,z)$. This simple motion allows the transported scalar field to be compared with the front-derived indicator without the additional effects of bubble deformation or a coupled flow solution.

The bubble has a diameter of $D=0.5$ and is initially centred at $(1,2,2)$ in a periodic domain of dimensions $[0,5]\times[0,4]\times[0,4]$; all quantities are dimensionless. The Eulerian mesh resolves the bubble diameter with 20 cells. The bubble interface is represented by a triangulated front that moves with the flow, and the indicator $I^{\mathrm{FT}}$, with $0\le I^{\mathrm{FT}}\le1$, is reconstructed from this front on the Eulerian mesh (Section~\ref{sec:ft}). An Eulerian scalar field $C$ is initialized as $C(\mathbf{x},0)=I^{\mathrm{FT}}(\mathbf{x},0)$ and subsequently transported on the Eulerian mesh, while the front is advected separately using the same velocity, $\mathbf{u}=(1,0,0)$. For Eulerian transport, we use the fifth-order WENO-Z scheme of \citet{Borges2008}, an improvement of the weighted essentially non-oscillatory scheme of \citet{Jiang1996}, to limit numerical dissipation with a high-order approximation, and integrate in time with the three-stage strong-stability-preserving Runge--Kutta method (SSP-RK3) of \citet{Shu1988} at a Courant--Friedrichs--Lewy (CFL) number of $0.2$. The bubble is followed until $t=3$, when its centre reaches $(4,2,2)$.

Figure~\ref{fig:motivation_indicator_profiles} compares the profiles of $I^{\mathrm{FT}}$ and $C$ along a line parallel to the $x$-axis passing through the bubble centre at $y=z=2$. Figure~\ref{fig:motivation_indicator_profiles}(a) shows the initially coincident profiles at $t=0$, when the bubble is centred at $x=1$. Figure~\ref{fig:motivation_indicator_profiles}(b) shows the profiles at $t=3$, when the bubble has translated to $x=4$ (six bubble diameters); the transported scalar $C$ now has a wider interfacial transition and longer tails than the front-derived indicator $I^{\mathrm{FT}}$. The transition midpoints remain nearly aligned, so the difference is mainly in the shape of the profiles rather than their position. As the bubble travels, this mismatch increases, even though the volume integral of $C$ remains conserved to round-off accuracy. Thus, using a high-order conservative transport scheme does not by itself keep the transported scalar in agreement with the phase distribution defined by the tracked front. Refining the mesh reduces this mismatch only slowly, because the error of a capturing scheme for a jump that the flow carries decreases below first order with the cell size \citep{Banks2008}; at any finite resolution the transported scalar remains smeared relative to the front, and the mismatch keeps growing with the transport distance.

Although $C$ is initialized to equal the indicator in this example, it is transported as a scalar quantity. The example therefore shows that conservative Eulerian advection preserves the total scalar mass but smears the scalar relative to the indicator: the numerical diffusion of the advection scheme accumulates in the transported scalar, whereas the indicator is rebuilt from the tracked front at every step. Similar behaviour can occur for more general initial scalar distributions; for a scalar confined to one phase, this smearing can also lead to loss of consistency with the phase region defined by the tracked front, and leakage of the scalar into the other phase.

\begin{figure}[htbp]
\centering
\includegraphics[width=\textwidth]{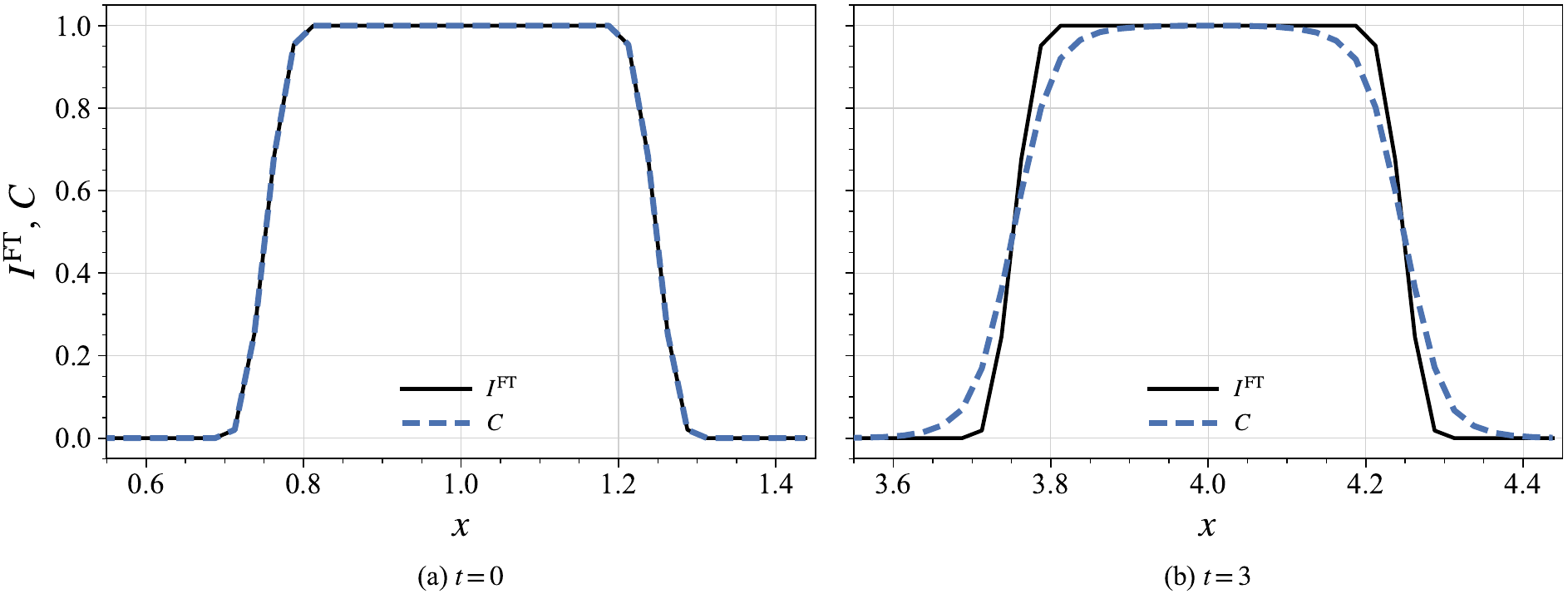}
\caption{\textcolor{black}{Centreline indicator and scalar profiles for a uniformly translating bubble: (a) initially coincident fields $C=I^{\mathrm{FT}}$ at $t=0$ and (b) profiles at $t=3$ after translation through six bubble diameters. The profiles are taken along a line parallel to the $x$-axis passing through the bubble centre at $y=z=2$. The solid black line denotes the front-derived indicator $I^{\mathrm{FT}}$ and the dashed blue line the WENO-Z-transported scalar $C$.}}
\label{fig:motivation_indicator_profiles}
\end{figure}
}

\section{Governing equations}
\label{sec:governing}

The incompressible and isothermal flows considered in this work are governed by the continuity and momentum equations,
\begin{equation}
\nabla \cdot \mathbf{u} = 0,
\label{eq:continuity}
\end{equation}
\begin{equation}
\frac{\partial (\rho\mathbf{u})}{\partial t} + \nabla \cdot \left( \rho\mathbf{u} \otimes \mathbf{u} \right) = -\nabla p + \nabla \cdot \boldsymbol{\tau} + \mathbf{S}_\sigma + \mathbf{S}_g .
\label{eq:momentum}
\end{equation}
where $\rho$ is the fluid density, $\mathbf{u}\equiv (u~v~w)^\mathrm{T}$ is the velocity vector, $t$ denotes time, $p$ is pressure, $\boldsymbol{\tau} = \mu \left(\nabla \mathbf{u} + \nabla \mathbf{u}^{\mathrm{T}} \right)$ is the viscous stress tensor, $\mu$ is the dynamic viscosity,  $\mathbf{S}_\sigma$ is the interfacial force source term, and $\mathbf{S}_g=\rho \mathbf{g}$ is the gravitational source term with $\mathbf{g}$ as the gravitational acceleration.

In addition to the flow equations, we solve the transport of a scalar that is confined to one of the two phases, the carrying phase. Let $S$ denote the support of the carrying phase, equal to one in the carrying phase and zero elsewhere, and let $q$ denote the concentration within the carrying phase. The scalar is represented by the cell-averaged concentration $C=Sq$. On the mesh, these quantities are cell values, defined in Section~\ref{sec:scalar_variables}: $S_i$ is the fraction of cell $i$ occupied by the carrying phase, obtained from the indicator, $q_i$ the phase-averaged concentration and $C_i=S_iq_i$ the cell-averaged concentration. The scalar obeys the conservative transport equation
\begin{equation}
\frac{\partial C}{\partial t} + \nabla \cdot \left(C\mathbf{u}\right) = \nabla\cdot\left(\mathcal{D}\,S\,\nabla q\right) + \dot{m}_C+\dot{m}_{\mathrm{sorp}},
\label{eq:conservative_transport}
\end{equation}
where $\mathcal{D}$ is the molecular diffusivity of the scalar in the carrying phase, $\dot{m}_C$ is a prescribed volumetric source term, with units of $C$ per unit time, and $\dot{m}_{\mathrm{sorp}}$ is the exchange of the scalar with the interface by sorption, the adsorption and desorption of a surfactant; it is located at the interface. The diffusive flux $-\mathcal{D}\,S\,\nabla q$ acts only within the carrying phase and vanishes outside it, so that diffusion does not carry the scalar across the interface. Front-tracking studies of soluble surfactants typically write this flux as $-\mathcal{D}S\nabla C$, with the diffusivity multiplied by the indicator of the carrying phase \citep{Muradoglu2008,Muradoglu2014}. The two forms agree inside the phase, where $S=1$; across the discrete interface, however, where $0<S<1$, $S\nabla C=S^2\nabla q+Sq\nabla S$ contains a flux driven by the phase fraction rather than by the concentration, which moves the scalar towards the cells of small support even when $q$ is uniform. Writing the flux with $\nabla q$ removes it. The diffusivity and the source terms of each test case are stated in Section~\ref{sec:results} and \ref{app:surfactant_sphere}.

We consider two immiscible fluids occupying a computational domain denoted by $\Omega$. The subdomains occupied by phases $\mathrm a$ and $\mathrm b$ are denoted by $\Omega_{\mathrm a}$ and $\Omega_{\mathrm b}$, respectively, so that $\Omega=\Omega_{\mathrm a}\cup\Omega_{\mathrm b}$. The phases are separated by an interface with a constant surface-tension coefficient $\sigma$; in the surfactant calculations of Section~\ref{sec:surfactant_bubble} and \ref{app:surfactant_sphere}, the surfactant does not change $\sigma$.

\section{Numerical framework}
\label{sec:numerics}

The numerical framework couples a Lagrangian representation of the interface to the Eulerian flow and scalar fields. We first describe the front-tracking method, which supplies the interface geometry and interfacial forces. We next define the scalar variables and phase support, before introducing the indicator projection and the transport of the phase-averaged concentration, which restore phase consistency while preserving the phase volume and scalar mass. The spatial and temporal discretization of the flow equations and their coupled solution follow. Finally, these components are brought together in the complete time-step procedure.

\subsection[Front-tracking implementation]{Front-tracking implementation}
\label{sec:ft}

The front-tracking implementation provides the interface geometry used by the phase-concentration transport and the interfacial forces required by the flow solver. The reader is referred to \citet{Gorges2022} for the full implementation. The main steps of front advection, indicator reconstruction, and force spreading are summarized here, while details of remeshing and interface-quality control are given in that reference.

In front tracking \citep{Unverdi1992,Tryggvason2001}, the interface between the two immiscible fluids is represented explicitly by a triangulated Lagrangian surface mesh $\mathcal{F}$. Each front vertex $i$, located at $\mathbf{x}_i$, is advected according to
\begin{equation} \frac{\mathrm{d}\mathbf{x}_i}{\mathrm{d}t}=\overline{\mathbf{u}}(\mathbf{x}_i,t), \label{eq:front_vertex_advection} \end{equation}
where $\overline{\mathbf{u}}(\mathbf{x}_i,t)$ is the fluid velocity interpolated from the Eulerian mesh to the front. The interpolation employs the compact Peskin cosine kernel \citep{Peskin1977,Peskin2003},
\begin{equation} \overline{\mathbf{u}}(\mathbf{x}_i,t)=\sum_{P\in\mathcal{N}_i}\mathbf{u}_P\prod_{\alpha=1}^{3}\mathcal{K}\left(\frac{x_{\alpha,i}-x_{\alpha,P}}{h}\right), \label{eq:front_velocity_interpolation} \end{equation}
where $P$ denotes an Eulerian cell, $\mathbf{u}_P$ is its cell-centred velocity, and $\mathcal{N}_i$ is the compact interpolation stencil associated with vertex $i$. Here, $h$ denotes the cell spacing of the locally uniform Cartesian stencil used in this expression. The one-dimensional kernel is
\begin{equation} \mathcal{K}(r)=\begin{cases} \dfrac{1}{4}\left[1+\cos\left(\dfrac{\pi r}{2}\right)\right], & |r|<2,\\[4pt] 0, & |r|\ge2. \end{cases} \label{eq:peskin_kernel_front_tracking} \end{equation}
The front-vertex trajectories are integrated using the classical fourth-order Runge--Kutta method.

The two phases are identified by the sharp indicator function
\begin{equation}
I(\mathbf{x})=
\begin{cases}
0, & \mathbf{x}\in\Omega_{\mathrm a},\\
1, & \mathbf{x}\in\Omega_{\mathrm b}.
\end{cases}
\label{eq:sharp_phase_indicator}
\end{equation}
The corresponding cell-centred indicator reconstructed from the
tracked front is denoted by $I^{\mathrm{FT}}$. Unlike the sharp
indicator, it varies smoothly between zero and unity across the
numerical interfacial region.

The front normals point from the enclosed phase ($I=1$) towards the ambient phase ($I=0$), so the sharp-indicator jump in the normal direction is $-1$. Each triangle $e$, with unit normal $\mathbf n_e$ and area $A_e$, contributes $-\mathbf n_e A_e$ to the regularized indicator gradient $\mathbf G$. This contribution is spread to neighbouring Eulerian cells using the tensor-product Peskin kernel, with weights normalized to preserve its volume-integrated value. Following \citet{Tryggvason2011a} and \citet{Gorges2022}, the Eulerian indicator is obtained from the Poisson equation
\begin{equation} \nabla^2 I^{\mathrm{FT}}=\nabla\cdot\mathbf{G}. \label{eq:indicator_poisson} \end{equation}
After this equation has been solved, $I^{\mathrm{FT}}$ is bounded to $[0,1]$.

The reconstructed indicator then determines the density and viscosity fields used in the flow equations through
{\color{black}
\begin{equation}
\phi = \phi_\mathrm{a} + I^{\mathrm{FT}} \left( \phi_\mathrm{b}-\phi_\mathrm{a} \right),
\qquad \phi\in\{\rho,\mu\}.
\label{eq:material_interpolation}
\end{equation}
}

Surface tension is evaluated on the Lagrangian triangulation with the Frenet-element formulation of \citet{Gorges2022}, in which the capillary force on a triangular element $e$ is
\begin{equation}
\mathbf{F}_{\sigma,e}=\sigma\sum_{m=1}^{3}l_m\mathbf{m}_m,
\qquad \mathbf{m}_m=\mathbf{n}_m\times\mathbf{t}_m,
\label{eq:frenet_element_force}
\end{equation}
where $l_m$ is the length of edge $m$, $\mathbf{t}_m$ its unit tangent and $\mathbf{n}_m$ the front normal at that edge. The element forces are spread to the Eulerian cells with the Peskin kernel and enter the momentum equation as the source $\mathbf{S}_\sigma$; the details are given by \citet{Gorges2022}.

\subsection[Scalar variables and phase support]{Scalar variables and phase support}
\label{sec:scalar_variables}

{\color{black}
The transported scalar is confined to one of the two fluid phases. We therefore define its \emph{phase support} as the part of the computational domain in which the scalar is defined and allowed to exist. At the cell level, the support $S_i$ represents the fraction of cell $i$ occupied by the scalar-carrying phase. Thus, $S_i=1$ means that the complete cell belongs to the carrying phase, $S_i=0$ means that the carrying phase is absent from the cell, and $0<S_i<1$ denotes an interfacial cell that is only partially occupied by the carrying phase.

Since the phase geometry is already represented by the indicator field $I(\mathbf{x})$, no additional geometrical field is required for the scalar: the scalar support is obtained directly from $I$ according to which phase carries the scalar,
\begin{equation} S(\mathbf{x})=\begin{cases}1-I(\mathbf{x}), & \text{scalar carried by }\Omega_{\mathrm a},\\[2mm] I(\mathbf{x}), & \text{scalar carried by }\Omega_{\mathrm b}.\end{cases} \label{eq:scalar_support_definition}\end{equation}
Since the indicator is reconstructed from the front with the kernel of Eq.~\eqref{eq:peskin_kernel_front_tracking}, $S$ changes from zero to one over about four cells, and a phase-confined scalar is confined to this support rather than to the sharp front.

\begin{figure}[t]
\centering
\includegraphics[width=\textwidth]{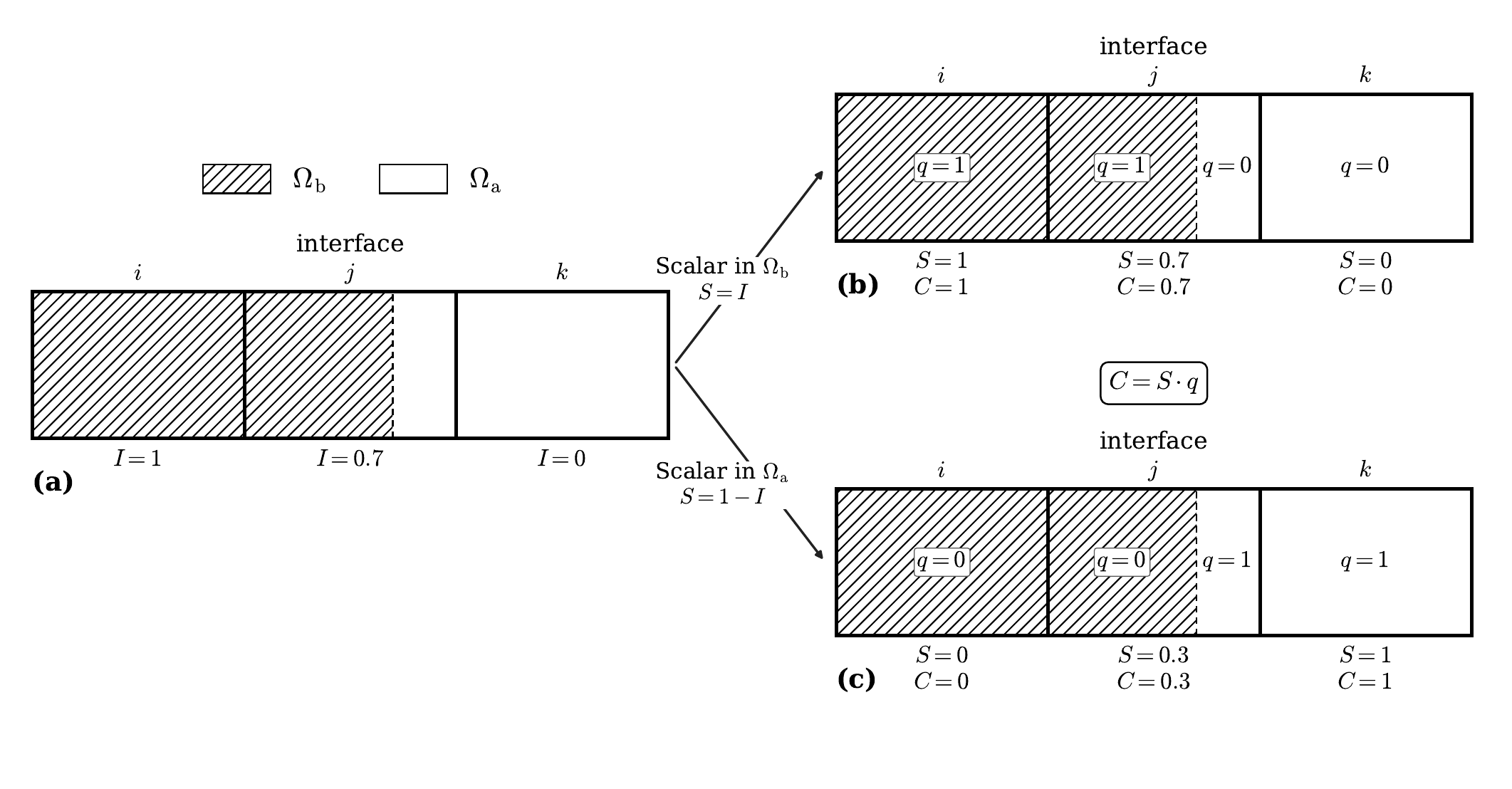}
\caption{Definition of the scalar phase support and its relation to the cell-averaged transported quantity. (a) Indicator field $I$ for three representative cells: a bulk $\Omega_{\mathrm b}$ cell, an interfacial cell, and a bulk $\Omega_{\mathrm a}$ cell. (b) If the scalar is carried by $\Omega_{\mathrm b}$, its support is $S=I$. (c) If the scalar is carried by $\Omega_{\mathrm a}$, its support is $S=1-I$. The example uses $q=1$ in the scalar-carrying phase and $q=0$ in the opposite phase only to illustrate the relation $C=S q$; in general, $q$ is not restricted to these values.}
\label{fig:scalar_support_definition}
\end{figure}

Figure~\ref{fig:scalar_support_definition} illustrates this construction using three adjacent computational cells, labelled $i$, $j$, and $k$. In Fig.~\ref{fig:scalar_support_definition}(a), the interface cuts the middle cell $j$, separating phase $\Omega_{\mathrm b}$ on the left (hatched region) from phase $\Omega_{\mathrm a}$ on the right (unhatched region). Cell $i$ lies entirely in $\Omega_{\mathrm b}$, whereas cell $k$ lies entirely in $\Omega_{\mathrm a}$. In cell $j$, the two phases occupy $70\%$ and $30\%$ of the cell volume, respectively, giving the indicator values $I_i=1$, $I_j=0.7$, and $I_k=0$. A scalar carried by $\Omega_{\mathrm b}$ therefore has $S_i=1$, $S_j=0.7$, and $S_k=0$, as shown in Fig.~\ref{fig:scalar_support_definition}(b). If the scalar is instead carried by $\Omega_{\mathrm a}$, the same indicator field gives $S_i=0$, $S_j=0.3$, and $S_k=1$ because $S=1-I$, as illustrated in Fig.~\ref{fig:scalar_support_definition}(c). The detail in Fig.~\ref{fig:middle_cell_volume} highlights the middle cell for the case in Fig.~\ref{fig:scalar_support_definition}(b): let $V_{j,q}$ denote the volume occupied by the scalar-carrying phase and $V_j$ the total cell volume; thus, $V_{j,q}=S_jV_j=0.7V_j$. The support therefore identifies both \emph{where} the scalar may exist and, in an interfacial cell, the fraction of the cell volume available to it.

Let $m_i$ denote the mass of transported scalar contained in cell $i$, and let $V_i$ denote the total cell volume. Since the scalar is restricted to the carrying phase, the volume available to it in cell $i$ is $S_iV_i$. The concentration within the carrying phase is therefore
\begin{equation}
q_i=\frac{m_i}{V_{i,q}}=\frac{m_i}{S_iV_i},
\qquad S_i>0.
\label{eq:phase_concentration_definition}
\end{equation}
Thus, $q_i$ is the \emph{phase-averaged concentration}: it represents the concentration within the scalar-carrying fluid itself, averaged over its portion of the cell, as in phase-specific species transport in VOF methods \citep{Haroun2010a,Marschall2012,Deising2016,Tourbier2024}; in the terminology of volume averaging, it is the intrinsic phase average. Its denominator is the actual carrying-phase volume $S_iV_i$, not the full computational-cell volume. In contrast, the finite-volume transported quantity $C_i$ is the \emph{cell-averaged concentration}, defined using the complete cell volume,
\begin{equation} C_i=\frac{m_i}{V_i}. \label{eq:cell_averaged_scalar_definition}\end{equation}
Both $q_i$ and $C_i$ are masses per unit volume, but they use different averaging volumes: carrying-phase volume for $q_i$ and total cell volume for $C_i$. The same scalar mass must be recovered from either representation, $m_i=q_iS_iV_i=C_iV_i$, and therefore
\begin{equation} C_i=S_iq_i. \label{eq:scalar_amount_definition}\end{equation}
Equation~\eqref{eq:scalar_amount_definition} shows directly the role of the phase support. The phase-averaged concentration $q_i$ specifies how much scalar is present per unit volume of the carrying phase, whereas multiplication by $S_i$ converts this quantity into the concentration averaged over the whole computational cell. In the following, ``the scalar'' refers to the cell-averaged concentration $C$, the field that is transported and whose mass is conserved, and ``the concentration'' on its own to the phase-averaged concentration $q$.

\begin{figure}[htbp]
\centering
\includegraphics[width=0.32\textwidth]{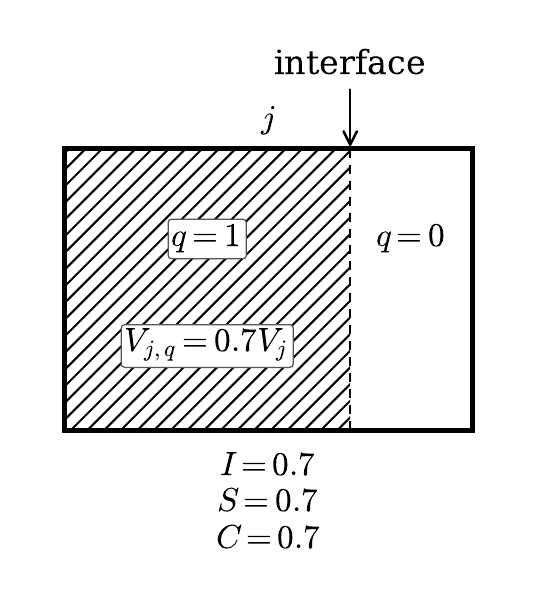}
\caption{Detail of the interfacial cell $j$ in Fig.~\ref{fig:scalar_support_definition}(b). The scalar-carrying phase, phase $\mathrm b$, occupies the volume $V_{j,q}=S_jV_j=0.7V_j$, whereas $V_j$ denotes the whole cell volume. The phase-averaged concentration $q_j=m_j/V_{j,q}$ and the cell-averaged concentration $C_j=m_j/V_j$ therefore use different averaging volumes. For the illustrative value $q_j=1$, this gives $C_j=0.7$.}
\label{fig:middle_cell_volume}
\end{figure}

For example, in the interfacial cell $j$ of Fig.~\ref{fig:scalar_support_definition}(b), the carrying phase occupies $V_{j,q}=0.7V_j$. The two averaging volumes give $q_j=m_j/V_{j,q}$ and $C_j=m_j/V_j$, so $C_j=(V_{j,q}/V_j)q_j=S_jq_j$. With the illustrative phase-averaged concentration $q_j=1$, the cell-averaged concentration is $C_j=0.7\times1=0.7$. In Fig.~\ref{fig:scalar_support_definition}(c), the other phase carries the scalar and occupies only $0.3V_j$, giving $C_j=0.3\times1=0.3$ for the same $q_j=1$. In both cases, the concentration within the carrying fluid remains unity; the smaller value of $C_j$ reflects averaging over the entire cell, not dilution within the carrying phase.

For a cell completely occupied by the carrying phase, $S_i=1$ and therefore $C_i=q_i$. In an interfacial cell, $0<S_i<1$, only a fraction of the cell is available to the scalar and consequently $C_i=S_iq_i<q_i$ for $q_i>0$. The scalar mass is finally recovered as $m_i=C_iV_i=S_iq_iV_i$.
}

\subsection[Discretization of the scalar transport]{Discretization of the scalar transport}
\label{sec:scalar_transport_discretisation}

Integrating the advective part of Eq.~\eqref{eq:conservative_transport} over cell $i$ and applying the divergence theorem gives the semi-discrete finite-volume equation
\begin{equation}
V_i\frac{\mathrm{d}C_i}{\mathrm{d}t}
+\sum_{f\in\partial V_i}F_fC_f
=0,
\label{eq:scalar_transport_discrete}
\end{equation}
where $C_i$ is the cell-averaged concentration, $F_f=(\mathbf{u}_f\cdot\mathbf{n}_f)A_f$ is the outward volumetric face flux, and $C_f$ is the face value obtained with the selected advection scheme, the WENO-Z reconstruction of Section~\ref{sec:wenoz} in this work. The advective operator is integrated with the three-stage strong-stability-preserving Runge--Kutta method of \citet{Shu1988} (see also \citet{Gottlieb1998}). Applied directly to $C$, this conservative form is referred to as the standard method. In the proposed method the scalar is instead transported through its phase-averaged concentration, as described in Section~\ref{sec:scalar_correction}.

Each time step is split into the advection, integrated over the three Runge--Kutta stages, and the source, sorption and diffusion terms, which are applied once per step in this order for both methods. The prescribed source is added first: in the standard method to the advected scalar, $\Delta t\,\dot{m}_{C,i}$, and in the proposed method to the rebuilt scalar before the mass projection, whose target mass includes it (Section~\ref{sec:scalar_correction}). The exchange with the interface and the diffusion then act on the resulting scalar $\tilde C$: in the standard method the advected scalar with the source, in the proposed method the rebuilt scalar of Eq.~\eqref{eq:scalar_projection_shift}. With $\tilde q=\tilde C/S$, the exchange and diffusion update giving the scalar at the new time level is
\begin{equation}
C_i^{n+1}=\tilde C_i+\Delta t\,\dot{m}_{\mathrm{sorp},i}+\frac{\Delta t}{V_i}\sum_{f\in\partial V_i}\mathcal{D}_fA_f\,\frac{\tilde q_j-\tilde q_i}{\Delta s_f},
\qquad
\mathcal{D}_f=\frac{2\,(\mathcal{D}S_i)(\mathcal{D}S_j)}{\mathcal{D}S_i+\mathcal{D}S_j},
\label{eq:scalar_diffusion_discrete}
\end{equation}
where $j$ is the neighbour of cell $i$ across face $f$, $\Delta s_f$ is the distance between the two cell centres, and $S$ is the support of the scalar. As in \citet{Tourbier2024}, the diffusive flux is driven by the difference of the phase-averaged concentration rather than of the cell-averaged concentration: in a cell cut by the interface, $C$ is smaller than in a neighbouring bulk cell even when $q$ is uniform, and a flux based on $C$ would diffuse the scalar artificially. The face diffusivity is the harmonic mean \citep{Patankar1980} of the phase-weighted diffusivities $\mathcal{D}S$, which vanishes when either cell lies outside the carrying phase; faces next to cells whose support falls below $10^{-9}$ carry no diffusive flux, so the scalar does not diffuse beyond its support. Because the flux is driven by $\tilde q=\tilde C/S$, the diffusion update keeps the scalar non-negative if $\Delta t\sum_f\mathcal{D}_fA_f/\Delta s_f\le S_iV_i$ in every cell. As the harmonic mean gives $\mathcal{D}_f\le2\mathcal{D}S_i$, a sufficient condition on a uniform mesh of spacing $h$ is $\mathcal{D}\Delta t/h^2\le1/12$. This is half the stability limit of the explicit scheme for a uniform diffusivity \citep{OBrien1950}. The condition guarantees a non-negative scalar for any field; above it, negative values can appear only at a sharp local maximum of the concentration. It is exceeded in the sphere of \ref{app:surfactant_sphere} at $D/h=40$ ($0.128$) and in the first two steps of the surfactant bubble at $D/h=40$ ($0.16$), while its concentration is still uniform; the scalar remained non-negative in every step of every calculation. The sorption term is explicit as well and keeps the scalar non-negative as long as it removes no more than the scalar present in a cell.

\subsection[WENO-Z face reconstruction]{WENO-Z face reconstruction}
\label{sec:wenoz}

\begin{figure}[htbp]
\centering
\includegraphics[width=\textwidth]{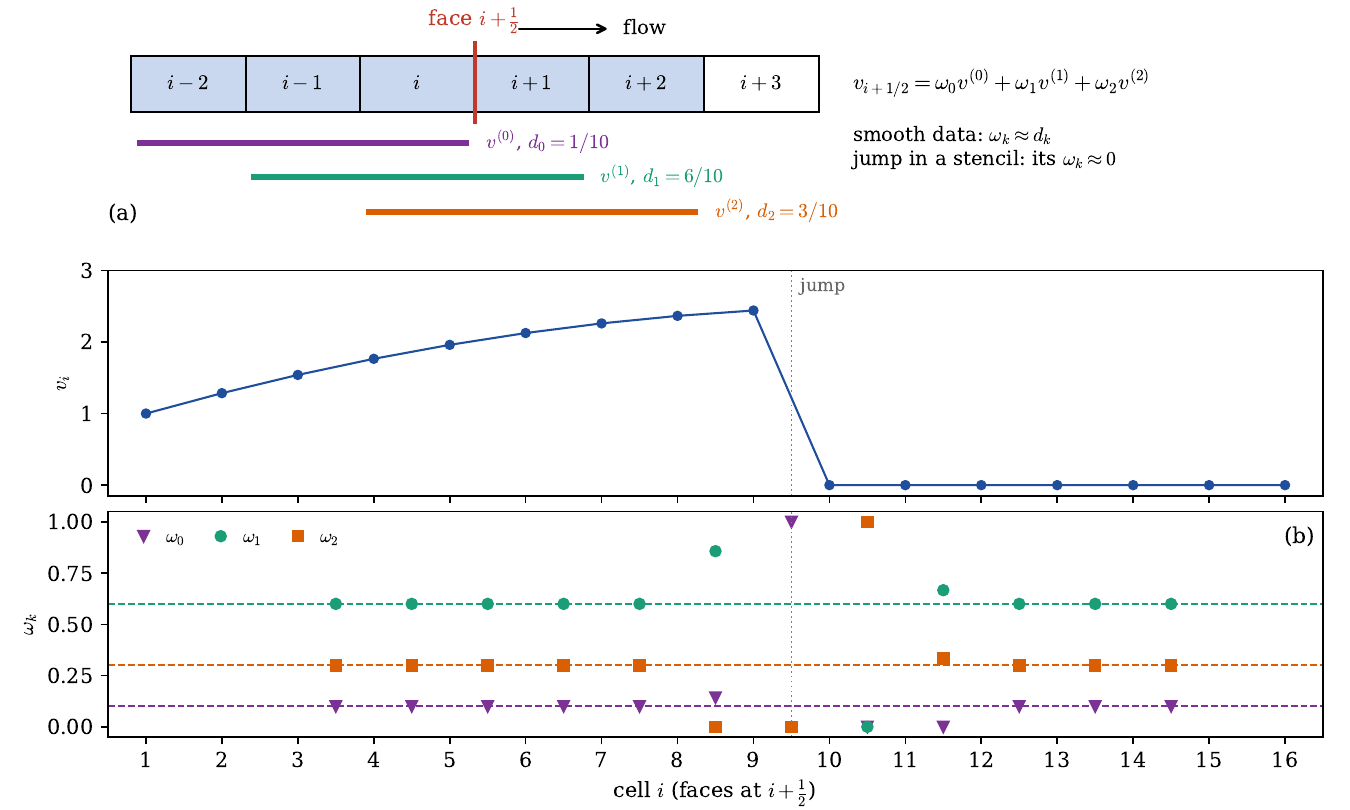}
\caption{WENO-Z reconstruction: (a) the five-cell stencil of the face $i+\tfrac12$, its three candidate stencils and their linear weights; (b) cell values with a jump at the face $9\tfrac12$ (top) and the weights $\omega_k$ at the faces $i+\tfrac12$ for a flow towards increasing $i$ (bottom), with the linear weights dashed.}
\label{fig:weno_z}
\end{figure}

The face values are reconstructed with the fifth-order WENO-Z scheme of \citet{Borges2008}, a variant of the weighted essentially non-oscillatory (WENO) scheme of \citet{Jiang1996}. A WENO scheme combines several low-order reconstructions so that it is of high order where the data are smooth and does not oscillate at steep gradients. For the face $i+\tfrac12$ of a grid line, with the flow from cell $i$ to cell $i+1$, WENO-Z uses the five cells $i-2,\dots,i+2$. They contain three candidate stencils of three cells each (Fig.~\ref{fig:weno_z}(a)), and on each of them the quadratic polynomial with the given cell averages $v$ gives a third-order face value,
\begin{equation}
v^{(0)}=\frac{2v_{i-2}-7v_{i-1}+11v_i}{6},\qquad
v^{(1)}=\frac{-v_{i-1}+5v_i+2v_{i+1}}{6},\qquad
v^{(2)}=\frac{2v_i+5v_{i+1}-v_{i+2}}{6}.
\label{eq:weno_candidates}
\end{equation}
With the linear weights $d_0=1/10$, $d_1=6/10$ and $d_2=3/10$, the combination $\sum_kd_kv^{(k)}$ is the fifth-order reconstruction from all five cells. WENO-Z replaces them by nonlinear weights,
\begin{equation}
v_{i+1/2}=\sum_{k=0}^{2}\omega_kv^{(k)},\qquad
\omega_k=\frac{\alpha_k}{\sum_{l=0}^{2}\alpha_l},\qquad
\alpha_k=d_k\left[1+\left(\frac{\tau_5}{\beta_k+\epsilon}\right)^2\right],\qquad
\tau_5=|\beta_0-\beta_2|,
\label{eq:wenoz_weights}
\end{equation}
where the smoothness indicators of \citet{Jiang1996},
\begin{equation}
\begin{aligned}
\beta_0&=\tfrac{13}{12}(v_{i-2}-2v_{i-1}+v_i)^2+\tfrac14(v_{i-2}-4v_{i-1}+3v_i)^2,\\
\beta_1&=\tfrac{13}{12}(v_{i-1}-2v_i+v_{i+1})^2+\tfrac14(v_{i-1}-v_{i+1})^2,\\
\beta_2&=\tfrac{13}{12}(v_i-2v_{i+1}+v_{i+2})^2+\tfrac14(3v_i-4v_{i+1}+v_{i+2})^2,
\end{aligned}
\label{eq:weno_smoothness}
\end{equation}
measure the variation of the data on each candidate stencil, and $\epsilon$, $10^{-12}$ times the largest squared value on the five cells, only prevents a division by zero. Where the data are smooth, the three $\beta_k$ are nearly equal, $\tau_5$ is much smaller than each of them, and the weights approach the linear ones, so that the reconstruction is fifth-order accurate. Where a candidate stencil contains a jump, its $\beta_k$ is large and its weight nearly vanishes, and the face value is taken from the candidate stencils on one side of the jump. Figure~\ref{fig:weno_z}(b) shows this for data with a jump: away from the jump the weights equal the linear ones, and at the faces whose five-cell stencils reach across it the weight moves to the candidate stencils that do not. At the face $10\tfrac12$, for instance, the linear weights would give the face value $-0.45$ between the cell values $0$ and $0$, whereas WENO-Z gives $0$. The scheme avoids such oscillations at the price of a lower order next to the jump and of numerical diffusion, which smears the jump over a few cells. The reconstruction is the same in all three directions; for a flow in the negative direction the stencil is mirrored, and a face whose five-cell stencil does not lie on one uniform level of the adaptive mesh takes the upwind cell value.

\subsection[Volume-preserving indicator projection for front tracking]
{Volume-preserving indicator projection for front tracking}
\label{sec:ft_support_projection}

The scalar and the tracked interface are advanced through different numerical representations. The scalar $C$ is transported on the Eulerian mesh, whereas the phase geometry is obtained from the Lagrangian front. Front tracking does not conserve the phase volume exactly. Front advection and remeshing can change the enclosed volume. In addition, the cell sum of $I^{\mathrm{FT}}$ is not exactly the volume enclosed by the front: it depends on the position of the front relative to the cells, and it changes again when $I^{\mathrm{FT}}$ is bounded to $[0,1]$ after Eq.~\eqref{eq:indicator_poisson} is solved. For the constant phase densities considered here, a change in phase volume also implies a change in phase mass. We therefore define a bounded corrected indicator $I^c$ that remains as close as possible to the front-derived indicator $I^{\mathrm{FT}}$ while preserving the phase volume, and hence phase mass. Its role is to provide a conservative updated phase support for the subsequent scalar transport. If $I^{\mathrm{FT}}$ already preserves the phase volume and satisfies the required bounds, the entire volume-preserving indicator projection is unnecessary: we set $I^c=I^{\mathrm{FT}}$ and use the front-derived support directly.

Initially, $I^c=I^{\mathrm{FT}}$. At each subsequent time step, the projection must preserve the total phase volume of the corrected indicator of the previous time step, $I^{c,n}$. Let $V_i$ denote the volume of cell $i$. In the discrete sums and cell sets below, $i\in\Omega$ denotes a cell index in the mesh discretizing the physical domain $\Omega$. The target phase volume, denoted by $M_I^{\mathrm{target}}$, is
\begin{equation}
M_I^{\mathrm{target}}=\sum_{i\in\Omega}I_i^{c,n}V_i.
\label{eq:indicator_target_volume}
\end{equation}

\paragraph{Bounded indicator projection}

The objective of this projection is to recover the target phase volume while keeping the corrected indicator as close as possible to the front-derived indicator. The result is a corrected indicator bounded between zero and unity, whose volume integral matches the target and whose spatial distribution remains close to that defined by the tracked front. This field supplies the updated phase support for the subsequent scalar transport.

To retain the pure-phase regions defined by the front, we initially keep the corrected indicator at zero in cells classified as pure phase zero and at unity in cells classified as pure phase one. Only mixed cells are initially allowed to accommodate the difference between the front-derived and target phase volumes. A tolerance $\varepsilon^{\mathrm{pure}}$ specifies how close an indicator value must be to zero or unity to be treated as pure. We denote the cells classified as phase zero by $\Omega_0$, those classified as phase one by $\Omega_1$, and the remaining mixed cells by $\Omega_m$:
\begin{equation}
\begin{aligned}
\Omega_0&=\{i\in\Omega:I_i^{\mathrm{FT}}
\le\varepsilon^{\mathrm{pure}}\},\\
\Omega_1&=\{i\in\Omega:I_i^{\mathrm{FT}}
\ge1-\varepsilon^{\mathrm{pure}}\},\\
\Omega_m&=\{i\in\Omega:\varepsilon^{\mathrm{pure}}
<I_i^{\mathrm{FT}}<1-\varepsilon^{\mathrm{pure}}\}.
\end{aligned}
\label{eq:indicator_cell_sets}
\end{equation}

If the mixed cells cannot accommodate the target volume within the bounds, further cells are made adjustable: if the target is smaller than the volume of the cells in $\Omega_1$, the cells of $\Omega_1$ next to the mixed cells are released first and, if they do not suffice, all cells of $\Omega_1$; if the target exceeds what the mixed cells can hold, all cells are released. The projection then has the same form on the enlarged set (\ref{app:indicator_expansion}). In each mixed cell, the corrected indicator must remain between zero and unity so that it represents an admissible phase volume fraction. Among the fields satisfying these bounds and the target volume, we select the one closest to the front-derived field, which remains unchanged as the reference. Closeness is measured by the sum of squared differences between the corrected and front-derived indicator values over the mixed cells, weighted by cell volume. The projection is therefore defined by the constrained minimization problem
\begin{equation}
\begin{aligned}
\underset{I^c}{\operatorname{minimize}}\quad
&\frac12\sum_{i\in\Omega_m}(I_i^c-I_i^{\mathrm{FT}})^2V_i,\\
\text{subject to}\quad
&\sum_{i\in\Omega}I_i^cV_i=M_I^{\mathrm{target}},
\qquad 0\le I_i^c\le1\quad(i\in\Omega_m),\\
&I_i^c=0\quad(i\in\Omega_0),
\qquad I_i^c=1\quad(i\in\Omega_1).
\end{aligned}
\label{eq:indicator_projection_minimisation}
\end{equation}

The Karush--Kuhn--Tucker conditions \cite{Nocedal2006} give a simple form for this minimizer: the same constant shift is added to the front-derived value of every mixed cell, after which each result is clipped to $[0,1]$. We denote this shift by $\lambda_I$; it is the volume-constraint multiplier, with the sign convention used below, and is chosen to recover the target volume. Clipping leaves a value inside its interval unchanged and replaces a value outside it by the nearest bound. We denote this operation on a value $x$ over an interval $[a,b]$ by $\mathcal{C}_{[a,b]}(x)$; the corrected field is then
\begin{equation}
I_i^c(\lambda_I)=\begin{cases}
0,&i\in\Omega_0,\\
1,&i\in\Omega_1,\\
\mathcal{C}_{[0,1]}(I_i^{\mathrm{FT}}+\lambda_I),&i\in\Omega_m,
\end{cases}
\qquad \mathcal{C}_{[a,b]}(x)=\min(\max(x,a),b).
\label{eq:indicator_general_projection}
\end{equation}

It remains to choose $\lambda_I$ so that the corrected field has the required total phase volume. Increasing the shift cannot decrease any corrected indicator value, although values that reach unity stop increasing. The total volume is therefore a continuous, non-decreasing function of the shift. If the target lies within the attainable volume range, the multiplier is obtained by a safeguarded semismooth Newton iteration \citep{Qi1993} between two shifts whose volumes bracket the target. \ref{app:correction_implementation} gives the numerical solution procedure and tolerances.

The resulting indicator is bounded and preserves the target phase volume:
\begin{equation}
0\le I_i^c\le1,
\qquad \sum_{i\in\Omega}I_i^cV_i=M_I^{\mathrm{target}}.
\label{eq:indicator_projection_properties}
\end{equation}
The volume equality is satisfied to the prescribed numerical tolerance. The support obtained from this indicator is used by the phase-concentration transport in the following subsection. \ref{app:correction_implementation} gives the release of further cells and a guard that tightens the bounds where the projection would create a jump from zero to unity between neighbouring cells; neither changes the target phase volume.

\subsection[Phase-concentration transport and conservative rebuild]{Phase-concentration transport and conservative rebuild}
\label{sec:scalar_correction}

The indicator projection supplies, at every time step, a bounded phase support that follows the tracked front and preserves the phase volume. The scalar itself must be transported onto this support. The cell-averaged concentration $C=Sq$ is a poor candidate for transport: across the interfacial cells it falls from $q$ to zero together with the support, and a high-order reconstruction of this jump broadens the transported profile relative to the front-derived support, as shown in Section~\ref{sec:motivation}. For such a jump, which the flow carries without steepening it, the $L_1$ error of a capturing scheme of order $p$ decreases only as $h^{p/(p+1)}$, where $h$ is the cell size, below first order for every $p$ \citep{Banks2008}. The phase-averaged concentration $q$ does not share this jump behaviour. Within the carrying phase it varies only with the physical distribution of the scalar, and the interfacial variation of $C$ is carried entirely by $S$. We therefore advect $q$ rather than $C$, and recover $C$ from the corrected support after every advection step.

Substituting $C=Sq$ into Eq.~\eqref{eq:conservative_transport} and using the fact that the phase support is a material quantity of the incompressible flow, $\partial S/\partial t+\mathbf{u}\cdot\nabla S=0$, gives within the carrying phase
\begin{equation}
S\left(\frac{\partial q}{\partial t}+\mathbf{u}\cdot\nabla q\right)=\nabla\cdot\left(\mathcal{D}\,S\,\nabla q\right)+\dot{m}_C+\dot{m}_{\mathrm{sorp}}.
\label{eq:phase_concentration_full}
\end{equation}
Each time step is split into an advection step and a subsequent update by the source, diffusion and the exchange with the interface (Section~\ref{sec:scalar_transport_discretisation}). The advection step solves
\begin{equation}
\frac{\partial q}{\partial t}+\mathbf{u}\cdot\nabla q=0,
\label{eq:phase_concentration_transport}
\end{equation}
so that the phase-averaged concentration is only carried with the flow; its result is the advected concentration $q^\ast$. The prescribed source $\dot{m}_C$ is then added to the scalar rebuilt from $q^\ast$, and the exchange with the interface and the diffusion act on the rebuilt scalar $\tilde C$, in the order of Section~\ref{sec:scalar_transport_discretisation}. Outside the carrying phase, $q$ has no direct physical meaning; it is required only by the reconstruction stencils of the cells next to the interface, and is supplied there by an extension of $q$ across the interface.

Each time step from $t^n$ to $t^{n+1}$ consists of five operations. The phase-averaged concentration is first recovered from the scalar where the support is resolved and then extended into the neighbouring cells outside the carrying phase. The extended field is advected over the time step. The scalar is then rebuilt on the corrected support of the new time level, and a non-negative mass projection finally restores the scalar mass. Because $C=S^{n+1}q$ holds by construction after the rebuild, the scalar cannot be placed outside the corrected support.

The individual steps are described below, and Algorithm~\ref{alg:scalar_correction} at the end of this section summarizes the procedure.

\paragraph{Recovery of the phase-averaged concentration}

Let $C^n$ and $S^n=S(I^{c,n})$ denote the scalar and the corrected support at the start of the time step; initially, $S^0=S(I^{\mathrm{FT}})$. Before the scalar is transported, we record the scalar mass that the corrected field must carry,
\begin{equation}
M_C^n=\sum_{i\in\Omega}C_i^nV_i,
\label{eq:scalar_start_mass}
\end{equation}
retaining the signed cell values. The phase-averaged concentration is recovered by division by the support only where the support is resolved. With a threshold $\varepsilon^{\mathrm{rec}}$, the resolved set and the recovered concentration are
\begin{equation}
\mathcal{R}^n=\{i\in\Omega:S_i^n>\varepsilon^{\mathrm{rec}}\},\qquad
q_i^n=\frac{\max\!\left(C_i^n,0\right)}{S_i^n},\quad i\in\mathcal{R}^n.
\label{eq:recovered_concentration}
\end{equation}
The threshold prevents division by small supports. In cells outside $\mathcal{R}^n$, the phase-averaged concentration is not recovered but extended, as described next. Unless stated otherwise, $\varepsilon^{\mathrm{rec}}=0.03$.

\paragraph{Extension across the interface}

The WENO-Z reconstruction at a face uses a five-cell stencil, so the face values of a resolved cell next to the interface require concentrations up to three cells beyond it. These cells lie outside the carrying phase or in its thin interfacial region, where no meaningful concentration can be recovered. We therefore extend $q$ across the interface. The same need arises in ghost-fluid methods, where quantities defined on one side of the interface are extrapolated into the other \citep{Fedkiw1999,Aslam2004}. \citet{Tanguy2007}, for instance, extrapolate the liquid temperature and the vapour mass fraction of an evaporating droplet linearly in the normal direction, by the partial differential equations of \citet{Aslam2004}, so that the convective terms can be computed without smearing these fields across the interface; see also \citet{RuedaVillegas2016}. The extension used here serves the same purpose, but it is algebraic: it extrapolates linearly along the grid lines over a fixed number of cell layers, and therefore requires neither a normal field on the mesh nor the iteration of extrapolation equations to a steady state.

\begin{figure}[htbp]
\centering
\includegraphics[width=\textwidth]{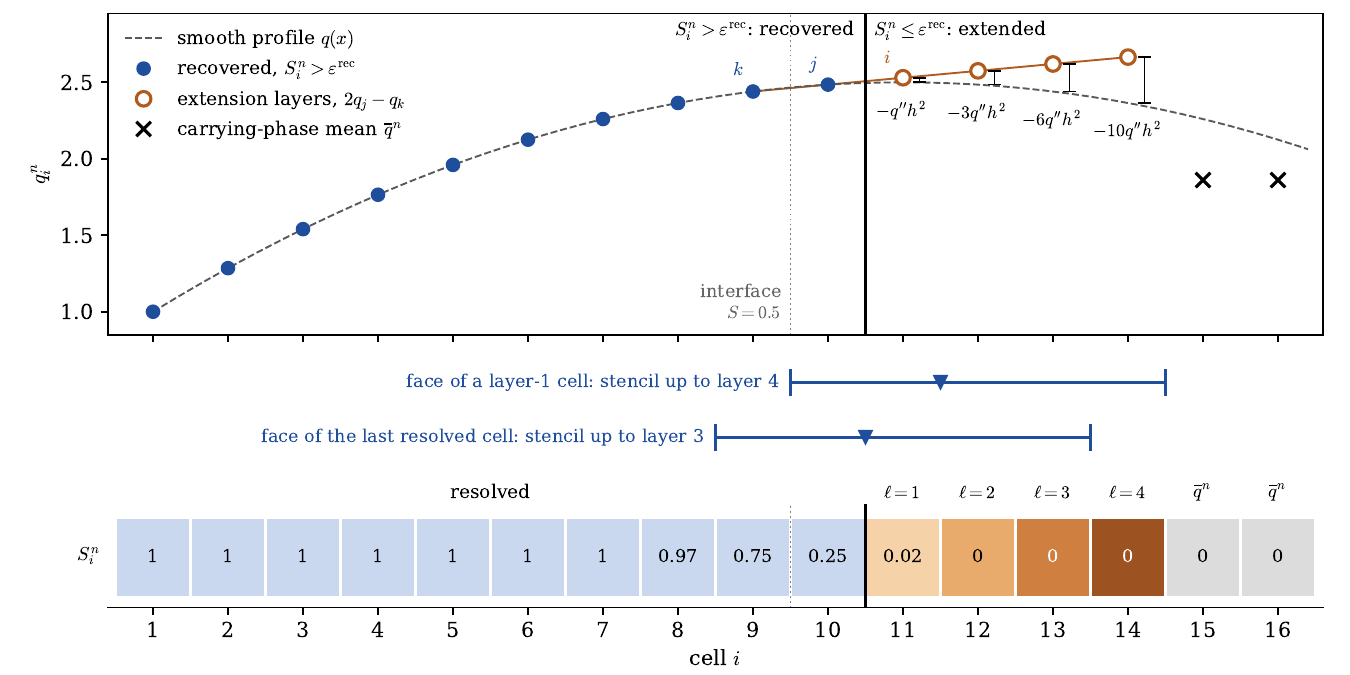}
\caption{Extension of the phase-averaged concentration along one grid line, for the support written in the cells (bottom) and the curved concentration $q(x)=1+0.3(x-1)-0.015(x-1)^2$, with $h=1$. Top: concentration recovered in the resolved cells, $S_i^n>\varepsilon^{\mathrm{rec}}$ (filled circles), the four extension layers (open circles), which lie on the straight line through the last two recovered values $q_k$ and $q_j$, and the carrying-phase mean concentration beyond the last layer (crosses). The bars give the difference between the extended values and the smooth profile, $-\tfrac12\ell(\ell+1)q''h^2$ in layer $\ell$. The solid vertical line separates the resolved cells from the others: the extension starts at the threshold $\varepsilon^{\mathrm{rec}}$, not at the interface $S=0.5$ (dotted line), so that the interfacial cells $8$--$10$, with $S_i^n$ down to $0.25$, are recovered, and cell $11$, with $S_i^n=0.02$, is extended. Bottom: the support $S_i^n$ of each cell, the resolved cells, the extension layers and the cells that receive the carrying-phase mean, with the five-cell WENO-Z stencils of the face of the last resolved cell and of the face of a first-layer cell (triangles) for a flow towards the carrying phase; they reach the third and the fourth layer.}
\label{fig:extension_layers}
\end{figure}

The extension proceeds layer by layer from the resolved cells $\mathcal{R}^n$. The first layer consists of the unknown cells that share a face with a resolved cell, the second of the unknown cells that share a face with the first layer, and so on, up to layer $N_\ell$. For a cell $i$ of a layer, every grid direction $d$ in which its face neighbour $j$ is already known gives an estimate $e_{i,d}$ of $q_i$, the continuation of the straight line through $j$ and the next cell $k$ beyond it,
\begin{equation}
e_{i,d}=2q_j-q_k,
\label{eq:concentration_extension}
\end{equation}
or $e_{i,d}=q_j$ if $k$ is not known, and $q_i$ is the average of $e_{i,d}$ over these directions. A cell for which no grid line provides a known neighbour, such as a cell at a refinement transition, takes the mean of its known face neighbours. Each layer is computed from the values of the previous layers only, and the parallel processes exchange the values of their boundary cells between two layers, so that the extended field does not depend on the order in which the cells are visited or on the domain decomposition. The extension is computed once per time step, not per Runge--Kutta stage, and costs $N_\ell$ sweeps over the cells. Figure~\ref{fig:extension_layers} shows the layers along one grid line. In several dimensions, a cell next to a curved interface often has known neighbours in more than one direction and takes the average of the corresponding continuations.

Along a grid line with uniform spacing $h$, Eq.~\eqref{eq:concentration_extension} continues the straight line through the last two known values, so that layer $\ell$ receives $q_j+\ell\,(q_j-q_k)$. A linear profile is therefore continued exactly. For a smooth profile, the error in layer $\ell$ is, to leading order, $-\tfrac12\ell(\ell+1)\,q''h^2$, that is $-q''h^2$, $-3q''h^2$, $-6q''h^2$ and $-10q''h^2$ in the four layers, where $q''$ is the second derivative along the line (Fig.~\ref{fig:extension_layers}, top). The first layer, which is closest to the resolved cells and enters the most reconstruction stencils, is second-order accurate, in agreement with the second-order convergence of linear extrapolation reported by \citet{Aslam2004}. Constant extrapolation, $q_i=q_j$, is only first-order accurate and introduces a kink into the profile at the interface. Quadratic extrapolation is third-order accurate for smooth profiles, but it amplifies the curvature of concentration structures next to the interface, so that the extended values can leave the range of the recovered ones. In several dimensions, every direction continues a linear field exactly, since the three cell centres involved are equally spaced on a grid line, and so does the average over the directions if each of them has two known cells. The constant estimate $e_{i,d}=q_j$ for a missing second cell and the mean of the face neighbours are exact only for a uniform concentration. In the vortex and the source of Section~\ref{sec:results}, on average over the stored fields at most $1.2\%$ of the extended cells use the constant estimate, each in a single direction, and none takes the mean of its face neighbours.

With $N_\ell=4$ layers, the extension covers the five-cell stencils of the faces of the resolved cells, which reach three cells beyond them, and the fourth layer covers the stencils of the cells of the first layer, which the interface can enter during the step (Fig.~\ref{fig:extension_layers}, bottom). The cells that remain unknown after the last layer lie outside every stencil that reaches a resolved cell, but the indicator reconstructed from the front carries small positive values far from the interface, and the scalar held there must still be represented by a meaningful concentration. These cells are therefore assigned the mean concentration of the carrying phase
\begin{equation}
\overline{q}^n=\frac{\max\!\left(0,\sum_{i\in\Omega}C_i^nV_i\right)}{\sum_{i\in\Omega}S_i^nV_i}.
\label{eq:phase_mean_concentration}
\end{equation}

\paragraph{Advection of the phase-averaged concentration}

The extended concentration is advanced according to the advective form of Eq.~\eqref{eq:phase_concentration_transport}. Written with the same volumetric face fluxes as Eq.~\eqref{eq:scalar_transport_discrete}, its semi-discretized form is
\begin{equation}
V_i\frac{\mathrm{d}q_i}{\mathrm{d}t}+\sum_{f\in\partial V_i}F_fq_f-q_i\sum_{f\in\partial V_i}F_f=0,
\label{eq:concentration_advection_discrete}
\end{equation}
where $q_f$ is the WENO-Z face value of the extended field. The last term is the cell value multiplied by the net outward volume flux of the cell. It vanishes for a discretely divergence-free flux field, and otherwise ensures that a uniform concentration remains exactly uniform, independently of the residual divergence left by the iterative flow solution. The equation is integrated with the same three-stage Runge--Kutta method as the scalar; the extension is computed once at the start of the step, and every cell, including the extended ones, is advanced. No phase support is transported. We denote the advected concentration by $q^\ast$.

\paragraph{Rebuild on the corrected support}

At the end of the step, the indicator projection of Section~\ref{sec:ft_support_projection} gives the corrected indicator $I^{c,n+1}$, whose phase volume equals that of $I^{c,n}$, and hence the corrected support $S^{n+1}=S(I^{c,n+1})$. The provisional scalar, i.e.\ the rebuilt scalar before its mass is corrected, is obtained by adding the volumetric source of the step to the non-negative advected concentration and multiplying the result by this support,
\begin{equation}
C_i^{\mathrm{prov}}=\begin{cases}
S_i^{n+1} q_i^{\mathrm{prov}}, & S_i^{n+1}>0,\\
0, & S_i^{n+1}=0,
\end{cases}
\qquad q_i^{\mathrm{prov}}=\max\!\left(q_i^\ast,0\right)+\frac{\Delta t\,\dot{m}_{C,i}}{S_i^{n+1}}.
\label{eq:provisional_reattached_scalar}
\end{equation}
The source thus enters the provisional concentration, so that the mass projection below keeps the scalar as close as possible to the advected concentration plus the source. The target mass of the scalar is the step-start mass increased by the same source contribution,
\begin{equation}
M_C^{\mathrm{target}}=M_C^n+\Delta t\sum_{i\in\Omega}\dot{m}_{C,i}V_i.
\label{eq:scalar_target_mass}
\end{equation}
The provisional field is non-negative and vanishes wherever the corrected carrying phase is absent. Its total mass need not equal $M_C^{\mathrm{target}}$: the advective form of Eq.~\eqref{eq:concentration_advection_discrete} does not conserve $\sum_iC_iV_i$ by construction, and the extension and the change of support from $S^n$ to $S^{n+1}$ also change the mass. The remaining global discrepancy is removed by the mass projection below.

The extended values serve only as input to the face reconstruction. Because the scalar is rebuilt as $S^{n+1}q^\ast$, they can affect only cells that contain the carrying phase at the new time level. This happens in three places: (i) in the face values of the cells next to the interface, which is their purpose; (ii) at the rear of a moving drop, where the flow carries them into the cells the drop is leaving; and (iii) in cells at the very edge of the phase, $0<S_i^{n+1}\le\varepsilon^{\mathrm{rec}}$, whose concentration comes mostly from the extension. In (ii) and (iii), an error of the extension therefore enters the scalar, but only in a thin layer at the interface. Everywhere else, the extended values are multiplied by $S_i^{n+1}=0$ and have no effect.

\paragraph{Non-negative mass projection}

The mass projection changes concentration while holding the corrected support $S_i^{n+1}$ fixed. Its objective is to recover the target mass while remaining as close as possible to the provisional concentrations. The target mass of Eq.~\eqref{eq:scalar_target_mass} contains the mass added by the volumetric source during the step, so that the projection conserves the scalar mass together with this source. We first compute the mass of the provisional scalar and the difference from the target mass that the projection must restore:
\begin{equation}
M_C^{\mathrm{prov}}=\sum_{i\in\Omega}C_i^{\mathrm{prov}}V_i,
\qquad
\Delta M_C=M_C^{\mathrm{target}}-M_C^{\mathrm{prov}}.
\label{eq:scalar_mass_residual_after_reattachment}
\end{equation}
All cells containing the corrected carrying phase are adjustable:
\begin{equation}
\mathcal{A}=\{i\in\Omega:S_i^{n+1}>0\}.
\label{eq:scalar_projection_supported_region}
\end{equation}
Cells outside $\mathcal{A}$ remain at zero, so redistribution cannot place scalar outside the corrected carrying phase. In the adjustable cells, we minimize the squared change from the provisional concentration, weighted by the corrected carrying-phase volume $S_i^{n+1}V_i$. The candidate concentrations must remain non-negative and recover the target mass:
\begin{equation}
\begin{aligned}
\underset{\{q_i:i\in\mathcal{A}\}}{\operatorname{minimize}}\quad
&\frac12\sum_{i\in\mathcal{A}}S_i^{n+1}V_i
\left(q_i-q_i^{\mathrm{prov}}\right)^2,
\\
\text{subject to}\quad
&\sum_{i\in\mathcal{A}}S_i^{n+1} q_iV_i
=M_C^{\mathrm{target}},
\qquad q_i\ge0.
\end{aligned}
\label{eq:scalar_concentration_projection}
\end{equation}

The optimality conditions give a common concentration shift $\lambda_C$ in all adjustable cells. A shifted concentration that would be negative is truncated at zero. Multiplication by $S_i^{n+1}$ then gives the rebuilt scalar $\tilde C$, while cells outside $\mathcal{A}$ remain at zero:
\begin{equation}
\tilde C_i(\lambda_C)=
\begin{cases}
\max\!\left(0,C_i^{\mathrm{prov}}+\lambda_CS_i^{n+1}\right),
&i\in\mathcal{A},\\
0,
&i\notin\mathcal{A}.
\end{cases}
\label{eq:scalar_projection_shift}
\end{equation}
Before truncation, each adjustable phase-averaged concentration changes by the same value $\lambda_C$, whereas its cell-averaged concentration changes by $\lambda_CS_i^{n+1}$. The algorithm therefore shifts concentration directly and does not obtain the increment by dividing a scalar increment by a small phase support. The shift is chosen so that the resulting total mass equals the target. Defining the mass residual as
\begin{equation}
G(\lambda_C)=
\sum_{i\in\mathcal{A}}
\max\!\left(0,C_i^{\mathrm{prov}}+\lambda_CS_i^{n+1}\right)V_i
-M_C^{\mathrm{target}},
\label{eq:scalar_projection_multiplier}
\end{equation}
we require $G(\lambda_C)=0$. Increasing $\lambda_C$ cannot decrease the resulting mass, so $G$ is monotone and piecewise linear, and the multiplier is obtained by a semismooth Newton iteration \citep{Qi1993}. The solver details and tolerances are given in~\ref{app:scalar_solver}.

The rebuilt scalar satisfies
\begin{equation}
\tilde C_i\ge0,\qquad \tilde C_i=0\ \text{if }S_i^{n+1}=0,\qquad \sum_{i\in\Omega}\tilde C_iV_i=M_C^{\mathrm{target}},
\label{eq:final_scalar_correction_requirements}
\end{equation}
where the mass equality is enforced to the prescribed numerical tolerance. The exchange with the interface and the diffusion, where present, then act on this scalar and give $C^{n+1}$ (Eq.~\eqref{eq:scalar_diffusion_discrete}); without them, $C^{n+1}=\tilde C$. They keep it non-negative under the conditions given there, and in all calculations of Section~\ref{sec:results} the scalar of the proposed method remained non-negative after every complete step.

The transport of $q$ is not conservative locally: the mass that the rebuild removes near the interface is returned by the single shift $\lambda_C$, uniformly over the carrying phase (Table~\ref{tab:scheme_study} gives its size for seven schemes). Summed over the steps, the mass moved in this way is zero for a uniform concentration and at most $2.2\%$ of the scalar mass in the prescribed flows of Sections~\ref{sec:grid_aligned_translation}--\ref{sec:source_transport}, decreasing with refinement. Besides the pure-cell tolerance of the indicator projection, the method has two parameters, the recovery threshold $\varepsilon^{\mathrm{rec}}$ and the number of extension layers $N_\ell$.

\paragraph{Choice of the advection scheme}

The procedure does not depend on the advection scheme: any face reconstruction can be used to advect the extended concentration, and the number of extension layers follows from the width of its stencil. The gain over the standard method, however, grows with the order of the scheme. Numerical diffusion spreads the concentration from the carrying phase into extension cells where the corrected support vanishes; the rebuild removes this part of the scalar mass, and the mass projection returns it through a uniform shift of the concentration over the whole phase rather than where it was lost. In the following sections, the procedure is demonstrated with WENO-Z; Section~\ref{sec:scheme_dependence} compares it with six other face reconstructions, from the second-order MUSCL and QUICK schemes to the fifth-order targeted essentially non-oscillatory (TENO) scheme.

\paragraph{Summary of the procedure}
Algorithm~\ref{alg:scalar_correction} summarizes one time step of the phase-confined scalar, from the scalar and support at the start of the step to the rebuilt, phase-confined, mass-conservative scalar $\tilde C$, before the sorption and diffusion of Eq.~\eqref{eq:scalar_diffusion_discrete}.

\begin{algorithm}[t]
\small
\DontPrintSemicolon
\SetAlgoLined
\caption{Phase-concentration transport with conservative rebuild}
\label{alg:scalar_correction}

\KwIn{\(C^n\), \(I^{c,n}\), the face fluxes \(F_f\), and \(I^{\mathrm{FT},n+1}\)}
\KwOut{\(I^{c,n+1}\), \(S^{n+1}\), and the rebuilt scalar \(\tilde C\), from which Eq.~\eqref{eq:scalar_diffusion_discrete} gives \(C^{n+1}\)}

\textbf{1. Recovery}\;
Set \(S^n\) from \(I^{c,n}\); compute \(M_C^n=\sum_i C_i^nV_i\)\;
Set \(q_i^n=\max(C_i^n,0)/S_i^n\) for \(S_i^n>\varepsilon^{\mathrm{rec}}\)\;

\textbf{2. Extension}\;
Extend \(q\) over \(N_\ell\) layers by Eq.~\eqref{eq:concentration_extension}\;
Assign \(\overline{q}^n\) to the cells beyond the last layer\;

\textbf{3. Advection}\;
Advance \(q\) with Eq.~\eqref{eq:concentration_advection_discrete} over the three Runge--Kutta stages, obtaining \(q^\ast\)\;

\textbf{4. Rebuild}\;
Project \(I^{\mathrm{FT},n+1}\) to obtain \(I^{c,n+1}\) with \(\sum_iI_i^{c,n+1}V_i=\sum_iI_i^{c,n}V_i\); set \(S^{n+1}\)\;
Set \(C_i^{\mathrm{prov}}=S_i^{n+1}\max(q_i^\ast,0)\); add the source; compute \(M_C^{\mathrm{target}}\)\;

\textbf{5. Non-negative mass projection}\;
Set \(\mathcal{A}=\{i:S_i^{n+1}>0\}\)\;
Shift the concentration by \(\lambda_C\) over \(\mathcal{A}\) to obtain \(\tilde C\) with mass \(M_C^{\mathrm{target}}\)\;
\(\tilde C_i\geq0\) and \(\tilde C_i=0\) for \(S_i^{n+1}=0\) by construction;
report the final mass residual \(\sum_i \tilde C_iV_i-M_C^{\mathrm{target}}\)\;

\end{algorithm}

\subsection[Flow discretization and coupled solution]{Flow discretization and coupled solution}
\label{sec:flow_discretisation}

{\color{black}
The indicator projection and the phase-concentration transport described above are implemented within the in-house finite-volume solver \texttt{MultiFlow}. We now describe the associated flow discretization, which uses a collocated, second-order finite-volume method on a Cartesian Eulerian mesh with adaptive refinement around the front. The continuity and momentum equations are solved simultaneously using the coupled pressure--velocity algorithm of \citet{Denner2020}, with momentum-weighted interpolation (MWI) for the face velocity following \citet{Bartholomew2018}. The corresponding linearized flow system is written as
\begin{equation}
\mathbf{A}\cdot\boldsymbol{\chi} = \mathbf{b},
\end{equation}
where $\boldsymbol{\chi}$ contains the three velocity components and pressure as the implicitly sought solution variables.

In the discretizations below, the iteration counter $k$ denotes nonlinear iterations within each time step. Superscript $(k)$ denotes lagged quantities evaluated from the previous nonlinear iteration, while superscript $(k+1)$ denotes quantities treated implicitly at the current nonlinear iteration. The index $n$ is reserved for physical time levels.
}

\subsubsection[Continuity equation]{Continuity equation}
\label{sec:continuity}

With the divergence theorem, the continuity equation, Eq.~\eqref{eq:continuity}, is discretized over a control volume $P$ as
\begin{equation}
\iiint_{V_P} \nabla \cdot \mathbf{u} \, \mathrm{d}V \approx \sum_f F_f^{(k+1)} = 0,
\label{eq:continuity_disc}
\end{equation}
where the summation is over all faces $f$ bounding the control volume. The volumetric flux through face $f$ is defined as
\begin{equation}
F_f = \vartheta_f A_f, \qquad \vartheta_f = \mathbf{u}_f \cdot \mathbf{n}_f,
\label{eq:flux_simple}
\end{equation}
where $A_f$ is the face area, $\mathbf{n}_f$ is the outward unit normal vector of face $f$, and $\vartheta_f$ is the advecting velocity at the face.

Since the solver uses a collocated variable arrangement, $\vartheta_f$ is evaluated using a momentum-weighted interpolation (MWI) \citep{Rhie1983,Bartholomew2018}. This interpolation couples pressure and velocity at the cell faces and prevents pressure--velocity decoupling on the collocated finite-volume mesh. The MWI formulation follows the unified formulation of \citet{Bartholomew2018}, including the transient correction required for time-step-independent face velocities. In the rising-bubble calculations, the pressure terms entering the MWI formulation are treated semi-implicitly. This introduces an implicit pressure dependence into the continuity equation, allowing pressure and velocity to be solved in a fully coupled manner \citep{Denner2020}.

\subsubsection[Momentum equations]{Momentum equations}
\label{sec:numerics_momentum}

The momentum equations are solved implicitly in velocity and pressure. We first describe the spatial discretization of the advection, pressure, viscous, and source terms, and then the time discretization of the conservative momentum.

The advection term is discretized using a lagged volumetric face flux and a central interpolation of the transported velocity to the cell faces:
\begin{equation}
\iiint_{V_P} \nabla \cdot \left( \rho \mathbf{u} \otimes \mathbf{u} \right) \, \mathrm{d}V \approx \sum_f F_f^{(k)} \left( \rho \mathbf{u} \right)_f^{(k+1)},
\label{eq:momentum_advection_central}
\end{equation}
where $F_f^{(k)}=\vartheta_f^{(k)}A_f$ is the volumetric face flux evaluated from the previous nonlinear iteration. The face value $(\rho\mathbf{u})_f^{(k+1)}$ is obtained using central differencing and is treated implicitly.

The pressure-gradient term is discretized using linearly interpolated face values:
\begin{equation}
\iiint_{V_P} \nabla p \, \mathrm{d}V \approx \sum_f \overline{p}_f^{(k+1)} \mathbf{n}_f A_f .
\label{eq:momentum_pressuregrad}
\end{equation}
The pressure $p$ is treated implicitly, as indicated by the superscript $(k+1)$. Here, $\overline{p}_f$ denotes the pressure at face $f$, obtained
by linear interpolation with inverse-distance weighting from the
cell-centred pressures in the adjacent cells $P$ and $Q$.

The viscous-stress divergence is discretized using standard Cartesian finite-volume gradients as
\begin{equation}
\iiint_{V_P} \nabla \cdot \boldsymbol{\tau} \, \mathrm{d}V \approx \sum_f \breve{\mu}_f \left[ \frac{\mathbf{u}_Q^{(k+1)}-\mathbf{u}_P^{(k+1)}}{\Delta s_f} + \overline{\nabla \mathbf{u}}_f^{\mathrm{T},(k+1)} \cdot \mathbf{n}_f \right] A_f .
\label{eq:momentum_shear_all}
\end{equation}
Here, $\breve{\mu}_f$ is the harmonically interpolated face viscosity and $\Delta s_f$ is the distance between the centres of cells $P$ and $Q$.

The momentum source terms, which include $\mathbf{S}_\sigma$ and $\mathbf{S}_g$, are discretized consistently with the pressure-gradient term to avoid imbalance between pressure and source contributions \citep{Mencinger2007,Denner2014b,Bartholomew2018}. Following \citet{Bartholomew2018}, the discretized source term is constructed as
\begin{equation}
\mathbf{S}^{\star}_P = \frac{1}{V_P} \sum_f \left( \overline{\mathbf{S}}_f \cdot \mathbf{s}_f \right) {\Delta s_f} \mathbf{n}_f A_f .
\label{eq:momentum_source_disc}
\end{equation}
Here, $\overline{\mathbf{S}}_f$ is the source term evaluated at face $f$, and $\mathbf{s}_f$ is the unit vector directed from the centre of cell $P$ to the centre of its neighbour $Q$ across face $f$.

{\color{black}
To complete the momentum discretization, the transient term is approximated using a variable-time-step second-order backward differentiation formula (BDF2) \citep{Moukalled2016}. This time discretization applies only to the flow equations, not to the scalar or indicator transport. BDF2 is applied to the conservative momentum $\rho\mathbf{u}$, with density evaluated at the corresponding time level. For a momentum component $\phi\in\{\rho u,\rho v,\rho w\}$, it reads
\begin{equation}
\left. \frac{\partial \phi}{\partial t} \right|_P \approx \left( \frac{1}{\Delta t_1} + \frac{1}{\Delta t_1 + \Delta t_2} \right) \phi_P^{(t)} - \left( \frac{1}{\Delta t_1} + \frac{1}{\Delta t_2} \right) \phi_P^{(t-\Delta t_1)} + \frac{\Delta t_1}{\Delta t_1 \Delta t_2 + \Delta t_2^2} \phi_P^{(t-\Delta t_1-\Delta t_2)},
\label{eq:bdf2}
\end{equation}
where $\Delta t_1$ is the current time step and $\Delta t_2$ is the previous time step. The superscripts $(t)$, $(t-\Delta t_1)$, and $(t-\Delta t_1-\Delta t_2)$ denote the current, previous, and previous-previous time levels, respectively.
}
{\color{black}
The scalar is passive and one-way coupled to the flow. The velocity field and
the Lagrangian interface determine scalar transport and the evolution of the
phase support, whereas the scalar does not enter the continuity
or momentum equations, the interfacial force, or the front-advection procedure.
Consequently, the scalar transport cannot modify the flow field or interface
dynamics.}

\subsection[Solution procedure]{Solution procedure}
\label{sec:solution}

\begin{figure}[t]
  \begin{center}
  \begin{small}
      \begin{tikzpicture}
      \node (pro0) [process] {Update previous time levels: \\[0.25em] $\chi^{(t-\Delta t_1-\Delta t_2)}\leftarrow \chi^{(t-\Delta t_1)}$ \\[0.25em] $\chi^{(t-\Delta t_1)}\leftarrow \chi^{(k+1)\phantom{...)}}$  \\[0.25em] $F_f^{(t-\Delta t_1)} \leftarrow F_f^{(k+1)\phantom{....}}$};

      \node (pro0b) [process, below of=pro0, yshift=-1cm]
      {Advect the Lagrangian front, update its \mbox{geometry}, and reconstruct $I^{\mathrm{FT}}$};

      \node (pro0c) [process, below of=pro0b, yshift=-0.45cm]
      {Recover, extend and \mbox{advect} $q$; rebuild $C$ on the corrected support};

      \node (pro0d) [process, below of=pro0c, yshift=-0.45cm]
      {Restore the scalar mass; apply diffusion and interfacial exchange};

      \node (pro1a) [process, below of=pro0d, yshift=-0.45cm]
      {\mbox{Gather the coefficients} and assemble \mbox{$\mathbf{A}$ and $\mathbf{b}$}};

      \node (pro1b) [process, below of=pro1a, yshift=-0.45cm]
      {Solve $\mathbf{A} \cdot \boldsymbol{\chi} = \mathbf{b}$};

      \node (pro3) [process, below of=pro1b, yshift=-0.45cm]
      {Compute $F_f^{(k+1)}$};

      \node (dec1) [decision, below of=pro3, yshift=-0.95cm]
      {Conservation satisfied?};

      \node (pro4) [process, below of=dec1, yshift=-1.1cm]
      {\mbox{Adapt the fluid mesh,} if applicable};

      \draw [arrow] (pro0) -- (pro0b);
      \draw [arrow] (pro0b) -- (pro0c);
      \draw [arrow] (pro0c) -- (pro0d);
      \draw [arrow] (pro0d) -- (pro1a);
      \draw [arrow] (pro1a) -- (pro1b);
      \draw [arrow] (pro1b) -- (pro3);
      \draw [arrow] (pro3) -- (dec1);
      \draw [arrow] (dec1) --+(-3cm,0) |- (pro1a);
      \draw [arrow] (dec1) -- (pro4);
      \draw [arrow] (pro4) --+(+3cm,0) |- (pro0);

      \node at (-2.7,-11.05) {no};
      \node at (0.35,-12.55) {yes};
      \node [rotate=90] at (-3.2,-8.8) {$k \leftarrow k+1$};
      \node [rotate=90] at (3.2,-6.8) {$t \leftarrow t+\Delta t$};
      \end{tikzpicture}
  \end{small}
  \caption{Flow chart of the solution procedure for one time step. The Lagrangian interface is first advanced and used to compute the interfacial forcing. The phase-averaged concentration of the phase-confined scalar is then recovered, extended and advected on the fixed mesh, and the scalar is rebuilt on the corrected support with the mass projection. The discretized flow equations are then assembled and solved. Here, $k$ denotes the nonlinear iteration counter, $\boldsymbol{\chi}=\{u,v,w,p\}$ denotes the flow variables, which form the solution vector, $F_f$ is the volumetric flux through mesh face $f$, $\mathbf{A}$ is the coefficient matrix, and $\mathbf{b}$ contains deferred nonlinear terms and contributions from previous time levels.}
  \label{fig:flowchart}
  \end{center}
\end{figure}
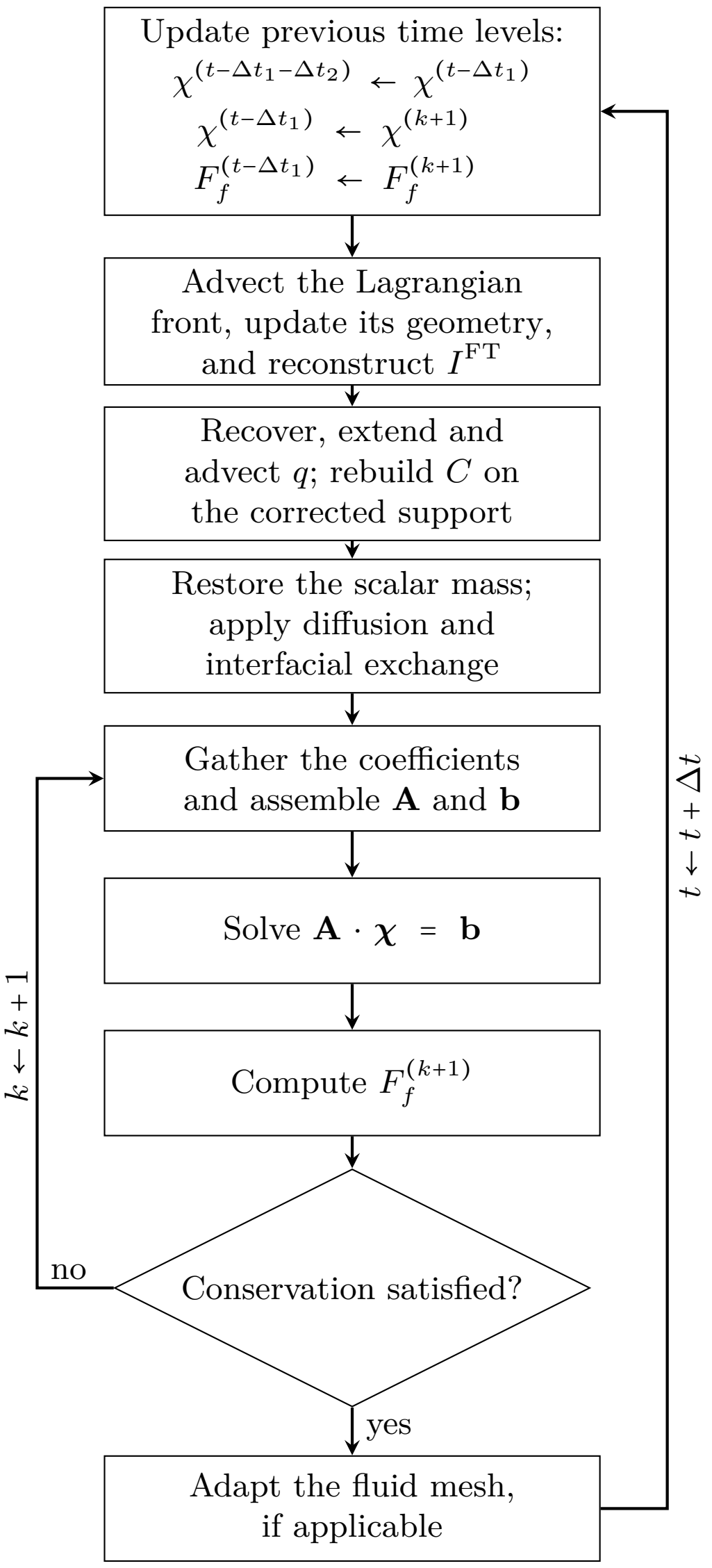

During each time step from $t^n$ to $t^{n+1}$, the interface motion, indicator projection, phase-confined scalar transport, and flow solution are advanced in a partitioned sequence. First, the Lagrangian front is advected using the selected front-advection scheme, and remeshing is applied when enabled. The Eulerian indicator associated with the updated front is then reconstructed, and the interfacial forces are computed and spread to the Eulerian mesh.

After the front update, the phase-averaged concentration of the phase-confined scalar is recovered from the scalar, extended across the interface, and advected on the fixed Eulerian grid. The corrected indicator of the new time level is obtained from the indicator projection of the updated front-derived indicator, the scalar is rebuilt on the corresponding support together with any volumetric source, and the non-negative mass projection restores the scalar mass. The exchange with the interface and the diffusion, where present, then act on the rebuilt scalar.

Finally, the flow equations are assembled and solved using the updated material distribution, interfacial forcing, and scalar fields. The nonlinear iteration is repeated until the prescribed conservation criterion is satisfied, after which the fluxes and previous time-level quantities are updated and the computation proceeds to the next time step. The front update, phase-concentration transport, and coupled flow solution are combined in the time-step sequence shown in Fig.~\ref{fig:flowchart}.

\section{Results and validation}
\label{sec:results}

The phase-concentration transport of Section~\ref{sec:scalar_correction} is assessed in five three-dimensional cases of increasing complexity. A uniform concentration translated along one axis (Section~\ref{sec:grid_aligned_translation}) isolates the transport from deformation and from the flow solution. A linear concentration gradient carried through a reversible vortex (Section~\ref{sec:reversible_vortex}) adds a moderate interface deformation and a non-uniform concentration, with an exact solution at every time, and is repeated with the scalar carried by the phase outside the drop. A localized source in an obliquely translating drop (Section~\ref{sec:source_transport}) tests a concentration that changes in time and a total mass that is not conserved. A rising bubble (Section~\ref{sec:rising_bubble}) carries a uniform concentration through an interface whose motion follows from the coupled flow solution. A soluble surfactant in the rising bubble adds bulk diffusion and adsorption at the interface and shows their interplay with the transport (Section~\ref{sec:surfactant_bubble}); the adsorption and desorption are validated on a sphere at rest against a converged reference solution. Section~\ref{sec:ablation} replaces one ingredient of the method at a time in the vortex and the source. Section~\ref{sec:scheme_dependence} repeats the source and the rising bubble with six other advection schemes. The dependence on the parameters of the method and on the time step is examined in Section~\ref{sec:sensitivity}, and the computational cost in Section~\ref{sec:cost}.

Apart from the computing times of the cases reported in Section~\ref{sec:cost}, which are wall-clock times in seconds, all quantities in this section are dimensionless. The first three cases have a prescribed velocity and are defined by their geometry, velocity and resolution; their lengths, velocities, times and concentrations are given in arbitrary consistent units, and the volumetric source in concentration per unit time. The others are characterized by the dimensionless groups
\begin{equation}
\mathrm{E\ddot{o}}=\frac{\rho_\ell gD^2}{\sigma},\quad
\mathrm{Ga}=\frac{\rho_\ell\sqrt{gD}\,D}{\mu_\ell},\quad
\mathrm{Pe}_c=\frac{UL}{\mathcal{D}},\quad
\mathrm{Pe}_s=\frac{UL}{\mathcal{D}_s},\quad
K=\frac{k_aq_0}{k_b},\quad
\mathrm{Bi}=\frac{k_bL}{U},\quad
\mathrm{Da}=\frac{\Gamma_\infty}{Lq_0},
\label{eq:dimensionless_groups}
\end{equation}
together with the density and viscosity ratios $\rho_\ell/\rho_{\mathrm{g}}$ and $\mu_\ell/\mu_{\mathrm{g}}$ of the liquid ($\ell$) and the gas ($\mathrm{g}$). Here $L$ and $U$ are the length and velocity scales of the case, listed in Table~\ref{tab:dimensionless_groups}, and $q_0$ is the initial concentration; in these cases, lengths are given in units of the diameter $D$ and times as $t^*=tU/L$. The E\"otv\"os and Galilei numbers characterize the rising bubble. The surfactant groups are those of \citet{Muradoglu2008,Muradoglu2014}: the P\'eclet numbers of the bulk and the surface diffusion, the ratio $K$ of the adsorption rate $k_aq_0$ to the desorption rate $k_b$, the Biot number and the Damk\"ohler number, with the coefficients $k_a$, $k_b$, $\Gamma_\infty$ and $\mathcal{D}_s$ of Section~\ref{sec:surfactant_bubble}. Table~\ref{tab:dimensionless_groups} lists the scales and the groups of the two cases.

\begin{table}[htbp]
\centering
\caption{Length and velocity scales and dimensionless groups, Eq.~\eqref{eq:dimensionless_groups}, of the rising bubble (Section~\ref{sec:rising_bubble}) and the rising bubble with a soluble surfactant (Section~\ref{sec:surfactant_bubble}). Dashes mark groups that do not apply.}
\label{tab:dimensionless_groups}
\small
\begin{tabular}{lccrrrrrrrrr}
\toprule
case & $L$ & $U$ & E\"o & Ga & $\rho_\ell/\rho_{\mathrm{g}}$ & $\mu_\ell/\mu_{\mathrm{g}}$ & $\mathrm{Pe}_c$ & $\mathrm{Pe}_s$ & $K$ & Bi & Da \\
\midrule
rising bubble & $D$ & $\sqrt{gD}$ & 129 & 23.5 & 1005 & 100 & -- & -- & -- & -- & -- \\
surfactant, bubble & $D$ & $\sqrt{gD}$ & 129 & 23.5 & 1005 & 100 & 313 & 313 & 0.5 & 0.13 & 0.5 \\
\bottomrule
\end{tabular}
\end{table}

In every case the same scalar is transported twice, on the same mesh and with the same front and flow; without surfactant both copies are carried in the same simulation, and with the surfactant in two simulations with identical flow. The first copy uses the phase-concentration transport (Algorithm~\ref{alg:scalar_correction}), referred to below as the proposed method. The second copy, referred to as the standard method, transports $C$ with Eq.~\eqref{eq:scalar_transport_discrete} and the same WENO-Z reconstruction and SSP-RK3 integration, without any projection; Section~\ref{sec:scheme_dependence} replaces WENO-Z in both by other reconstructions. Neither scalar acts on the flow, so the differences between them result from the scalar transport alone. Diffusion is set to zero in Eq.~\eqref{eq:scalar_diffusion_discrete} in the first four cases, so that exact concentrations are available.

The front indicator is obtained from Eq.~\eqref{eq:indicator_poisson}, solved to a residual norm of $10^{-12}$, and the indicator projection uses $\varepsilon^{\mathrm{pure}}=10^{-5}$. Unless stated otherwise, $\varepsilon^{\mathrm{rec}}=0.03$ and $N_\ell=4$. Three resolutions, $D/h=10$, $20$ and $40$, are considered, where $D$ is the diameter of the drop or bubble and $h$ the finest cell size of the adaptive mesh. The mesh is a Cartesian base mesh refined in two or three levels around the front: the cells next to the front and all cells inside the bounding box of the front, enlarged by $50$ cells along $x$ and $10$ cells along $y$ and $z$, are on the finest level. After the first step, the cells with $S>\varepsilon^{\mathrm{rec}}$ lie at least eight finest cells from any coarser cell in every stored field, so that the interface cells, the extension layers and the reconstruction stencils around them lie on the finest level, where the mesh is uniform. Table~\ref{tab:settings} lists the mesh, the time step and the front settings of each case. In all calculations, the coupled flow solution is iterated within each step until the $L_2$ norm of the continuity error is below $10^{-5}$, and the linear solver uses an absolute tolerance of $10^{-5}$. Where the front is remeshed, edges are split, collapsed and flipped towards a target edge length of $0.75h$, with lengths kept between $0.25$ and $1.5$ times this target.

In the translation and the source case, the flow equations are solved with equal properties of both fluids and without surface tension, so that the prescribed uniform velocity is preserved to round-off ($10^{-14}$). The front moves with the velocity interpolated by Eq.~\eqref{eq:front_velocity_interpolation}, which reproduces a uniform field exactly, since the kernel weights sum to one. In the vortex, the front moves with the analytical velocity (Section~\ref{sec:reversible_vortex}). The rising bubbles have a no-slip wall at $x=0$ and, at the top, zero-gradient velocities and a zero pressure; gravity enters as the source $\rho\mathbf{g}$, discretized consistently with the pressure gradient (Section~\ref{sec:numerics_momentum}), so that the hydrostatic pressure is part of the computed pressure.

\begin{table}[htbp]
\centering
\caption{Settings of the calculations: domain, base mesh at $D/h=10$, $20$ and $40$, number of refinement levels, number of cells at $D/h=40$ over the stored fields after the first step, time step, front velocity and front remeshing. Lengths and times in the units of each case (Section~\ref{sec:results}).}
\label{tab:settings}
\footnotesize
\setlength{\tabcolsep}{2.5pt}
\begin{tabular}{lllcclll}
\toprule
case & domain & base mesh & levels & cells ($10^6$) & time step & front velocity & remeshing \\
\midrule
translation & $[0,5]\times[0,4]^2$ & $25{\times}20^2$, $25{\times}20^2$, $50{\times}40^2$ & 2, 3, 3 & 1.15 & $\Delta t=0.01$, $0.005$, $0.0025$ & interpolated & every 10 steps \\
vortex & $[0,4]^3$ & $10^3$, $20^3$, $40^3$ & 3 & 1.02--1.10 & CFL $\le0.1$ & analytical & none \\
source & $[0,4]^3$ & $20^3$, $20^3$, $40^3$ & 2, 3, 3 & 1.02--1.04 & CFL $0.1$ & interpolated & every 10 steps \\
rising bubble & $[0,8]\times[0,4]^2$ & $10{\times}5^2$, $20{\times}10^2$, $40{\times}20^2$ & 3 & 0.60--0.81 & CFL $0.1$, $\Delta t^*\le0.031$ & interpolated & every step \\
surfactant bubble & $[0,16]\times[0,4]^2$ & $20{\times}5^2$, $40{\times}10^2$, $80{\times}20^2$ & 3 & 0.74--0.96 & CFL $0.1$, $\Delta t^*\le0.031$ & interpolated & every step \\
sphere at rest & $[0,8]^3$ & $10^3$, $20^3$, $40^3$ & 3 & 1.02 & \ref{app:surfactant_sphere} & at rest & every 10 steps \\
\bottomrule
\end{tabular}
\end{table}

\begin{figure}[htbp]
\centering
\includegraphics[width=\textwidth]{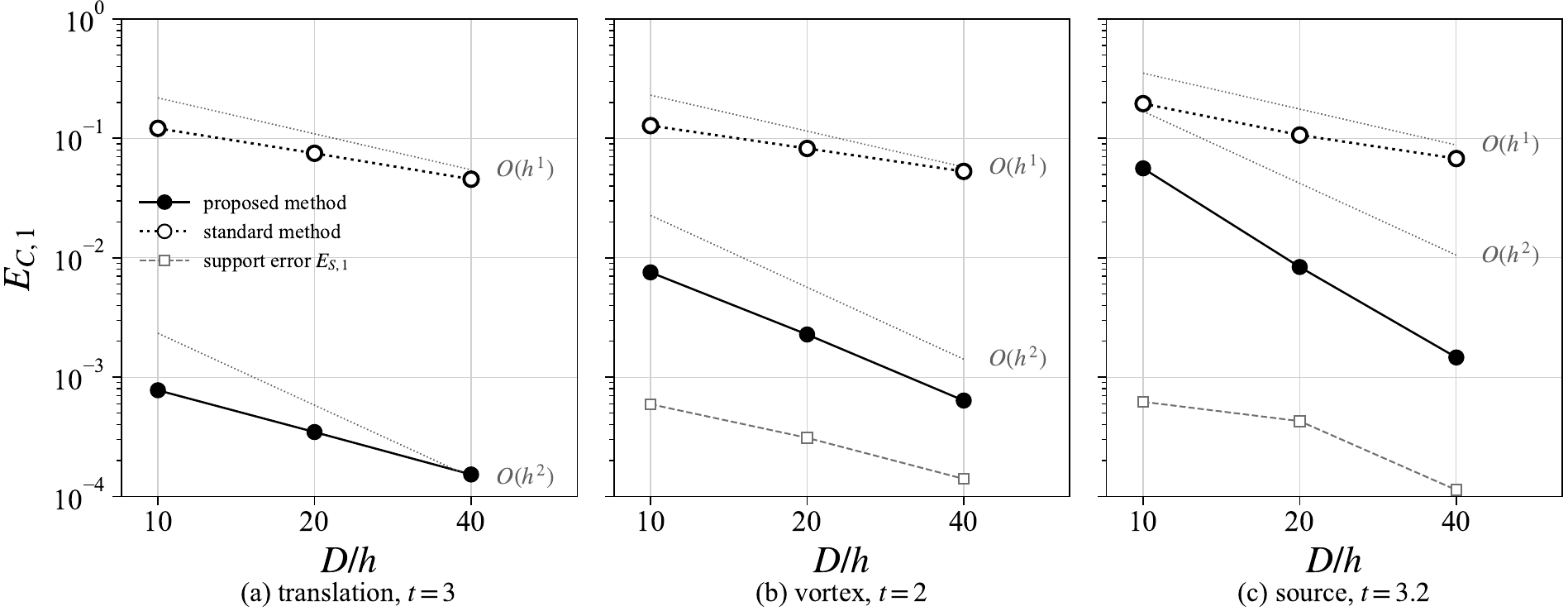}
\caption{Scalar--front error $E_{C,1}$ at the last stored field against the resolution for the three cases with a prescribed flow: (a) translation, (b) vortex, (c) source. Proposed method (solid line, filled symbols), standard method (dotted line, open symbols), and the support error $E_{S,1}$ of the proposed method (dashed) where it differs from $E_{C,1}$; for the uniform concentration of (a) the two coincide. Dotted grey lines indicate first- and second-order slopes.}
\label{fig:convergence}
\end{figure}

The support reconstructed from the tracked front is $S^{\mathrm{FT}}=S(I^{\mathrm{FT}})$, Eq.~\eqref{eq:scalar_support_definition}, and the exact cell-averaged concentration is the exact phase-averaged concentration of the cell, $q_i^{\mathrm{ex}}$ (its centre value; for the source of Section~\ref{sec:source_transport}, its cell average), distributed over it,
\begin{equation} C_i^{\mathrm{ex}}(t)=S_i^{\mathrm{FT}}(t)\,q_i^{\mathrm{ex}}(t). \label{eq:xtranslation_exact}\end{equation}
The scalar--front error is the normalized volume-weighted difference
\begin{equation} E_{C,1}=\frac{\sum_i|C_i-C_i^{\mathrm{ex}}|V_i}{\sum_i|C_i^{\mathrm{ex}}|V_i}. \label{eq:results_C_l1}\end{equation}
The phase-averaged concentration of every solution is evaluated against the same front-derived support,
\begin{equation} q_i=\frac{C_i}{S_i^{\mathrm{FT}}},\qquad
E_{q,1}=\frac{\sum_{i\in\mathcal{V}}S_i^{\mathrm{FT}}\,|q_i-q_i^{\mathrm{ex}}|\,V_i}{q_{\mathrm{ref}}\sum_{i\in\mathcal{V}}S_i^{\mathrm{FT}}V_i},\qquad
E_{q,\infty}=\max_{i\in\mathcal{V}}\frac{|q_i-q_i^{\mathrm{ex}}|}{q_{\mathrm{ref}}},\qquad
\mathcal{V}=\{i:S_i^{\mathrm{FT}}>10^{-2}\}, \label{eq:results_q_error}\end{equation}
where $q_{\mathrm{ref}}$ is the concentration scale of the case. The threshold of $\mathcal{V}$ excludes cells in which the quotient is not meaningful: an error $\delta C_i$ enters $q_i$ as $\delta C_i/S_i^{\mathrm{FT}}$, and below $S_i^{\mathrm{FT}}=10^{-2}$ it would be magnified more than a hundredfold. The concentration is plotted on the same set. Scalar carried into cells with negligible front-derived support is quantified by
\begin{equation} E_{\mathrm{leak}}=\frac{\sum_{i:S_i^{\mathrm{FT}}<\varepsilon^{\mathrm{leak}}}C_i^+V_i}{\sum_iC_i^+V_i},\qquad C_i^+=\max(C_i,0),\qquad \varepsilon^{\mathrm{leak}}=10^{-3}. \label{eq:results_leakage}\end{equation}
The indicator of Eq.~\eqref{eq:indicator_poisson} is not sharp, and a small part of the phase volume lies in cells with $S_i^{\mathrm{FT}}<\varepsilon^{\mathrm{leak}}$. The exact field $C^{\mathrm{ex}}$ therefore has a non-zero $E_{\mathrm{leak}}$ of its own, which is reported alongside: only the leakage beyond this share is scalar that has left the phase. All error measures thus use the front-derived support $S^{\mathrm{FT}}$ as the reference, not the sharp front. The support carried by the proposed method is compared with the front-derived support through
\begin{equation} E_{S,1}=\frac{\sum_i|S_i^{n+1}-S_i^{\mathrm{FT}}|V_i}{\sum_iS_i^{\mathrm{FT}}V_i}. \label{eq:results_support_error}\end{equation}
Finally, the total mass and its relative conservation error are
\begin{equation} M_C(t)=\sum_iC_i(t)V_i,\qquad E_M(t)=\frac{|M_C(t)-M_C(0)|}{M_C(0)}. \label{eq:results_scalar_mass_error}\end{equation}
All errors of a table row are evaluated at the same instant, the last stored field of the run, and no time averages are used. The errors and the leakage at that instant are listed for each case in its section (Tables~\ref{tab:translation_errors}--\ref{tab:bubble_errors}), with the convergence rates between $D/h=10$ and $20$ and between $20$ and $40$; the leakage is given together with the share of the exact field $C^{\mathrm{ex}}=S^{\mathrm{FT}}q^{\mathrm{ex}}$ in the same cells at the same instant. At $D/h=40$, the scalar--front error of the proposed method is $300$, $83$, $46$ and $1700$ times smaller than that of the standard method in the translation, the vortex, the source and the rising bubble. In the translation and the bubble the concentration is uniform, so that its transport is exact and these factors compare the support error with the smearing of the standard method; the factor of the bubble also depends on the instant (Section~\ref{sec:rising_bubble}). Figure~\ref{fig:convergence} shows the convergence of $E_{C,1}$ for the three cases with a prescribed flow. Both scalars conserve the total mass to round-off in every case: $E_M$ is at most $3.4\times10^{-12}$ without surfactant, measured for the source against the initial mass plus the injected one, and $3.4\times10^{-11}$ for the bulk and adsorbed surfactant together. These measures are used for the four cases without surfactant (Sections~\ref{sec:grid_aligned_translation}--\ref{sec:rising_bubble}), restricted to the interfacial band for the scalar carried by the phase outside the drop in Section~\ref{sec:reversible_vortex}, and for the sensitivity study (Section~\ref{sec:sensitivity}). The surfactant case (Section~\ref{sec:surfactant_bubble}) is assessed instead through the surfactant budget, defined in its section; the surfactant outside the phase is counted there with the same threshold $\varepsilon^{\mathrm{leak}}$ as $E_{\mathrm{leak}}$.

\FloatBarrier

The three prescribed flows converge at different rates (Fig.~\ref{fig:convergence}), because a different error dominates in each. For the uniform concentration of the translation, the transport of the concentration is exact, and the error is the support error, the difference between the projected and the front-derived support in the interfacial cells. It converges at about $1.2$. In the vortex, the concentration varies smoothly, but the deformation bends the initially linear profile during the cycle, so that the linear extension, which is exact only for a linear profile, makes errors of order $h^2$ (Section~\ref{sec:scalar_correction}). The error converges at $1.7$--$1.8$, close to second order. In the source case, the Gaussian is one cell wide at $D/h=10$, and the error there is dominated by its unresolved peak. At $D/h=10$ the peak concentration of the proposed method is $5.7$, against $9.0$ for the cell-averaged exact solution. As the peak becomes resolved at $D/h=20$ and $40$, the error falls by factors of $6.7$ and $5.7$, more than the factor of $4$ of second-order convergence. These rates therefore hold over the resolutions tested and are not yet asymptotic. Writing the error of the proposed method as $C-C^{\mathrm{ex}}=S^{n+1}(q-q^{\mathrm{ex}})+(S^{n+1}-S^{\mathrm{FT}})q^{\mathrm{ex}}$ shows what limits them (Table~\ref{tab:error_split}). The first part, the error of the concentration, converges at $1.8$--$1.9$ in the vortex and $2.6$--$2.8$ in the source. The second part, the mismatch between the projected and the front-derived support, converges at about first order, so that its share of $E_{C,1}$ grows with refinement, from $8$ to $22\%$ in the vortex and from $1$ to $7\%$ in the source. The offset from the cells of the indicator's tail in the vortex (Section~\ref{sec:reversible_vortex}) also converges more slowly than second order; without these cells the vortex error converges at $1.9$--$2.0$. The support part oscillates in time as the front crosses the cells, like the error of the bubble in Fig.~\ref{fig:bubble_errors}, so that its rates at a single instant scatter ($0.5$ and $1.9$ in the source); over $D/h=10$--$40$ together it converges at $1.0$ in the vortex and $1.2$ in the source. This first order comes from the pure-cell tolerance, not from the volume error of the front-derived indicator, which converges at about second order: the projection sets the cells of the indicator's far tail, $0<S^{\mathrm{FT}}\le\varepsilon^{\mathrm{pure}}$, to zero, and the shift $\lambda_I$ returns their volume to the mixed cells, a change that scales with the width of the interfacial band. Asymptotically, the indicator projection therefore limits the order of the proposed method to about one. This is still above the sub-linear rate of the standard method, so that its error, already one to two orders of magnitude smaller at $D/h=40$, keeps decreasing faster with refinement. Section~\ref{sec:ablation} shows that the order of the concentration part needs both the transport of the phase-averaged concentration and its linear extension.

\begin{table}[htbp]
\centering
\caption{Split of the scalar--front error of the proposed method into the concentration part $S^{n+1}(q-q^{\mathrm{ex}})$ and the support part $(S^{n+1}-S^{\mathrm{FT}})q^{\mathrm{ex}}$, both normalized as $E_{C,1}$, at the last stored field, and the share of the support part; order: convergence rates from $D/h=10$ to $20$ and from $20$ to $40$.}
\label{tab:error_split}
\small
\setlength{\tabcolsep}{4pt}
\begin{tabular}{lrrrrrrrr}
\toprule
 & \multicolumn{4}{c}{reversible vortex, $t=2$} & \multicolumn{4}{c}{localized source, $t=3.2$} \\
\cmidrule(lr){2-5}\cmidrule(lr){6-9}
$D/h$ & $E_{C,1}$ & concentration & support & share (\%) & $E_{C,1}$ & concentration & support & share (\%) \\
\midrule
10 & $7.6\times10^{-3}$ & $7.4\times10^{-3}$ & $5.8\times10^{-4}$ & 8 & $5.6\times10^{-2}$ & $5.6\times10^{-2}$ & $5.6\times10^{-4}$ & 1 \\
20 & $2.3\times10^{-3}$ & $2.2\times10^{-3}$ & $3.1\times10^{-4}$ & 14 & $8.4\times10^{-3}$ & $8.2\times10^{-3}$ & $3.8\times10^{-4}$ & 5 \\
40 & $6.4\times10^{-4}$ & $5.9\times10^{-4}$ & $1.4\times10^{-4}$ & 22 & $1.5\times10^{-3}$ & $1.4\times10^{-3}$ & $1.0\times10^{-4}$ & 7 \\
order & 1.7, 1.8 & 1.8, 1.9 & 0.9, 1.1 & & 2.7, 2.5 & 2.8, 2.6 & 0.5, 1.9 & \\
\bottomrule
\end{tabular}
\end{table}

The standard method converges below first order in all three cases, at $0.6$--$0.9$, because it transports the jump of $C$ at the interface. A jump that the flow only carries, without steepening it, is a linearly degenerate discontinuity, for which capturing schemes converge sub-linearly \citep{Banks2008}. The leading error of a scheme of order $p$ smears the jump into a layer that keeps widening in time, as $(ht)^{1/2}$ for first-order upwinding, whose leading error acts as a diffusion proportional to $h$ \citep{LeVeque2002}. The $L_1$ error therefore decreases only as $h^{p/(p+1)}$, from $1/2$ for $p=1$ to $5/6$ for $p=5$, and the $L_\infty$ error does not decrease at all. The standard method behaves accordingly: its rates, $0.6$--$0.9$, are of this size, and its largest concentration error grows with the resolution in Tables~\ref{tab:translation_errors}--\ref{tab:source_errors}. WENO-Z, whose order drops next to the jump (Section~\ref{sec:wenoz}), is not a linear scheme of fixed order, so that its rate does not follow the formula exactly, but it remains below first order. The proposed method transports the phase-averaged concentration, which has no jump at the interface, and is therefore not bound by this limit.

\subsection{Translation along one axis}
\label{sec:grid_aligned_translation}

The first case is the configuration of Section~\ref{sec:motivation}. A sphere of diameter $D=0.5$, initially centred at $(1,2,2)$ in the domain $[0,5]\times[0,4]\times[0,4]$, periodic in all three directions, is translated with the prescribed velocity $\mathbf{U}=(1,0,0)$ until $t=3$, when its centre reaches $(4,2,2)$, six diameters downstream. The advective CFL number is $0.2$ at every resolution. The phase-averaged concentration is uniform, $q^{\mathrm{ex}}=q_{\mathrm{ref}}=1$, so $C^{\mathrm{ex}}=S^{\mathrm{FT}}$.

By construction, the proposed method preserves a uniform concentration exactly: the shift $\lambda_C$ vanishes to round-off, and the rebuilt scalar is the projected support, $C^{n+1}=S^{n+1}$. Its scalar--front error is therefore the support error $E_{S,1}$ (Table~\ref{tab:translation_errors}). For a uniform concentration the advection of the concentration is exact with any consistent scheme, so that this case, like the uniform concentration in the rising bubble (Section~\ref{sec:rising_bubble}), tests the rebuild on the projected support rather than the transport of the concentration; the effect of transporting the phase-averaged concentration is shown by the non-uniform concentrations of Sections~\ref{sec:reversible_vortex} and~\ref{sec:source_transport}. The support error measures how closely the projected support, which keeps the transported phase volume, follows the support reconstructed from the front, and it converges at about $1.2$. The error of the standard method is $150$--$300$ times larger and converges at an order of $0.7$ (Fig.~\ref{fig:convergence}(a)).

\begin{table}[htbp]
\centering
\caption{Translation along one axis: errors of the proposed and standard methods at $t=3$. Order: convergence rates from $D/h=10$ to $20$ and from $20$ to $40$; for the uniform concentration, $E_{C,1}$ of the proposed method equals its support error $E_{S,1}$. Exact: the leakage of the exact field.}
\label{tab:translation_errors}
\small
\setlength{\tabcolsep}{4pt}
\begin{tabular}{lrrrrrrrrrr}
\toprule
 & \multicolumn{2}{c}{$E_{C,1}$} & \multicolumn{2}{c}{$E_{q,1}$} & \multicolumn{2}{c}{$E_{q,\infty}$} & $E_{S,1}$ & \multicolumn{3}{c}{$E_{\mathrm{leak}}$ (\%)} \\
\cmidrule(lr){2-3}\cmidrule(lr){4-5}\cmidrule(lr){6-7}\cmidrule(lr){9-11}
$D/h$ & proposed & standard & proposed & standard & proposed & standard & & exact & proposed & standard \\
\midrule
10 & $7.8\times10^{-4}$ & 0.121 & $2.5\times10^{-5}$ & 0.095 & $9.5\times10^{-4}$ & 3.95 & $7.8\times10^{-4}$ & 0.35 & 0.36 & 2.01 \\
20 & $3.5\times10^{-4}$ & 0.075 & $1.3\times10^{-5}$ & 0.060 & $9.8\times10^{-4}$ & 12.81 & $3.5\times10^{-4}$ & 0.16 & 0.16 & 1.10 \\
40 & $1.5\times10^{-4}$ & 0.046 & $6.5\times10^{-6}$ & 0.036 & $1.1\times10^{-3}$ & 15.03 & $1.5\times10^{-4}$ & 0.06 & 0.06 & 0.68 \\
order & $=E_{S,1}$ & 0.7, 0.7 & & & & & 1.2, 1.2 & & & \\
\bottomrule
\end{tabular}
\end{table}

\begin{figure}[htbp]
\centering
\includegraphics[width=\textwidth]{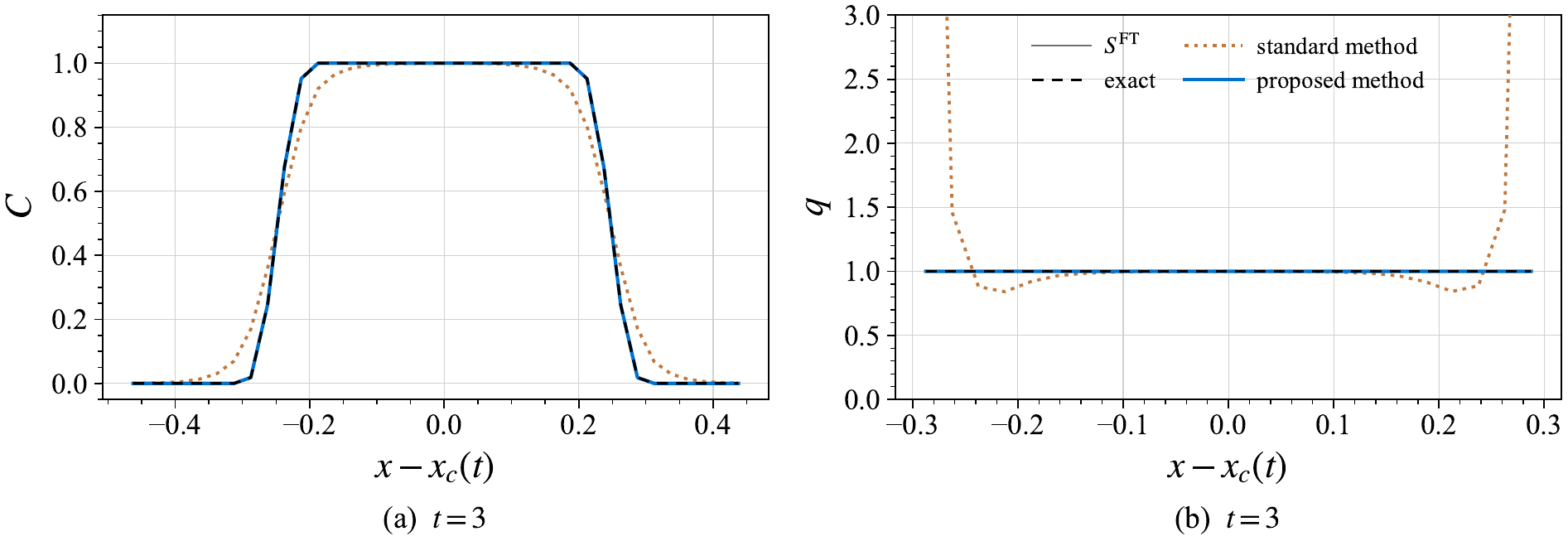}
\caption{Uniform translation at $D/h=20$ and $t=3$: (a) cell-averaged concentration $C$ and (b) phase-averaged concentration $q=C/S^{\mathrm{FT}}$ along the translation direction through the sphere centre. Values are taken at the centres of the cells crossed by the line, and $q$ is shown where $S^{\mathrm{FT}}>10^{-2}$.}
\label{fig:xtranslation_profiles}
\end{figure}

Figure~\ref{fig:xtranslation_profiles} shows the profiles at $t=3$. The scalar of the proposed method coincides with $S^{\mathrm{FT}}$ (Fig.~\ref{fig:xtranslation_profiles}(a)), and its concentration is unity across the sphere (Fig.~\ref{fig:xtranslation_profiles}(b)). The standard method smears the scalar beyond the front (Fig.~\ref{fig:xtranslation_profiles}(a)). Its concentration (Fig.~\ref{fig:xtranslation_profiles}(b)) falls below unity just inside the interface and rises steeply in the outer interfacial cells, where the smeared scalar exceeds the front-derived support; this peak of $q$ grows with refinement as the scalar reaches cells of smaller support ($E_{q,\infty}$ in Table~\ref{tab:translation_errors}). The standard method places $0.7$--$2.0\%$ of its mass outside the phase and produces small negative values, whereas the scalar of the proposed method matches the exact share outside the phase to within $0.012$ percentage points and is non-negative by construction.

\subsection{Reversible deformation of a concentration gradient}
\label{sec:reversible_vortex}

The second case carries a non-uniform concentration through a deforming interface; at the largest deformation, $t=1$, the drop is stretched to an aspect ratio of about $1.8$. A sphere of diameter $D=0.5$ centred at $\mathbf{x}_c=(2,3,2)$ in the domain $[0,4]^3$ is deformed by the single-vortex flow of \citet{LeVeque1996}, extruded in $z$,
\begin{equation} \mathbf{u}(\mathbf{x},t)=\sin\!\left(\frac{2\pi t}{T}\right)\left(-\frac{\partial\psi}{\partial y},\,\frac{\partial\psi}{\partial x},\,0\right),\qquad \psi=\frac{4}{\pi}\sin^2\!\left(\frac{\pi x}{4}\right)\sin^2\!\left(\frac{\pi y}{4}\right), \label{eq:vortex_velocity}\end{equation}
with $T=2$. The flow stretches the sphere until $t=T/2$ and returns it to its initial shape and position at $t=T$. The face fluxes are evaluated from the stream function at the face corners, so that they are exactly divergence-free, and the front points move with the analytical velocity at their own position. The initial concentration is a linear gradient across the sphere, oblique to the grid,
\begin{equation} q_0(\mathbf{x})=\min\!\left[0.9,\ \max\!\left(0.1,\ 0.5+0.4\,\frac{(\mathbf{x}-\mathbf{x}_c)\cdot\widehat{\mathbf{d}}}{D/2}\right)\right],\qquad \widehat{\mathbf{d}}=\frac{(0.5,0.3,0.2)}{\lVert(0.5,0.3,0.2)\rVert}, \label{eq:vortex_initial_q}\end{equation}
so that $q_0$ ranges from $0.1$ to $0.9$ inside the sphere and $q_{\mathrm{ref}}=0.5$. Without diffusion, $q$ is constant along particle paths. Since the time dependence of Eq.~\eqref{eq:vortex_velocity} is a common factor, a particle at $\mathbf{x}$ at time $t$ was at the point $\mathbf{X}$ reached by integrating the steady field backwards over the pseudo-time $\theta(t)=T(1-\cos(2\pi t/T))/(2\pi)$, and $q^{\mathrm{ex}}(\mathbf{x},t)=q_0(\mathbf{X})$. The backward integration uses the classical fourth-order Runge--Kutta method with steps of at most $2.5\times10^{-3}$; at $t=T$, $\theta=0$ and $q^{\mathrm{ex}}=q_0$. The CFL number is at most $0.1$ at every resolution.

\begin{figure}[htbp]
\centering
\includegraphics[width=\textwidth]{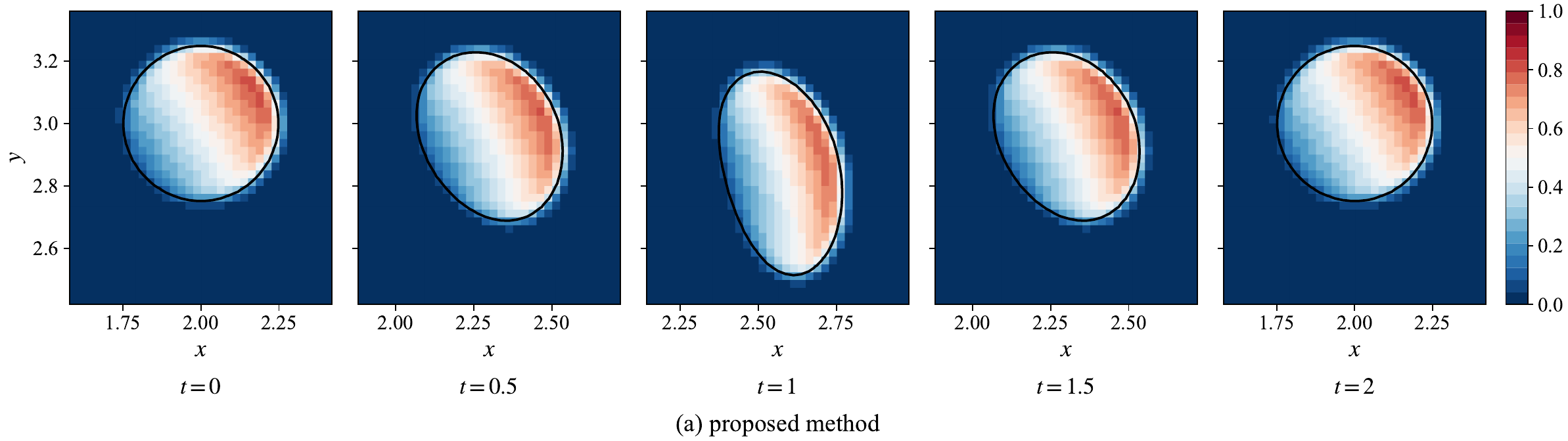}\\[2mm]
\includegraphics[width=\textwidth]{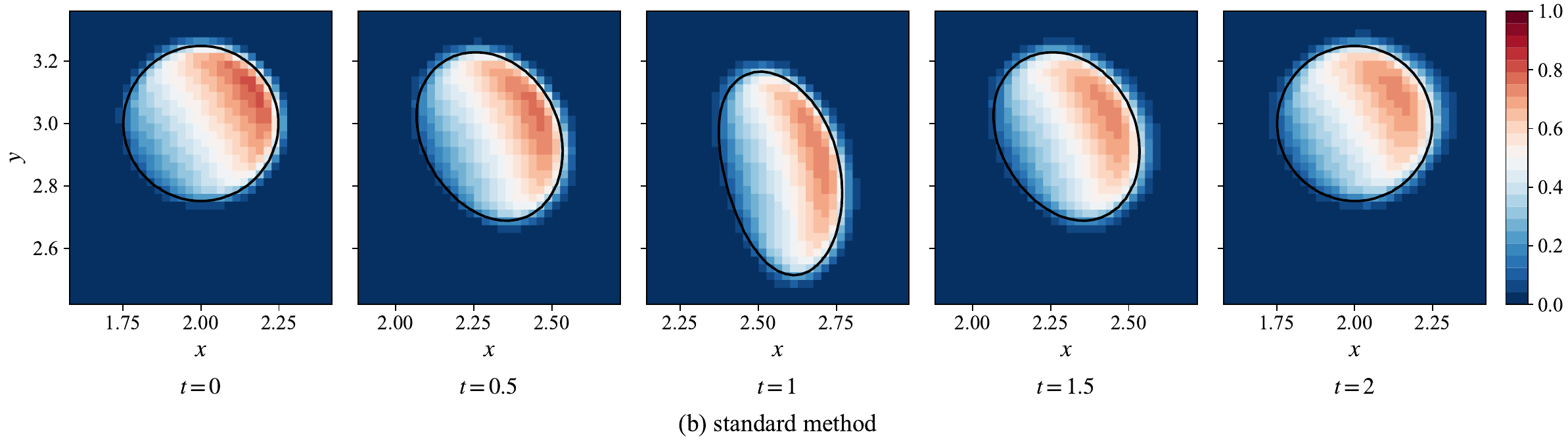}
\caption{Reversible vortex at $D/h=20$: scalar $C$ in the plane $z=2$ at $t=0$, $0.5$, $1$ (largest deformation), $1.5$ and $2$ (return, when the exact field is the initial one): (a) proposed method, (b) standard method. Each panel spans the same range in $y$ and is centred on the drop in $x$. The black line is $S^{\mathrm{FT}}=0.5$.}
\label{fig:vortex_fields}
\end{figure}

Figure~\ref{fig:vortex_fields} shows the scalar in the plane $z=2$ at $D/h=20$ over the cycle, from $t=0$ through the largest deformation, $t=1$, to the return, $t=2$, when the exact field is the initial one. The scalar of the proposed method follows the deformation and returns to the initial pattern (Fig.~\ref{fig:vortex_fields}(a)). The scalar of the standard method is smeared across the interface, increasingly with time (Fig.~\ref{fig:vortex_fields}(b)); at $t=2$ its high-concentration side is flattened, and in the outer interfacial cells its concentration lies far above the exact maximum. At $t=2$ the error of the proposed method converges at an order of $1.7$--$1.8$, against $0.6$ for the standard method, whose error is $17$--$83$ times larger (Table~\ref{tab:vortex_errors} and Fig.~\ref{fig:convergence}(b)). Unlike for a uniform concentration, the error of the proposed method exceeds the support error and is set by the transport of the concentration.

\begin{table}[htbp]
\centering
\caption{Reversible vortex: errors of the proposed and standard methods at $t=2$. Columns as in Table~\ref{tab:translation_errors}.}
\label{tab:vortex_errors}
\small
\setlength{\tabcolsep}{4pt}
\begin{tabular}{lrrrrrrrrrr}
\toprule
 & \multicolumn{2}{c}{$E_{C,1}$} & \multicolumn{2}{c}{$E_{q,1}$} & \multicolumn{2}{c}{$E_{q,\infty}$} & $E_{S,1}$ & \multicolumn{3}{c}{$E_{\mathrm{leak}}$ (\%)} \\
\cmidrule(lr){2-3}\cmidrule(lr){4-5}\cmidrule(lr){6-7}\cmidrule(lr){9-11}
$D/h$ & proposed & standard & proposed & standard & proposed & standard & & exact & proposed & standard \\
\midrule
10 & $7.6\times10^{-3}$ & 0.128 & $5.1\times10^{-3}$ & 0.099 & 0.169 & 11.69 & $5.9\times10^{-4}$ & 0.34 & 0.34 & 2.14 \\
20 & $2.3\times10^{-3}$ & 0.082 & $1.4\times10^{-3}$ & 0.065 & 0.103 & 21.54 & $3.1\times10^{-4}$ & 0.15 & 0.15 & 1.21 \\
40 & $6.4\times10^{-4}$ & 0.053 & $3.4\times10^{-4}$ & 0.042 & 0.068 & 25.06 & $1.4\times10^{-4}$ & 0.06 & 0.06 & 0.81 \\
order & 1.7, 1.8 & 0.6, 0.6 & & & & & 0.9, 1.1 & & & \\
\bottomrule
\end{tabular}
\end{table}

\begin{figure}[htbp]
\centering
\includegraphics[width=0.75\textwidth]{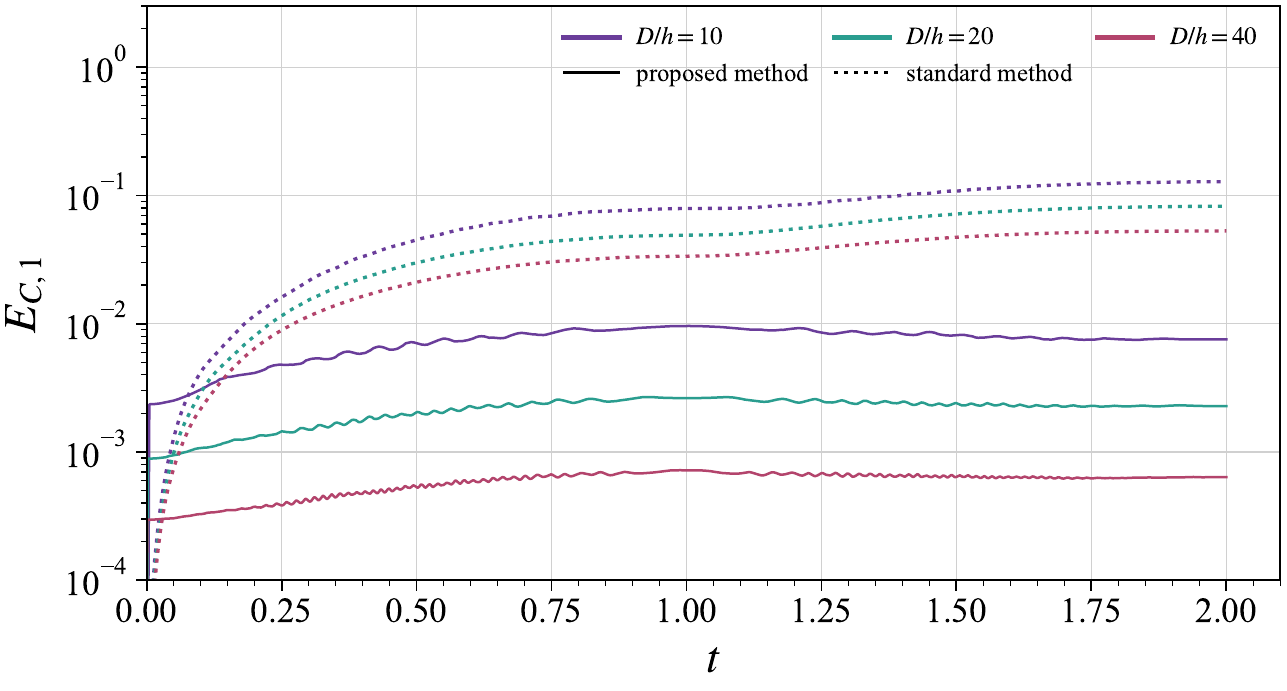}
\caption{Reversible vortex: scalar--front error $E_{C,1}$ in time of the proposed method (solid) and the standard method (dotted) at $D/h=10$, $20$ and $40$. The error of the proposed method starts from a non-zero value after the first step, from the tail of the indicator (see text).}
\label{fig:vortex_errors}
\end{figure}

Figure~\ref{fig:vortex_errors} shows the error in time. The error of the proposed method is non-zero already after the first step, before the flow, whose velocity vanishes at $t=0$, has moved the drop. This offset comes from the cells of the indicator's tail, $S^{\mathrm{FT}}\le\varepsilon^{\mathrm{rec}}$, where the method carries extended values while the exact field evaluates the initial gradient, and it accounts for $31$--$47\%$ of the error at $t=2$. Without these cells, which hold $1.3$, $0.5$ and $0.2\%$ of the mass at $D/h=10$, $20$ and $40$, $E_{C,1}$ at $t=2$ is $4.9\times10^{-3}$, $1.3\times10^{-3}$ and $3.4\times10^{-4}$ and converges at $1.9$ and $2.0$, while that of the standard method is $22$ to $127$ times larger. The largest concentration error, $E_{q,\infty}$, also lies in such cells, at the two ends of the gradient, where $q_0$ of Eq.~\eqref{eq:vortex_initial_q} is clipped; on $S^{\mathrm{FT}}>0.5$ it is six to seven times smaller but still converges more slowly than first order. The leakage of the proposed method equals that of the exact field, against $0.8$--$2.1\%$ for the standard method.

\begin{figure}[htbp]
\centering
\includegraphics[width=\textwidth]{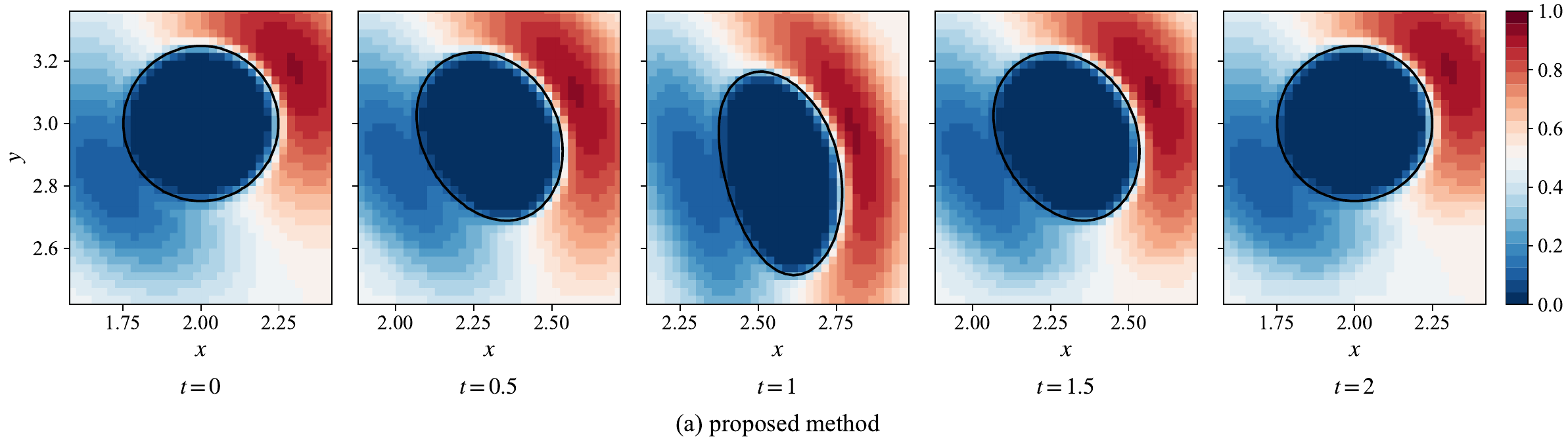}\\[2mm]
\includegraphics[width=\textwidth]{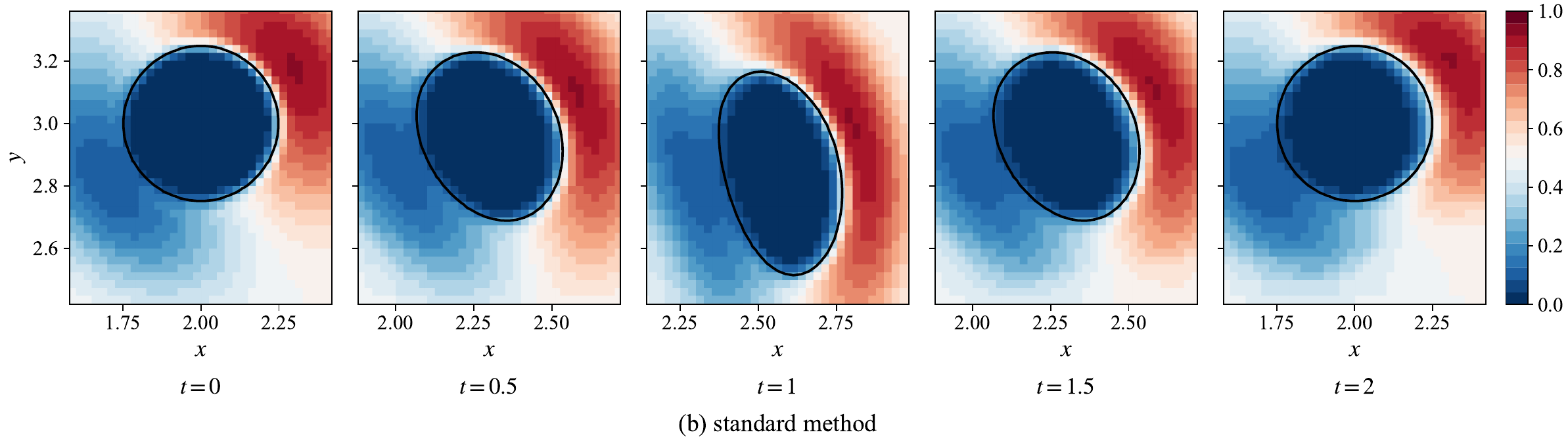}
\caption{Reversible vortex with the scalar carried by the phase outside the drop (gradient localized around the drop) at $D/h=20$: scalar $C$ in the plane $z=2$ at $t=0$, $0.5$, $1$ (largest deformation), $1.5$ and $2$: (a) proposed method, (b) standard method. Each panel spans the same range in $y$ and is centred on the drop in $x$. The black line is the interface, $I^{\mathrm{FT}}=0.5$.}
\label{fig:vortex_outer_fields}
\end{figure}

The method treats the two phases alike: for a scalar carried by the phase outside the drop, the support is $S=1-I$, Eq.~\eqref{eq:scalar_support_definition}, and the extension fills the interior of the drop. The vortex was therefore repeated at $D/h=20$ with the scalar in the phase outside the drop, initialized as $C=(1-I^{\mathrm{FT}})q_0$ with the gradient of Eq.~\eqref{eq:vortex_initial_q} localized around the drop, $q_0=0.5+0.4\,[(\mathbf{x}-\mathbf{x}_c)\cdot\widehat{\mathbf{d}}/(D/2)]\exp[-(|\mathbf{x}-\mathbf{x}_c|/D)^4]$, so that the concentration far from the drop is uniform (Fig.~\ref{fig:vortex_outer_fields}(a) for the proposed and (b) for the standard method). At the scale of the figure the two scalars look alike, since they differ mainly in the interfacial cells. This concentration also varies in the coarse cells away from the front, where both scalars carry the same transport error, so the two are compared in the interfacial band $10^{-3}<I^{\mathrm{FT}}<1-10^{-3}$, with the error in the band divided by $M_d=q_{\mathrm{ref}}\sum_iI_i^{\mathrm{FT}}V_i$, the mass of the drop filled with the concentration $q_{\mathrm{ref}}$. In the band, the error of the proposed method is $64$ and $141$ times smaller than that of the standard method at $t=1$ and $2$. The standard method carries a mass of $7.4\times10^{-3}M_d$ into the drop, whereas the proposed method places there only the share of the exact field. The method thus keeps a scalar carried by the phase outside the drop consistent with the front as well.

\subsection{Oblique translation with a localized scalar source}
\label{sec:source_transport}

The third case adds a source, so that the concentration changes in time and the total mass grows. A sphere of diameter $D=0.5$ is translated through the domain $[0,4]^3$, periodic in all three directions, with the constant velocity $\mathbf{U}=(0.5,0.3,0.2)$, oblique to the coordinate axes. Its centre moves as $\mathbf{x}_c(t)=\mathbf{x}_c(0)+\mathbf{U}t$ with $\mathbf{x}_c(0)=(1,1,1)$, and the calculations are advanced until the sphere has travelled four diameters, $T=4D/\lVert\mathbf{U}\rVert\approx3.244$; the comparisons use the last stored field, at $t=3.2$. The CFL number, based on the largest velocity component, is $0.1$ at every resolution. The concentration is initially uniform, $q_0=q_{\mathrm{ref}}=2$, and a localized source moving with the sphere is added,
\begin{equation} \dot{m}_C(\mathbf{x},t)=3.0822\,S(\mathbf{x},t)\exp\!\left[-\frac{|\mathbf{x}-\mathbf{x}_s(t)|^2}{0.005}\right], \label{eq:source_term}\end{equation}
where $\mathbf{x}_s(t)=\mathbf{x}_c(t)+0.7R\,\widehat{\mathbf{U}}$ lies between the centre and the leading surface of the sphere; $R=D/2$ is its radius and $\widehat{\mathbf{U}}=\mathbf{U}/\lVert\mathbf{U}\rVert$. The support $S$ is the corrected support at the start of the step, $S^n$, the same field for both methods, and the source is evaluated at the cell centres at the new time level. The Gaussian has a standard deviation of $0.05$, equal to the cell size at $D/h=10$, so that $0.005=2\times0.05^2$; the amplitude $3.0822=10/T$ makes the exact peak reach $12$ at $t=T$. The exact concentration is
\begin{equation} q^{\mathrm{ex}}(\mathbf{x},t)=2+3.0822\,t\,\exp\!\left[-\frac{|\mathbf{x}-\mathbf{x}_s(t)|^2}{0.005}\right], \label{eq:source_exact_q}\end{equation}
with the peak value $q^{\mathrm{ex}}_{\max}=12$ at $t=T$. Since the Gaussian is only one cell wide at $D/h=10$, the exact concentration is averaged over each cell in all comparisons, as the transported scalar is; over a cell, Eq.~\eqref{eq:source_exact_q} integrates to a product of error functions. At $t=3.2$, the cell-averaged peak is $9.0$, $11.3$ and $11.8$ at $D/h=10$, $20$ and $40$, against the point value $11.9$.

\begin{figure}[htbp]
\centering
\includegraphics[width=\textwidth]{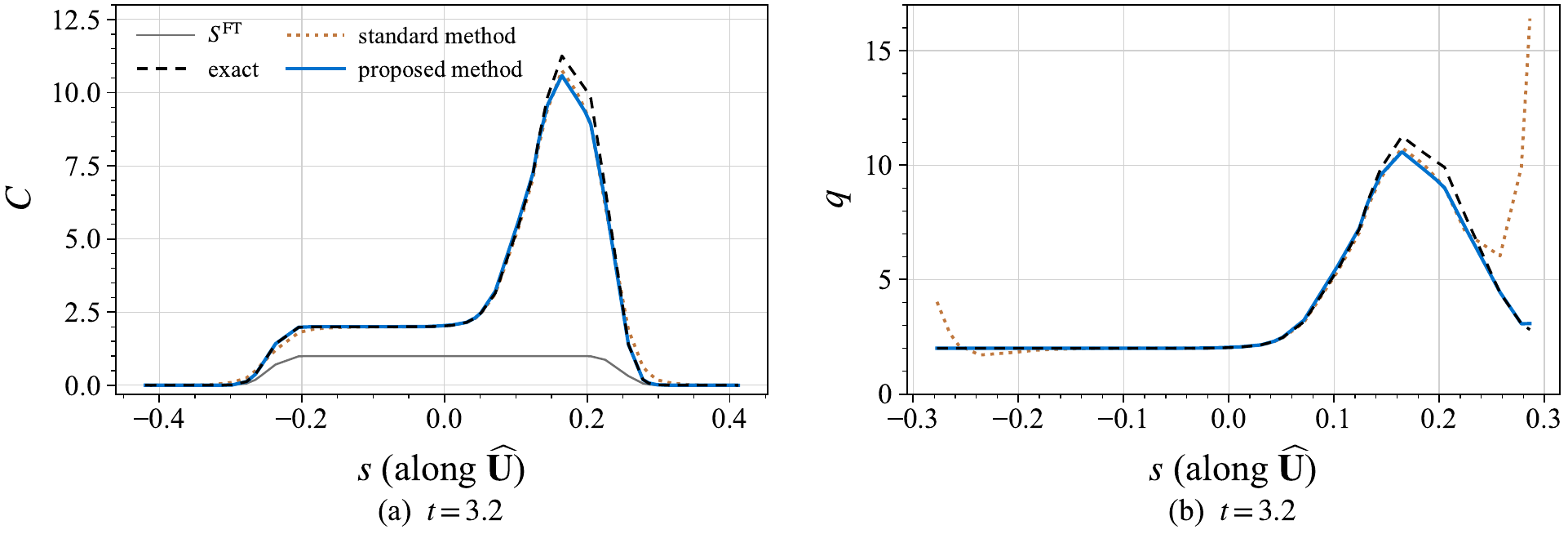}
\caption{Localized source at $D/h=20$ and $t=3.2$: (a) cell-averaged concentration $C$ and (b) phase-averaged concentration $q=C/S^{\mathrm{FT}}$ along the translation direction $\widehat{\mathbf{U}}$ through the source centre. The coordinate $s$ is measured from the sphere centre; $q$ is shown where $S^{\mathrm{FT}}>10^{-2}$.}
\label{fig:source_profiles}
\end{figure}

\begin{figure}[htbp]
\centering
\includegraphics[width=\textwidth]{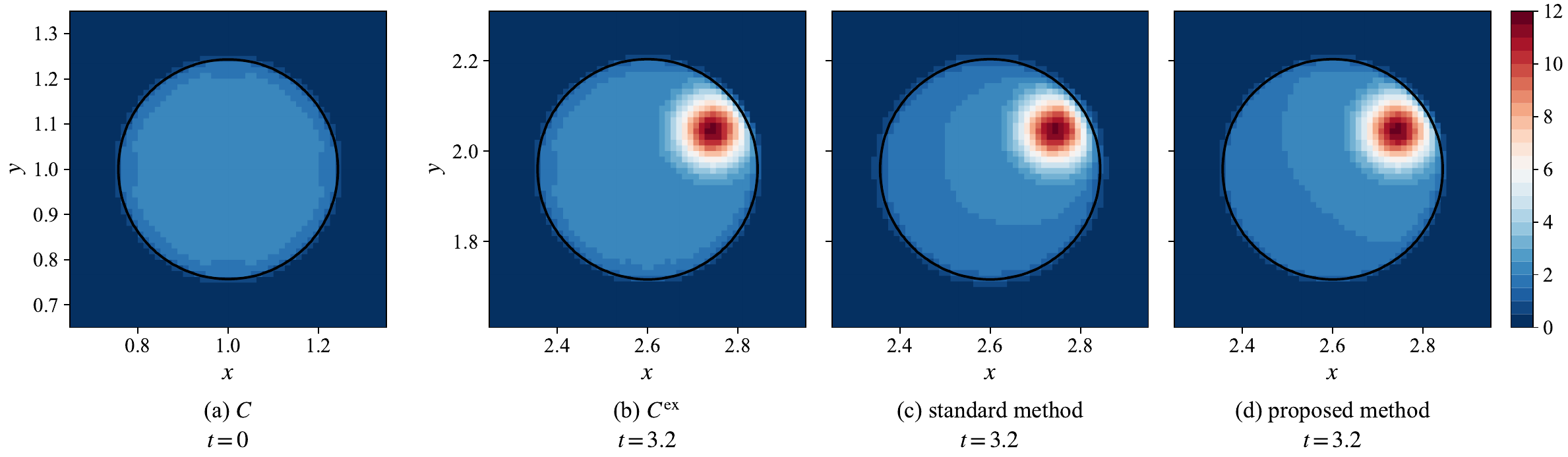}
\caption{Localized source at $D/h=40$: scalar $C$ in the plane $z=z_s(t)$ through the source centre, (a) at $t=0$, and (b) exact, (c) standard method and (d) proposed method at $t=3.2$. The black line is $S^{\mathrm{FT}}=0.5$.}
\label{fig:source_fields}
\end{figure}

Figure~\ref{fig:source_profiles} shows the scalar (a) and the concentration (b) along the direction of motion through the source centre at $D/h=20$, and Fig.~\ref{fig:source_fields} shows the scalar in the plane through the source centre at $D/h=40$, initially (a) and at $t=3.2$ for the exact solution (b), the standard method (c) and the proposed method (d). In the scalar, both methods carry the source peak with the sphere, slightly below the exact peak (Figs.~\ref{fig:source_profiles}(a) and \ref{fig:source_fields}(c,d)), and the concentration $q=C/S^{\mathrm{FT}}$ separates them at the interface (Fig.~\ref{fig:source_profiles}(b)). The concentration of the proposed method follows the exact one across the sphere (Fig.~\ref{fig:source_profiles}(b)), including the interfacial cells next to the source, with a peak limited by the resolution of the narrow source. The concentration of the standard method has spurious maxima in the interfacial cells (Fig.~\ref{fig:source_profiles}(b)) and is too low over a broad region inside the interface. The error of the proposed method converges at orders of $2.5$--$2.7$, against $0.7$--$0.9$ for the standard method (Table~\ref{tab:source_errors} and Fig.~\ref{fig:convergence}(c)): at $D/h=10$ the source spans one cell and is resolved by neither scalar, and the error of the proposed method then decreases faster than second order as the source becomes resolved. The support error is one to two orders of magnitude smaller than $E_{C,1}$, so that the error is set by the transport of the concentration. The difference between the methods lies at the interface: in the cells with $10^{-2}<S^{\mathrm{FT}}\le0.99$, the concentration error of the standard method remains near $0.2q_{\mathrm{ref}}$ at every resolution, whereas that of the proposed method decreases from $0.057q_{\mathrm{ref}}$ to $3\times10^{-3}q_{\mathrm{ref}}$. Only a few cells of small support at the leading interface next to the source inherit the extension error of the steep source profile (Section~\ref{sec:scalar_correction}).

\begin{table}[htbp]
\centering
\caption{Localized source: errors of the proposed and standard methods at $t=3.2$. Columns as in Table~\ref{tab:translation_errors}.}
\label{tab:source_errors}
\small
\setlength{\tabcolsep}{4pt}
\begin{tabular}{lrrrrrrrrrr}
\toprule
 & \multicolumn{2}{c}{$E_{C,1}$} & \multicolumn{2}{c}{$E_{q,1}$} & \multicolumn{2}{c}{$E_{q,\infty}$} & $E_{S,1}$ & \multicolumn{3}{c}{$E_{\mathrm{leak}}$ (\%)} \\
\cmidrule(lr){2-3}\cmidrule(lr){4-5}\cmidrule(lr){6-7}\cmidrule(lr){9-11}
$D/h$ & proposed & standard & proposed & standard & proposed & standard & & exact & proposed & standard \\
\midrule
10 & 0.056 & 0.196 & 0.062 & 0.173 & 1.68 & 6.95 & $6.2\times10^{-4}$ & 0.32 & 0.29 & 3.15 \\
20 & $8.4\times10^{-3}$ & 0.107 & $9.0\times10^{-3}$ & 0.095 & 0.450 & 12.59 & $4.3\times10^{-4}$ & 0.14 & 0.12 & 1.59 \\
40 & $1.5\times10^{-3}$ & 0.068 & $1.5\times10^{-3}$ & 0.061 & 0.256 & 14.55 & $1.1\times10^{-4}$ & 0.05 & 0.05 & 1.01 \\
order & 2.7, 2.5 & 0.9, 0.7 & & & & & 0.5, 1.9 & & & \\
\bottomrule
\end{tabular}
\end{table}

\subsection{Rising bubble with phase-confined scalars}
\label{sec:rising_bubble}

In the fourth case the interface motion is not prescribed but follows from the coupled flow solution. A bubble of diameter $D$ is released at $(1.5,2,2)$ in the domain $[0,8]\times[0,4]\times[0,4]$, periodic in $y$ and $z$, with gravity in the $-x$ direction, $\mathrm{E\ddot{o}}=129$, $\mathrm{Ga}=23.5$ and the density and viscosity ratios of Table~\ref{tab:dimensionless_groups}. The bubble rises along $x$ and deforms into a spherical cap; by $t^*\approx9.4$ it has risen by $4.7$--$5.0$ diameters. The time step follows from a CFL number of $0.1$, with an upper bound of $\Delta t^*=0.031$. The scalar has a uniform concentration, $q^{\mathrm{ex}}=q_0$, which also serves as the reference concentration $q_{\mathrm{ref}}$, and is carried by the proposed and standard methods in the same calculation. Since the scalar is passive, the value of $q_{\mathrm{ref}}$ has no physical significance; the fields are shown as $C/q_{\mathrm{ref}}$, which for the exact solution equals the support $S^{\mathrm{FT}}$.

\begin{figure}[htbp]
\centering
\includegraphics[width=\textwidth]{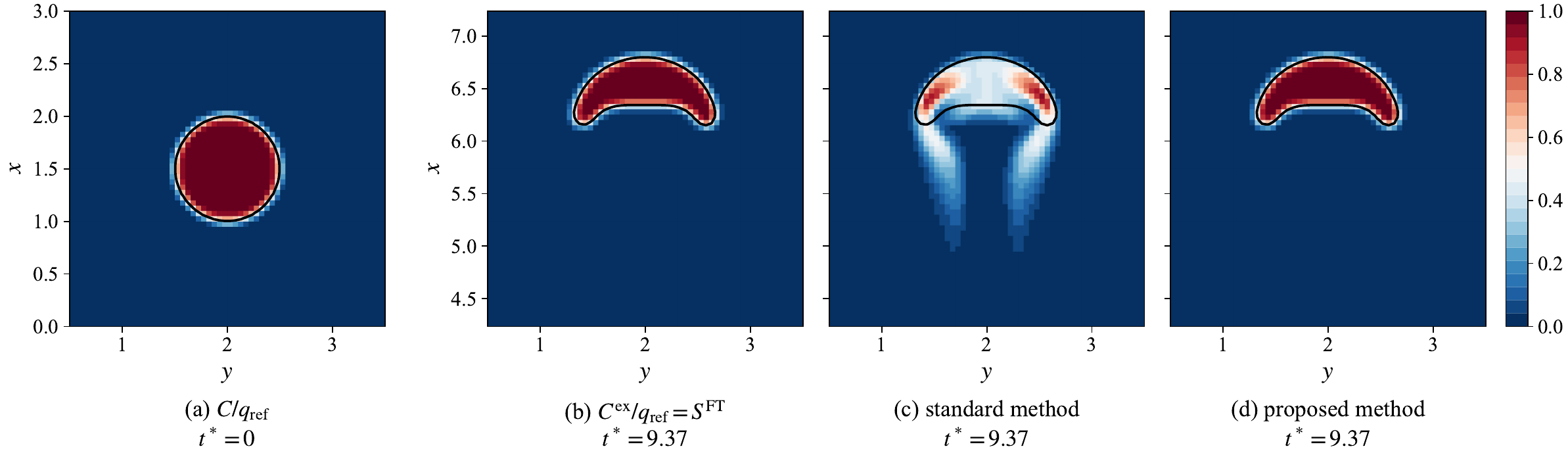}
\caption{Rising bubble at $D/h=20$, uniform concentration, scalar $C/q_{\mathrm{ref}}$ in the mid-plane $z=2$: (a) at $t^*=0$, and (b) exact, $C^{\mathrm{ex}}/q_{\mathrm{ref}}=S^{\mathrm{FT}}$, (c) standard method, which leaves a wake, and (d) proposed method at $t^*=9.37$. The rise direction $x$ points upwards. The black line is $S^{\mathrm{FT}}=0.5$.}
\label{fig:bubble_sections}
\end{figure}

\begin{figure}[htbp]
\centering
\includegraphics[width=0.75\textwidth]{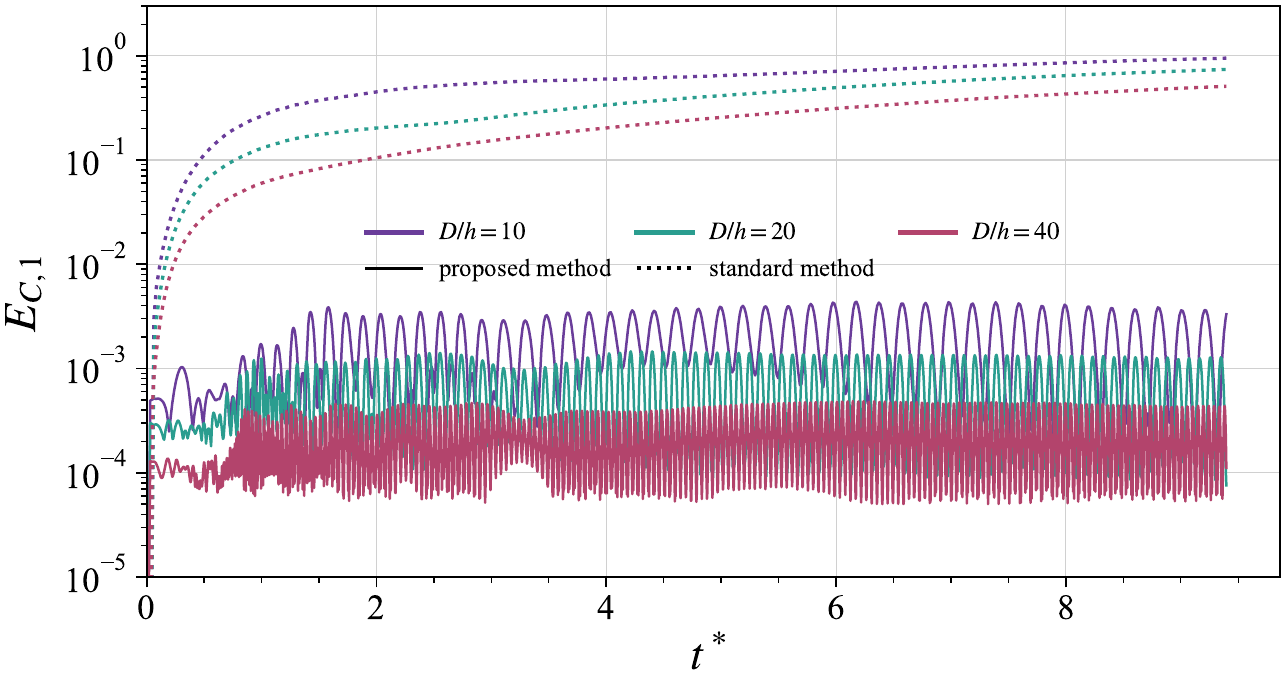}
\caption{Rising bubble with a uniform concentration: history of the scalar--front error $E_{C,1}$ of the proposed method (solid) and the standard method (dotted) at $D/h=10$, $20$ and $40$. The error of the proposed method equals the support error $E_{S,1}$ and oscillates with the position of the interface relative to the cells, so that values at a single instant are not compared across resolutions.}
\label{fig:bubble_errors}
\end{figure}

Figure~\ref{fig:bubble_sections} shows the scalar in the mid-plane $z=2$ at $t^*=0$ (a) and $t^*=9.37$, where the exact scalar is the support $S^{\mathrm{FT}}$ (b). The scalar of the standard method is left behind in a wake (Fig.~\ref{fig:bubble_sections}(c)) that holds $23$--$34\%$ of its mass at $t^*\approx9.4$ (Table~\ref{tab:bubble_errors}), and its concentration at the interface reaches tens of times $q_{\mathrm{ref}}$ and, at $D/h=40$, negative values. The scalar of the proposed method remains on the bubble, and its own concentration, $C/S^{n+1}$, stays uniform to round-off (Fig.~\ref{fig:bubble_sections}(d)); measured on the front-derived support, $q=C/S^{\mathrm{FT}}$ differs from $q_0$ only through the support error (Table~\ref{tab:bubble_errors}). Its scalar--front error is again the support error, about three orders of magnitude below that of the standard method (Table~\ref{tab:bubble_errors}). This error oscillates by up to an order of magnitude as the interface crosses the cells (Fig.~\ref{fig:bubble_errors}). The ratio of the errors of the two methods therefore depends on the instant: at $D/h=40$ it varies between $130$ and $9200$ once the bubble rises, $t^*>0.5$, with a median of about $1100$, against $1700$ at $t^*\approx9.4$. The projected support keeps the transported volume to round-off, whereas the cell sum of $S^{\mathrm{FT}}$ depends on the position of the front relative to the cell centres and is periodic in it with period $h$. The support error therefore oscillates with the period in which the bubble rises by one cell, and values at a single instant are not compared across resolutions.

\begin{table}[htbp]
\centering
\caption{Rising bubble, uniform concentration: errors of the proposed and standard methods at $t^*=9.33$, $9.37$ and $9.39$ for $D/h=10$, $20$ and $40$. Columns as in Table~\ref{tab:translation_errors}; no rate is given for $E_{S,1}$, which oscillates with the cell crossing (Fig.~\ref{fig:bubble_errors}).}
\label{tab:bubble_errors}
\small
\setlength{\tabcolsep}{4pt}
\begin{tabular}{lrrrrrrrrrr}
\toprule
 & \multicolumn{2}{c}{$E_{C,1}$} & \multicolumn{2}{c}{$E_{q,1}$} & \multicolumn{2}{c}{$E_{q,\infty}$} & $E_{S,1}$ & \multicolumn{3}{c}{$E_{\mathrm{leak}}$ (\%)} \\
\cmidrule(lr){2-3}\cmidrule(lr){4-5}\cmidrule(lr){6-7}\cmidrule(lr){9-11}
$D/h$ & proposed & standard & proposed & standard & proposed & standard & & exact & proposed & standard \\
\midrule
10 & $2.2\times10^{-4}$ & 0.942 & $3.1\times10^{-6}$ & 0.554 & $1.1\times10^{-4}$ & 18.14 & $2.2\times10^{-4}$ & 0.30 & 0.29 & 34.49 \\
20 & $9.5\times10^{-4}$ & 0.736 & $9.3\times10^{-5}$ & 0.411 & $5.8\times10^{-3}$ & 45.77 & $9.5\times10^{-4}$ & 0.16 & 0.08 & 30.62 \\
40 & $3.0\times10^{-4}$ & 0.508 & $3.0\times10^{-5}$ & 0.274 & $3.1\times10^{-3}$ & 38.53 & $3.0\times10^{-4}$ & 0.06 & 0.03 & 22.90 \\
order & $=E_{S,1}$ & 0.4, 0.5 & & & & & -- & & & \\
\bottomrule
\end{tabular}
\end{table}

\subsection{Rising bubble with a soluble surfactant}
\label{sec:surfactant_bubble}

The last case couples the scalar to the interface through adsorption. A soluble surfactant is dissolved in the bubble, with the bulk concentration $q$ and the cell-averaged concentration $C=Sq$, and adsorbs onto the interface, where its surface concentration $\Gamma$ obeys \citep{Muradoglu2008,Muradoglu2014}
\begin{equation}
\frac{\partial\Gamma}{\partial t}+\nabla_s\cdot(\Gamma\mathbf{u})=\mathcal{D}_s\nabla_s^2\Gamma+\dot{m}_\Gamma,\qquad \dot{m}_\Gamma=k_a\,c_s\,(\Gamma_\infty-\Gamma)-k_b\,\Gamma,
\label{eq:surfactant_interface}
\end{equation}
which is solved on the front elements. Here $\nabla_s$ is the surface gradient and $\mathcal{D}_s$ the surface diffusivity. The source of the surface concentration, $\dot{m}_\Gamma$, is the net adsorption rate per unit interfacial area; it follows Langmuir kinetics, with the adsorption and desorption coefficients $k_a$ and $k_b$ and the maximum packing $\Gamma_\infty$. The bulk concentration next to the interface, $c_s$, is the mean of $q_i=C_i/S_i$ over the cells around each front element on the side of the carrying phase, $S_i>0.5$, with the normalized kernel weights of Eq.~\eqref{eq:peskin_kernel_front_tracking}. The adsorbed mass leaves the bulk through the sorption term $\dot{m}_{\mathrm{sorp}}$ of Eq.~\eqref{eq:conservative_transport}. In the continuous problem, $\dot{m}_{\mathrm{sorp}}=-\dot{m}_\Gamma\,\delta_\Gamma$, with $\delta_\Gamma$ the surface delta function of the interface: what the interface gains per unit area, the bulk loses at the interface. Discretely, the source of each front element $e$, of area $A_e$, is spread to the cells around it with the kernel of Eq.~\eqref{eq:peskin_kernel_front_tracking}, restricted to the cells on the side of the carrying phase, $S_i>0.5$, and normalized,
\begin{equation}
\dot{m}_{\mathrm{sorp},i}=-\frac{1}{V_i}\sum_e \dot{m}_{\Gamma,e}\,A_e\,\frac{w_{ie}}{\sum_j w_{je}},
\label{eq:surfactant_bulk_source}
\end{equation}
where $w_{ie}$ is the kernel weight of cell $i$ for element $e$, so that $\sum_i\dot{m}_{\mathrm{sorp},i}V_i=-\sum_e\dot{m}_{\Gamma,e}A_e$ and the total mass of bulk and adsorbed surfactant is conserved. The derivation of this model and its front-tracking discretization are given by \citet{Muradoglu2008,Muradoglu2014}; as in these works, $c_s$ is read from the same cells to which the sorption term is applied. This sorption term and the diffusion of Eq.~\eqref{eq:scalar_diffusion_discrete} are applied once per step, after the mass projection, as described in Section~\ref{sec:scalar_transport_discretisation}. Surface tension is kept constant, so the surfactant does not act on the flow, and the scalars of the proposed and standard methods at a given resolution are computed with the same flow. Both scalars use the same interfacial model; they differ only in the transport of the bulk surfactant. The adsorption and desorption are validated in \ref{app:surfactant_sphere} against the exact solution for a sphere at rest: the error of the surface concentration decreases at first order, to $0.79\%$ at $D/h=40$.

The surfactant is dissolved in the rising bubble of Section~\ref{sec:rising_bubble}, with $q=q_0$ and $\Gamma=0$ at $t^*=0$; its groups, with $L=D$ and $U=\sqrt{gD}$ as for the Galilei number, are listed in Table~\ref{tab:dimensionless_groups}. The domain is lengthened to $[0,16]\times[0,4]\times[0,4]$, with the same cell sizes and refinement levels, and the calculations are compared at $t^*=15.7$, when the bubble has risen about eight diameters. No exact solution is available. The calculations are therefore compared through the surfactant budget: the masses adsorbed on the interface, dissolved in the bubble and carried into the wake,
\begin{equation}
M_\Gamma=\sum_e\Gamma_eA_e,\qquad
M_{\mathrm b}=\sum_{i:S_i^{\mathrm{FT}}\ge0.5}C_iV_i,\qquad
M_{\mathrm w}=\sum_{i:S_i^{\mathrm{FT}}<\varepsilon^{\mathrm{leak}}}C_iV_i,
\label{eq:surfactant_budget}
\end{equation}
given in percent of the initial mass $M_0=\sum_iC_i(0)V_i$. The sum over $e$ runs over the front elements, and the wake is defined by the threshold $\varepsilon^{\mathrm{leak}}$ of $E_{\mathrm{leak}}$. A further measure is the mean bulk concentration next to the interface, $\overline{c}_s=\sum_ec_{s,e}A_e/\sum_eA_e$, which drives the adsorption.

\begin{figure}[htbp]
\centering
\includegraphics[width=0.65\textwidth]{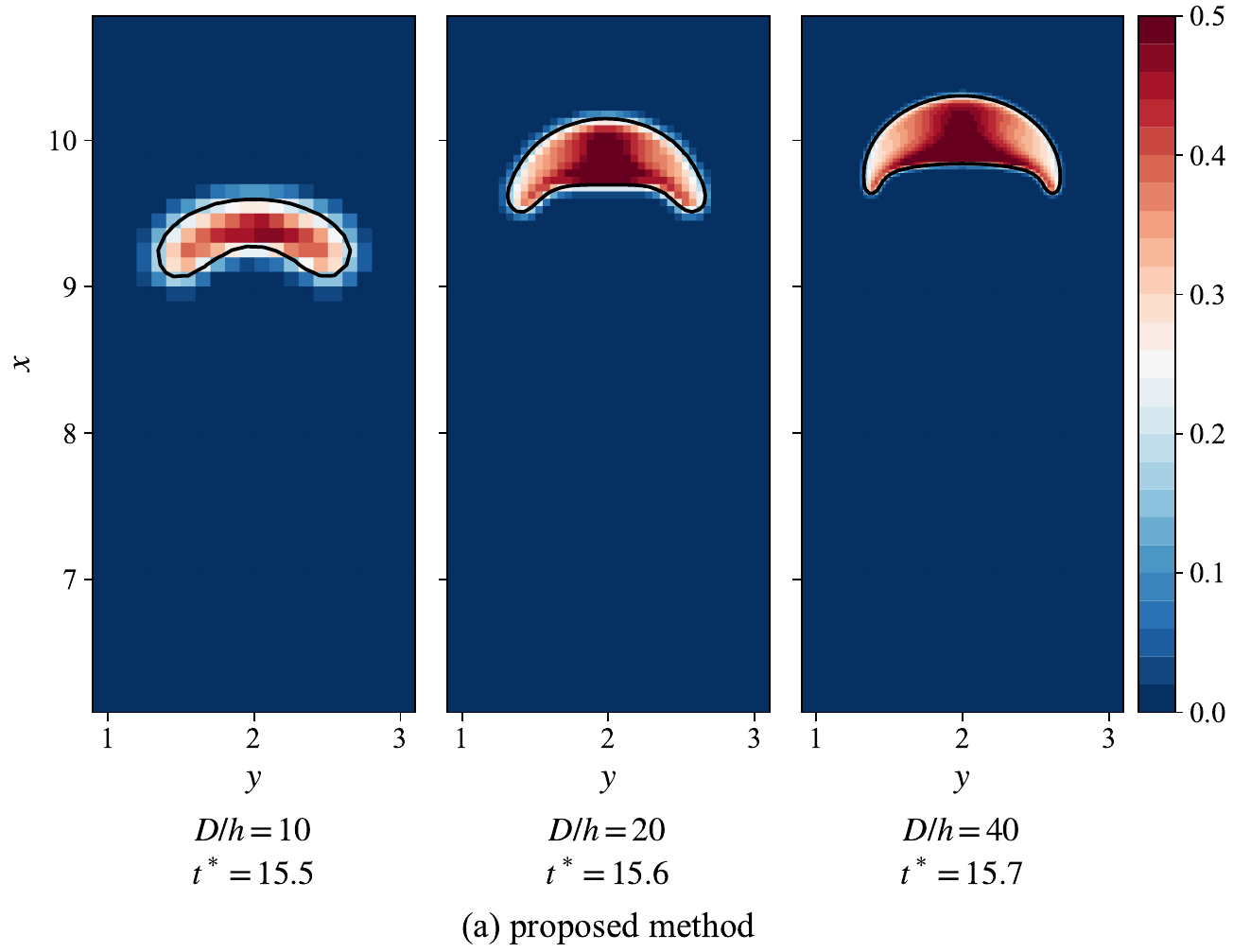}\\[2mm]
\includegraphics[width=0.65\textwidth]{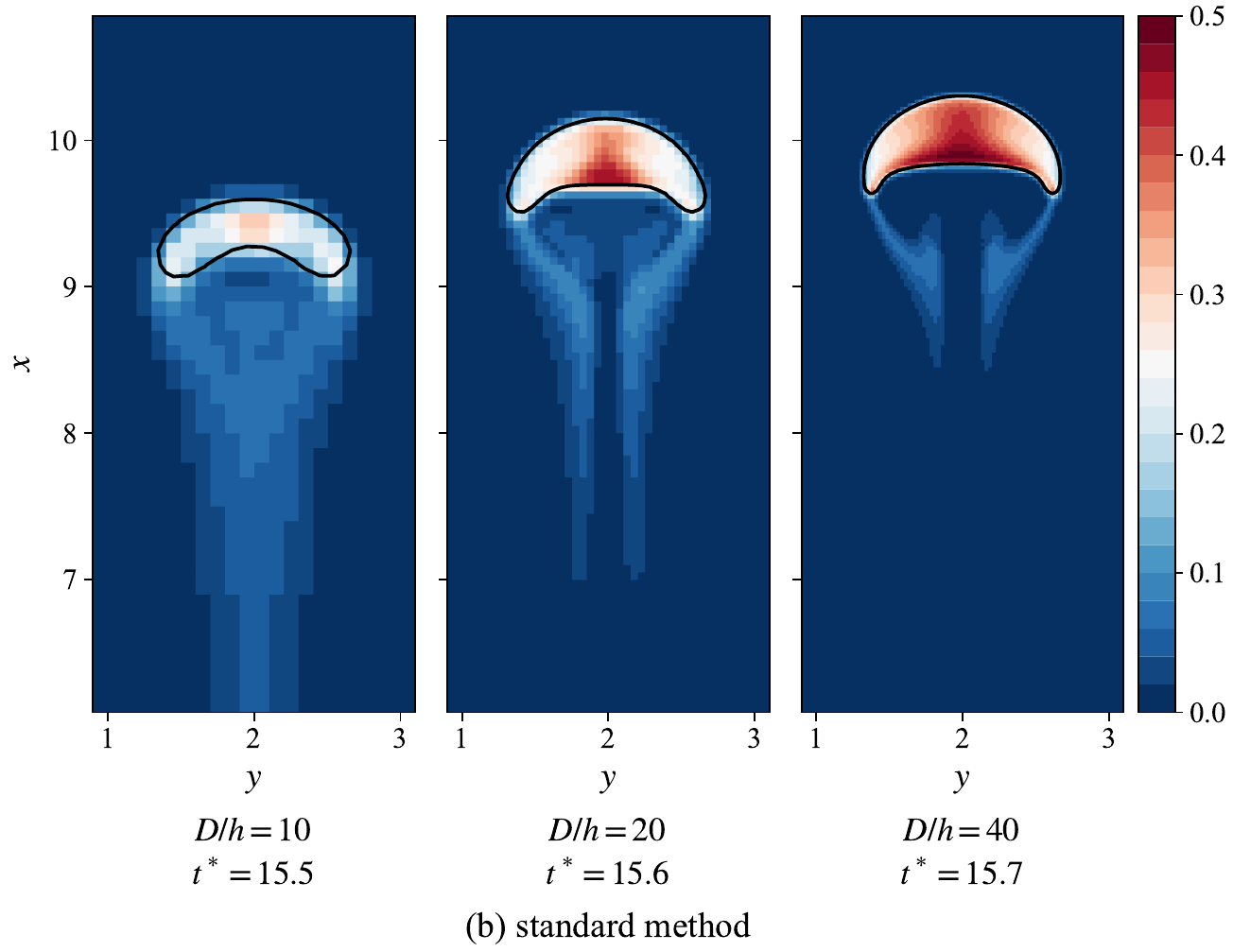}
\caption{Rising bubble with a soluble surfactant: dissolved surfactant $C/q_0$ in the mid-plane $z=2$ at $t^*\approx15.7$: (a) proposed method and (b) standard method, each at $D/h=10$, $20$ and $40$ from left to right; the rise direction $x$ points upwards. The black line is $S^{\mathrm{FT}}=0.5$.}
\label{fig:surfactant_bubble_wake}
\end{figure}

\begin{figure}[htbp]
\centering
\includegraphics[width=\textwidth]{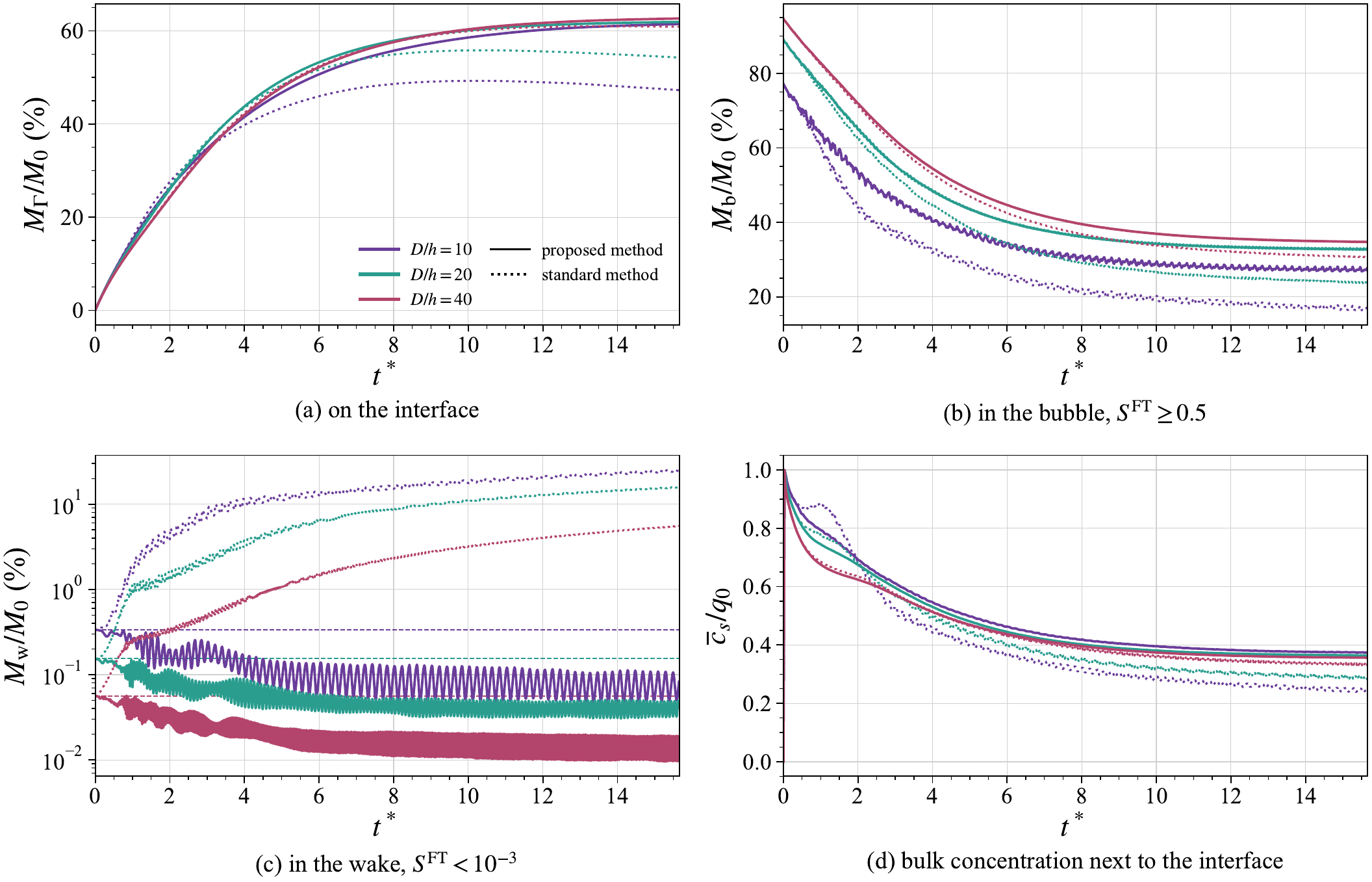}
\caption{Rising bubble with a soluble surfactant, proposed method (solid) and standard method (dotted) at $D/h=10$, $20$ and $40$, up to the comparison time $t^*=15.7$: the budget of Eq.~\eqref{eq:surfactant_budget}, (a) $M_\Gamma$, (b) $M_{\mathrm b}$ and (c) $M_{\mathrm w}$ in percent of $M_0$, with the share of the diffuse indicator's tail at $t^*=0$ in (c) (thin dashed), and (d) $\overline{c}_s/q_0$.}
\label{fig:surfactant_bubble}
\end{figure}

Figure~\ref{fig:surfactant_bubble_wake} shows the dissolved surfactant in the mid-plane at $t^*\approx15.7$, and Fig.~\ref{fig:surfactant_bubble} the budget in time. The scalar of the proposed method remains in the cap at every resolution (Fig.~\ref{fig:surfactant_bubble_wake}(a)), and its mass in the wake, $M_{\mathrm w}$, never exceeds the mass that the diffuse indicator places in the wake cells at $t^*=0$ by more than $1\%$ of that mass (dashed lines in Fig.~\ref{fig:surfactant_bubble}(c)). The scalar of the standard method trails a wake along the whole path of the bubble (Fig.~\ref{fig:surfactant_bubble_wake}(b)), which holds $5.5$--$24\%$ of the surfactant at $t^*=15.7$ (Fig.~\ref{fig:surfactant_bubble}(c) and Table~\ref{tab:surfactant_bubble}) and grows at an almost constant rate, $1.13$, $0.87$ and $0.42\%$ of $M_0$ per unit $t^*$ between $t^*\approx9.4$ and $15.7$ at $D/h=10$, $20$ and $40$; refining reduces it by about a quarter from $D/h=10$ to $20$ and halves it from $20$ to $40$, but does not stop the loss. The liquid passes the interface of the rising bubble and carries the escaped scalar into the wake.

The surfactant dissolved in the bubble decreases as it adsorbs (Fig.~\ref{fig:surfactant_bubble}(b)). At $t^*=15.7$, $M_{\mathrm b}$ of the proposed method is $28$, $33$ and $35\%$ at $D/h=10$, $20$ and $40$; the rest of its dissolved surfactant, $10.5$, $5.2$ and $2.6\%$, lies in the outer half of the interfacial cells, $S^{\mathrm{FT}}<0.5$, a band that narrows with refinement. The standard method keeps $17$, $24$ and $31\%$ in the bubble, since part of its surfactant has left for the wake.

The loss lowers the bulk concentration next to the interface (Fig.~\ref{fig:surfactant_bubble}(d)): at $t^*=15.7$, $\overline{c}_s$ is $0.356$--$0.375\,q_0$ with the proposed method and $0.249$--$0.332\,q_0$ with the standard method. This lowers the adsorption (Fig.~\ref{fig:surfactant_bubble}(a)): the adsorbed mass of the standard method stops rising at $t^*\approx10$ at $D/h=10$ and $20$ and at $t^*\approx14$ at $D/h=40$ and then decreases, as desorption exceeds adsorption, whereas that of the proposed method still rises by $0.6$--$0.9$ percentage points between $t^*=12.5$ and $15.7$. At the same resolution, the standard method adsorbs $14.2$, $7.6$ and $1.8$ percentage points of $M_0$ less than the proposed method, that is, $23$, $12$ and $3\%$ of its adsorbed mass (Table~\ref{tab:surfactant_bubble}). The adsorbed mass of the proposed method changes by only $1.2$ percentage points ($1.9\%$) between $D/h=10$ and $40$, from $61.5$ to $62.7\%$, so that its result at $D/h=10$ is closer to its result at $D/h=40$ than the result of the standard method at $D/h=40$, $60.9\%$. The mass projection keeps the dissolved surfactant in the bubble, but its shift $\lambda_C$ moves more mass than in the prescribed flows: summed over the steps up to $t^*=15.7$, $21.5$, $11.0$ and $4.4\%$ of $M_0$ at $D/h=10$, $20$ and $40$, against at most $0.3\%$ for the sphere at rest of \ref{app:surfactant_sphere}.

\begin{table}[htbp]
\centering
\caption{Rising bubble with a soluble surfactant at $t^*=15.7$: adsorbed mass, mass in the wake and bulk concentration next to the interface, Eq.~\eqref{eq:surfactant_budget}, for the proposed and standard methods.}
\label{tab:surfactant_bubble}
\small
\begin{tabular}{lrrrrrr}
\toprule
 & \multicolumn{2}{c}{$M_\Gamma/M_0$ (\%)} & \multicolumn{2}{c}{$M_{\mathrm w}/M_0$ (\%)} & \multicolumn{2}{c}{$\overline{c}_s/q_0$} \\
\cmidrule(lr){2-3}\cmidrule(lr){4-5}\cmidrule(lr){6-7}
$D/h$ & proposed & standard & proposed & standard & proposed & standard \\
\midrule
10 & 61.5 & 47.3 & 0.094 & 24.2 & 0.375 & 0.249 \\
20 & 61.9 & 54.3 & 0.037 & 15.7 & 0.364 & 0.288 \\
40 & 62.7 & 60.9 & 0.010 & 5.5 & 0.356 & 0.332 \\
\bottomrule
\end{tabular}
\end{table}

\subsection{Contribution of the ingredients of the method}
\label{sec:ablation}

The proposed method has three ingredients. It transports the phase-averaged instead of the cell-averaged concentration, so that the advected field has no jump at the interface. Before the advection it extends this concentration linearly across the interface. After the advection it rebuilds the scalar on the projected support, and a uniform shift of the concentration restores the mass.

Two variants each replace one of the first two ingredients and leave everything else unchanged. The first copies the value of the neighbouring cell instead of continuing the concentration linearly (constant extension). The second transports the cell-averaged concentration as the standard method does and then rebuilds it in the same way (transported $C$). Both variants are run for the two cases with a non-uniform concentration, the reversible vortex (Section~\ref{sec:reversible_vortex}) and the localized source (Section~\ref{sec:source_transport}). Since both keep the rebuild, they differ from the proposed method only in the accuracy of the concentration.

Table~\ref{tab:ablation} lists the scalar--front errors. In the vortex, where the concentration varies along the whole interface, both variants converge at a rate of only $1.3$, against $1.7$--$1.8$ for the proposed method. At $D/h=40$ the error is $12$ times larger with the constant extension and $8.7$ times larger with the transported $C$, although still about $7$ and $10$ times smaller than with the standard method. In the source, where the concentration is uniform over most of the interface and varies only near the leading edge, the errors of the variants are only $16\%$ larger at $D/h=10$. Their rates between $D/h=20$ and $40$ drop to $1.5$ and $1.7$, against $2.5$, so that at $D/h=40$ their errors are about three times larger.

The two variants lose accuracy for different reasons. A constant extension leaves an error of order $h$ in the extended values, where the linear extension leaves an error of order $h^2$ (Section~\ref{sec:scalar_correction}), and the reconstruction stencils of the interfacial cells reach into these values. Transporting $C$ brings back the jump at the interface that the proposed method avoids; rebuilding the scalar afterwards restores its mass and its support, but not its concentration. The accuracy of the proposed method therefore needs both the transport of the phase-averaged concentration and its linear extension; the rebuild alone keeps the scalar on its phase and conserves its mass, but leaves the concentration converging at close to first order.

\begin{table}[htbp]
\centering
\caption{Scalar--front error $E_{C,1}$ with one ingredient of the method replaced at a time, at the last stored field: proposed method, constant extension, transported $C$ and standard method, with the convergence rates from $D/h=10$ to $20$ and from $20$ to $40$.}
\label{tab:ablation}
\small
\setlength{\tabcolsep}{3pt}
\begin{tabular}{lrrrrrrrr}
\toprule
 & \multicolumn{4}{c}{reversible vortex, $t=2$} & \multicolumn{4}{c}{localized source, $t=3.2$} \\
\cmidrule(lr){2-5}\cmidrule(lr){6-9}
$D/h$ & proposed & constant & transported $C$ & standard & proposed & constant & transported $C$ & standard \\
\midrule
10 & $7.6\times10^{-3}$ & $4.6\times10^{-2}$ & $3.4\times10^{-2}$ & $1.3\times10^{-1}$ & $5.6\times10^{-2}$ & $6.5\times10^{-2}$ & $6.5\times10^{-2}$ & $2.0\times10^{-1}$ \\
20 & $2.3\times10^{-3}$ & $1.9\times10^{-2}$ & $1.4\times10^{-2}$ & $8.2\times10^{-2}$ & $8.4\times10^{-3}$ & $1.2\times10^{-2}$ & $1.4\times10^{-2}$ & $1.1\times10^{-1}$ \\
40 & $6.4\times10^{-4}$ & $7.7\times10^{-3}$ & $5.5\times10^{-3}$ & $5.3\times10^{-2}$ & $1.5\times10^{-3}$ & $4.2\times10^{-3}$ & $4.3\times10^{-3}$ & $6.8\times10^{-2}$ \\
order & 1.7, 1.8 & 1.3, 1.3 & 1.3, 1.3 & 0.6, 0.6 & 2.7, 2.5 & 2.4, 1.5 & 2.3, 1.7 & 0.9, 0.7 \\
\bottomrule
\end{tabular}
\end{table}

\subsection{Dependence on the advection scheme}
\label{sec:scheme_dependence}

The phase-concentration transport needs only a face reconstruction of the extended concentration, so it can be combined with any advection scheme. To show that it does not rely on WENO-Z, the localized source of Section~\ref{sec:source_transport} and the rising bubble of Section~\ref{sec:rising_bubble} are repeated at $D/h=20$ with seven face reconstructions. These are the second-order MUSCL scheme \citep{vanLeer1979} with the minmod and the van Leer limiter \citep{Roe1986,vanLeer1974}; the quadratic upstream interpolation QUICK \citep{Leonard1979}, which is not bounded and, for the cell averages of a finite-volume method, second-order accurate; and the fifth-order schemes WENO-JS (the WENO scheme of Jiang and Shu) \citep{Jiang1996}, WENO-Z \citep{Borges2008}, WENO-Z+ \citep{Acker2016} and TENO \citep{Fu2016}. The fifth-order schemes share the candidate polynomials and the smoothness indicators of WENO-JS; WENO-Z+ uses the weight parameter $\lambda=h^{2/3}$ of \citet{Acker2016}, and TENO the cut-off $C_T=10^{-5}$, below which a candidate stencil is discarded. In each calculation the proposed and standard methods use the same reconstruction, and all other settings are those of Sections~\ref{sec:source_transport} and~\ref{sec:rising_bubble}.

\begin{figure}[htbp]
\centering
\includegraphics[width=0.8\textwidth]{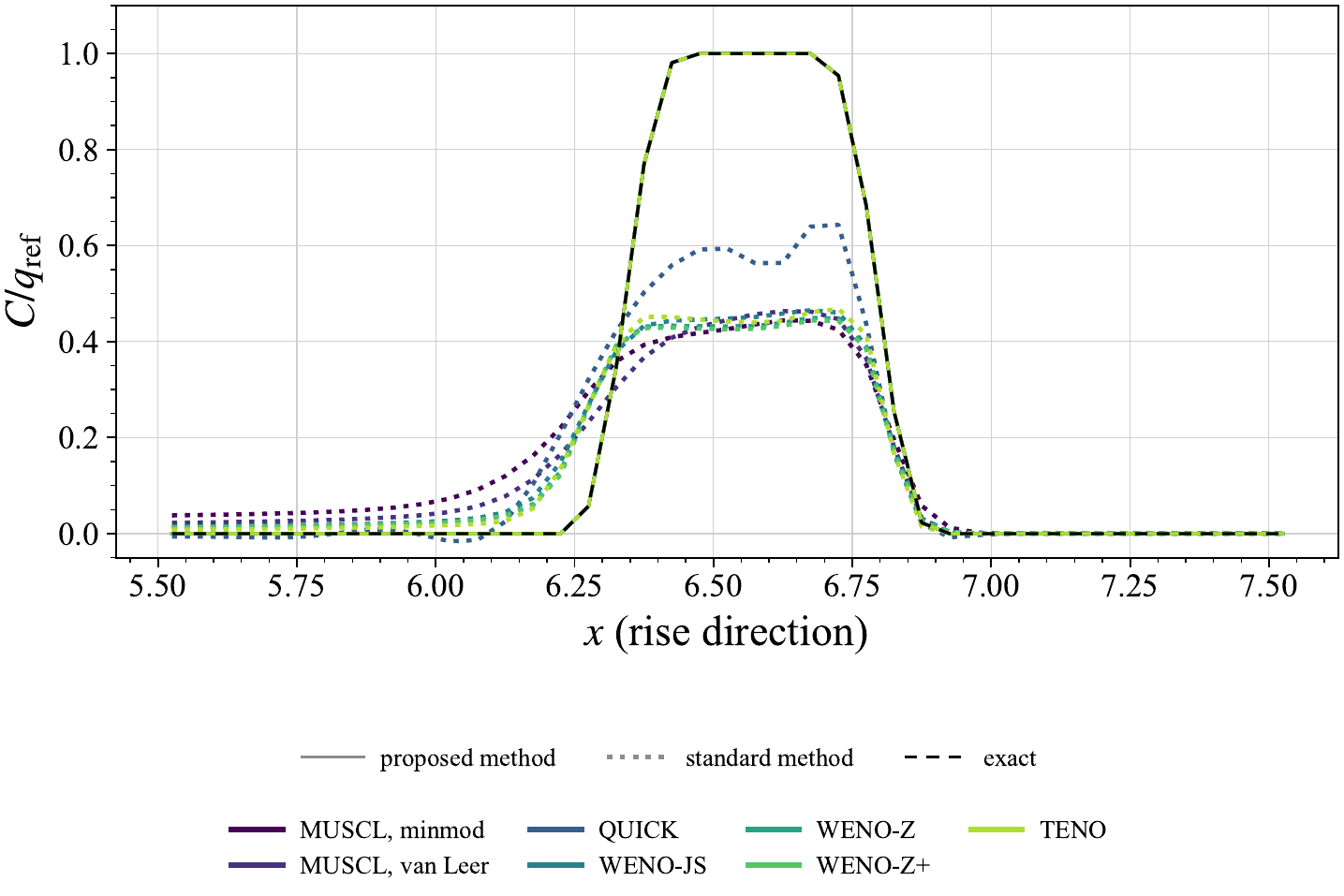}
\caption{Rising bubble at $D/h=20$ and $t^*=9.37$ with seven face reconstructions (colours): scalar $C/q_{\mathrm{ref}}$ along the rise axis ($y=z=2$) for the proposed method (solid) and the standard method (dotted). The exact scalar $S^{\mathrm{FT}}q_{\mathrm{ref}}$ (black dashed) coincides with the seven curves of the proposed method.}
\label{fig:scheme_profiles}
\end{figure}

Figure~\ref{fig:scheme_profiles} shows the scalar along the rise axis of the bubble, Fig.~\ref{fig:scheme_source} in \ref{app:scheme_source} the concentration along the direction of motion through the source, and Table~\ref{tab:scheme_study} the errors at the instants of Tables~\ref{tab:source_errors} and~\ref{tab:bubble_errors}. With every scheme, the proposed method keeps the scalar on its phase. In the bubble, its scalar is the same for all seven reconstructions and coincides with the exact one, whereas the standard method leaves $26$--$46\%$ of the scalar in the wake (Fig.~\ref{fig:scheme_profiles}). In the source case, the standard method produces concentration spikes at both interfaces with every scheme, while the concentration of the proposed method stays below the exact peak (Fig.~\ref{fig:scheme_source}); its scalar--front error is $5.5$ to $12.8$ times smaller, and its leakage remains below that of the exact field. QUICK oscillates at the interface, as \citet{Muradoglu2014} observed for a bulk surfactant: with it, the standard method produces negative values whose mass is $4.3\%$ of the scalar in the source case and $4.2\%$ in the bubble, whereas the scalar of the proposed method is non-negative with every scheme.

Inside the phase, the concentration is as accurate as the reconstruction: with MUSCL, both methods flatten the peak of the source to about half of its exact value. The ratio of the errors of the standard and proposed methods is therefore larger with the fifth-order schemes, $8.7$--$12.8$, than with the second-order ones, $5.5$--$6.4$. Numerical diffusion spreads the concentration from the carrying phase into extension cells whose support vanishes after the step; the rebuild removes this part of the scalar, and the concentration shift $\lambda_C$ of Eq.~\eqref{eq:scalar_projection_shift} returns it uniformly over the phase rather than where it was lost. This mass, relative to the scalar mass and averaged over the steps up to $t=3.2$, is the mass shift of Table~\ref{tab:scheme_study}; it is four to nine times smaller with the fifth-order schemes than with MUSCL, and as small with QUICK, which is dispersive rather than diffusive. WENO-Z, used in all other calculations, gives one of the largest ratios.

\begin{table}[htbp]
\centering
\caption{Dependence on the advection scheme at $D/h=20$: errors of the proposed and standard methods with the same face reconstruction. With every scheme, the proposed method has $E_{\mathrm{leak}}=0.12\%$ in the source case, and $E_{C,1}=9.5\times10^{-4}$ and $E_{\mathrm{leak}}=0.08\%$ in the bubble; the exact field has $E_{\mathrm{leak}}=0.14\%$ and $0.16\%$.}
\label{tab:scheme_study}
\small
\begin{tabular}{lcrrrrrrr}
\toprule
 & & \multicolumn{5}{c}{localized source, $t=3.2$} & \multicolumn{2}{c}{rising bubble, $t^*=9.37$} \\
\cmidrule(lr){3-7}\cmidrule(lr){8-9}
 & & \multicolumn{2}{c}{$E_{C,1}$} & & standard & mass shift & \multicolumn{2}{c}{standard} \\
\cmidrule(lr){3-4}\cmidrule(lr){8-9}
scheme & scheme order & proposed & standard & ratio & $E_{\mathrm{leak}}$ (\%) & ($10^{-6}$) & $E_{C,1}$ & $E_{\mathrm{leak}}$ (\%) \\
\midrule
MUSCL, minmod    & 2 & 0.074 & 0.467 & 6.4 & 10.9 & 23 & 1.06 & 45.6 \\
MUSCL, van Leer  & 2 & 0.051 & 0.282 & 5.5 & 3.6 & 13 & 0.886 & 37.6 \\
QUICK            & 2 & 0.044 & 0.280 & 6.4 & 2.4 & 3.0 & 0.727 & 28.7 \\
WENO-JS          & 5 & 0.016 & 0.153 & 9.7 & 2.1 & 3.5 & 0.771 & 32.3 \\
WENO-Z           & 5 & $8.4\times10^{-3}$ & 0.107 & 12.8 & 1.6 & 2.7 & 0.736 & 30.6 \\
WENO-Z+          & 5 & $7.6\times10^{-3}$ & 0.096 & 12.6 & 1.5 & 2.6 & 0.732 & 30.4 \\
TENO             & 5 & $8.2\times10^{-3}$ & 0.071 & 8.7 & 0.6 & 2.5 & 0.672 & 26.3 \\
\bottomrule
\end{tabular}
\end{table}

\subsection{Sensitivity to the parameters of the method}
\label{sec:sensitivity}

The method has two parameters of its own, the recovery threshold $\varepsilon^{\mathrm{rec}}$ and the number of extension layers $N_\ell$, and uses the pure-cell tolerance $\varepsilon^{\mathrm{pure}}$ of the indicator projection. Their influence, and that of the time step, is examined on the localized-source case at $D/h=20$, in which the concentration varies steeply next to the interface. Relative to the default settings, $\varepsilon^{\mathrm{rec}}=0.03$, $N_\ell=4$, $\varepsilon^{\mathrm{pure}}=10^{-5}$ and a CFL number of $0.1$, one parameter is changed at a time: $\varepsilon^{\mathrm{rec}}=0.01$ and $0.1$, $N_\ell=3$ and $6$, $\varepsilon^{\mathrm{pure}}=10^{-4}$ and $10^{-6}$, and the time step halved. Table~\ref{tab:sensitivity} lists the errors at the last stored field, $t=3.2$. The flow and the front are identical in the calculations with the default time step, and the scalar of the standard method agrees among them to within $0.02\%$ in $E_{C,1}$.

The number of layers has no significant influence, below $1\%$ in $E_{C,1}$: three layers already cover the stencils of the faces of the resolved cells, and the fourth covers the cells of the first layer, which the interface can enter during a step. The threshold has a moderate influence. Lowering it to $0.01$ reduces $E_{C,1}$ by $4\%$, since more interfacial cells carry a recovered concentration, and raising it to $0.1$ increases $E_{C,1}$ by $15\%$ and increases $E_{q,\infty}$ by a factor of $2.6$, as the cells between $0.03$ and $0.1$ are then extended from further inside. The pure-cell tolerance changes $E_{C,1}$ by at most $5.1\%$, and halving the time step by $1.9\%$. Halving the time step raises $E_{q,\infty}$ from $0.450$ to $0.595$ because its maximum moves from the source peak to a single cell of the indicator's tail outside the drop; on $S^{\mathrm{FT}}>\varepsilon^{\mathrm{rec}}$ it falls to $0.426$. In every setting, $E_{C,1}$ remains $11$--$13$ times smaller than that of the standard method, and the leakage stays below that of the exact field. The default setting is retained.

\begin{table}[htbp]
\centering
\caption{Sensitivity to the parameters of the method: localized source at $D/h=20$ and $t=3.2$, one parameter changed at a time from the default ($\varepsilon^{\mathrm{rec}}=0.03$, $N_\ell=4$, $\varepsilon^{\mathrm{pure}}=10^{-5}$, CFL number $0.1$). Errors and leakage, with the share of the exact field, as in Table~\ref{tab:source_errors}.}
\label{tab:sensitivity}
\small
\begin{tabular}{llrrrr}
\toprule
parameter & value & $E_{C,1}$ & $E_{q,1}$ & $E_{q,\infty}$ & $E_{\mathrm{leak}}$ (\%) \\
\midrule
default &  & $8.37\times10^{-3}$ & $9.03\times10^{-3}$ & 0.450 & 0.120 \\
$\varepsilon^{\mathrm{rec}}$ & 0.01 & $8.03\times10^{-3}$ & $8.66\times10^{-3}$ & 0.429 & 0.119 \\
 & 0.1 & $9.65\times10^{-3}$ & $1.04\times10^{-2}$ & 1.158 & 0.121 \\
$N_\ell$ & 3 & $8.35\times10^{-3}$ & $9.01\times10^{-3}$ & 0.449 & 0.121 \\
 & 6 & $8.40\times10^{-3}$ & $9.06\times10^{-3}$ & 0.449 & 0.117 \\
$\varepsilon^{\mathrm{pure}}$ & $10^{-4}$ & $8.80\times10^{-3}$ & $9.07\times10^{-3}$ & 0.450 & 0.112 \\
 & $10^{-6}$ & $8.37\times10^{-3}$ & $9.03\times10^{-3}$ & 0.449 & 0.120 \\
$\Delta t$ & halved & $8.21\times10^{-3}$ & $8.85\times10^{-3}$ & 0.595 & 0.120 \\
\midrule
\multicolumn{2}{l}{standard method} & $1.07\times10^{-1}$ & $9.46\times10^{-2}$ & 12.587 & 1.590 \\
\multicolumn{2}{l}{exact} & & & & 0.144 \\
\bottomrule
\end{tabular}
\end{table}

\subsection{Computational cost}
\label{sec:cost}

The cost of the method was measured on the oblique translation of Section~\ref{sec:source_transport} with a single scalar, carrying an off-centre Gaussian concentration, $q=0.7+0.6\exp[-|\mathbf{x}-\mathbf{x}_g|^2/(2(0.15D)^2)]$ with $\mathbf{x}_g$ offset from the drop centre by $(0.15,-0.10,0.10)D$, and no source. The calculations ran alone on one machine (AMD Ryzen Threadripper 3960X, 24 cores) with $32$ processes; all three resolutions are compared at $t=0.25$, the end of the calculation at $D/h=40$. The solver measures the time spent in the advection of the scalar, i.e.\ the three Runge--Kutta stages of the WENO-Z transport, separately from the time spent in the work of the method itself: the recovery and extension of the concentration, the indicator projection, the rebuild and the mass projection. Table~\ref{tab:cost} lists both with the total wall time, which also includes the front tracking and the flow solution.

The method's own work costs about as much as one advection of the scalar, $0.8$--$1.1$ times, so that a scalar carried with the proposed method costs about twice as much as one carried with the standard method; no phase support has to be transported, since it is taken from the front. The method takes $1.7$--$2.7\%$ of the total wall time. In calculations dominated by the interface the share is smaller still: in the surfactant bubble at $D/h=40$, the advection of the scalar took $0.16$\,s per time step, $0.2\%$ of the median step time of $94$\,s, most of which was spent on the front and its surfactant.

\begin{table}[htbp]
\centering
\caption{Computational cost of the proposed method: oblique translation with one scalar (Gaussian concentration), $32$ processes, at $t=0.25$. Wall time of the whole calculation, of the advection of the scalar (three Runge--Kutta stages) and of the method's own work (recovery, extension, indicator projection, rebuild and mass projection).}
\label{tab:cost}
\small
\begin{tabular}{rrrrrr}
\toprule
$D/h$ & total (s) & advection (s) & method (s) & method/advection & method/total (\%) \\
\midrule
10 & 46.3 & 1.12 & 1.25 & 1.12 & 2.7 \\
20 & 169.4 & 4.43 & 4.09 & 0.92 & 2.4 \\
40 & 1452.3 & 31.52 & 24.40 & 0.77 & 1.7 \\
\bottomrule
\end{tabular}
\end{table}

\FloatBarrier
\section{Conclusions}
\label{sec:conclusion}

We have presented a transport method for scalars confined to one phase in front-tracking simulations, which keeps the scalar mass-conservative and consistent with the phase region defined by the front. The method transports the phase-averaged concentration $q$ instead of the cell-averaged concentration $C=Sq$. In each time step, $q$ is recovered where the support is resolved, extended across the interface by linear continuation along grid lines over four cell layers, and advected with WENO-Z. The scalar is rebuilt on a phase support obtained by a bounded, volume-preserving projection of the front-derived indicator, and a single non-negative shift of the concentration restores the scalar mass. By construction, the scalar is non-negative and vanishes outside the corrected support, its total mass is conserved, a uniform concentration is preserved exactly, and a linear concentration is extended exactly where two known cells lie on the grid line. The mass is conserved globally; the mass projection returns the deficit of each step uniformly over the carrying phase.

The tests with exact solutions confirm these properties and quantify the accuracy. For a uniform concentration, the transported concentration remains exact to round-off, and the remaining scalar--front error is the difference between the projected and the front-derived supports, $1.5\times10^{-4}$ at $D/h=40$ in the translation test, against $0.046$ for the same scheme transporting $C$. For non-uniform concentrations, the error converges at orders of $1.7$--$1.8$ in the reversible vortex and $2.5$--$2.7$ for the moving source, against $0.6$--$0.9$ for the standard method, whose error in the interfacial cells does not decrease with refinement; these rates hold over the resolutions tested, $D/h=10$--$40$, and the first-order indicator projection limits them asymptotically to about one, still above the sub-linear rate of the standard method, so that the gain, one to two orders of magnitude at $D/h=40$, grows further with refinement. Replacing either the transport of $q$ or its linear extension lowers the order in the vortex to $1.3$ and increases the error at $D/h=40$ by factors of $8.7$--$12$ there and about three in the source: the rebuild on the projected support alone conserves the mass and keeps the scalar on its phase, but does not make the concentration accurate. Outside the phase, the proposed method leaves no more scalar than the exact solution, to within $0.012$ percentage points; the exact solution itself has $0.05$--$0.35\%$ of its mass there, only because the front-derived indicator is diffuse. The standard method places $0.7$--$3.2\%$ of its mass there in the prescribed flows and $23$--$34\%$ in the rising bubble. The total mass is conserved to round-off. With the scalar carried by the phase outside the drop, the vortex behaves alike: at $D/h=20$ the error at the interface is $64$--$141$ times smaller than with the standard method.

In the rising bubble, whose interface motion follows from the flow solution, the method keeps a uniform concentration exact. With a soluble surfactant in the bubble, the standard method loses the dissolved surfactant into the wake at a rate that no longer decays at late times at $D/h=20$ and $40$, is reduced by only about a quarter from $D/h=10$ to $20$ and is halved from $20$ to $40$, and after a rise of about eight diameters it underestimates the adsorbed mass by $23$, $12$ and $3\%$ of that of the proposed method at $D/h=10$, $20$ and $40$. The adsorbed mass of the proposed method varies by $1.9\%$ ($1.2$ percentage points) between $D/h=10$ and $40$, so that at $D/h=10$ it is closer to its result at $D/h=40$ than the standard method at $D/h=40$. The adsorption and desorption themselves are validated on a sphere at rest against a converged reference solution: the error of the surface concentration decreases at first order, to $0.79\%$ at $D/h=40$, and translating the sphere changes it by at most $0.5$ percentage points.

The results depend moderately on the parameters: varying the recovery threshold between $0.01$ and $0.1$ changes the error by $-4$ to $+15\%$, the number of extension layers between three and six by less than $1\%$, and the pure-cell tolerance between $10^{-6}$ and $10^{-4}$ by at most $5.1\%$, and halving the time step by $1.9\%$. The method is not tied to WENO-Z: with seven face reconstructions, from the second-order MUSCL and the unbounded QUICK scheme to fifth-order TENO, it keeps the scalar on its phase and non-negative, and reduces the scalar--front error of the standard method with the same scheme by factors of $5.5$--$6.4$ with the second-order schemes and $8.7$--$12.8$ with the fifth-order schemes for the moving source. The method's own work costs about one advection of the scalar, $1.7$--$2.7\%$ of the total wall time in the cost benchmark. Exchange between both phases with a jump condition such as Henry's law, sharp indicator reconstructions, and a bounded extension for under-resolved interfacial regions are natural extensions of this work.

\section*{Data Availability Statement}
The data that support the findings of this study are reproducible and openly available in the
repository with DOI 10.5281/zenodo.24391269, available at \href{https://doi.org/10.5281/zenodo.24391269}{https://doi.org/10.5281/zenodo.24391269}

\section*{Declaration of Generative AI and AI-assisted technologies in the writing process}

The authors used Anthropic's Claude to assist with writing, editing and proofreading this manuscript. The authors then reviewed and revised the text as necessary and assume full responsibility for the final content of the publication.

\section*{Acknowledgements}
\noindent This research was funded by the Deutsche Forschungsgemeinschaft (DFG, German Research Foundation), grant numbers 420239128 and 458610925.

\appendix

\section{Implementation details of the \texorpdfstring{indicator}{indicator} projection}
\label{app:correction_implementation}

The main text formulates the projection for the case in which only the mixed cells are adjusted. In general, the mesh is partitioned into the adjustable cells $\mathcal{A}_I$ and the cells $\mathcal{P}_0$ and $\mathcal{P}_1$ whose corrected values are fixed at zero and unity, respectively. Initially, $\mathcal{A}_I=\Omega_m$, $\mathcal{P}_0=\Omega_0$ and $\mathcal{P}_1=\Omega_1$, and Eqs.~\eqref{eq:indicator_projection_minimisation} and \eqref{eq:indicator_general_projection} hold with $\Omega_m$, $\Omega_0$ and $\Omega_1$ replaced by these sets. This appendix gives the numerical feasibility, cell-set expansion, multiplier solution, and neighbour guard used to implement that projection. For a selected adjustable set $\mathcal{A}_I$, the lower and upper bounds $L_i$ and $U_i$ satisfy $0\le L_i\le U_i\le1$. The basic projection uses $L_i=0$ and $U_i=1$; the neighbour guard of \ref{app:indicator_guard} may tighten these bounds without changing the prescribed target volume. For fixed sets and bounds, the corrected field is
\begin{equation}
I_i^c(\lambda_I)=
\begin{cases}
0, & i\in\mathcal{P}_0,\\
1, & i\in\mathcal{P}_1,\\
\mathcal{C}_{[L_i,U_i]}\!\left(I_i^{\mathrm{FT}}+\lambda_I\right), & i\in\mathcal{A}_I.
\end{cases}
\label{eq:indicator_bounded_projection_implementation}
\end{equation}
The corrected field is therefore the closest admissible field, in the sense of Eq.~\eqref{eq:indicator_projection_minimisation}, for the final adjustable set and bounds. Expanding the adjustable set can give an initially pure cell an intermediate corrected value; exact recovery of the target phase volume takes priority over retaining the initial pure-cell classification.

\subsection{Feasibility and adjustable-set expansion}
\label{app:indicator_expansion}

Let $V_{\mathrm{fix},I}=\sum_{i\in\mathcal{P}_1}V_i$ denote the contribution from cells fixed at unity. For the selected bounds, the minimum and maximum attainable volumes are
\begin{equation}
M_I^{\mathrm{minimum}}=V_{\mathrm{fix},I}+\sum_{i\in\mathcal{A}_I}L_iV_i,
\qquad
M_I^{\mathrm{maximum}}=V_{\mathrm{fix},I}+\sum_{i\in\mathcal{A}_I}U_iV_i.
\label{eq:indicator_general_feasibility}
\end{equation}
The projection is feasible if $M_I^{\mathrm{minimum}}\le M_I^{\mathrm{target}}\le M_I^{\mathrm{maximum}}$. For the initial choice, $\mathcal{A}_I=\Omega_m$, $\mathcal{P}_0=\Omega_0$ and $\mathcal{P}_1=\Omega_1$, with $L_i=0$ and $U_i=1$ in the adjustable cells. If $V_1=\sum_{i\in\Omega_1}V_i$ and $V_m=\sum_{i\in\Omega_m}V_i$, its attainable interval is
\begin{equation}
V_1\le M_I^{\mathrm{target}}\le V_1+V_m.
\label{eq:indicator_projection_feasibility}
\end{equation}

No expansion is required when the target lies in this interval. If the target is below $V_1$, eligible interface-adjacent pure-one cells are released first. A cell is eligible when it shares a face with an originally mixed cell and has no face neighbour in the original pure-one set: releasing a pure-one cell next to one that stays fixed could otherwise create a jump from zero to unity between two pure-one cells, away from the interface. If these cells do not provide sufficient capacity, all originally pure-one cells are released, including interior cells, while the pure-zero cells remain fixed. If the target exceeds $V_1+V_m$, all cells are made adjustable. None of these releases was needed in any calculation of Section~\ref{sec:results} or of the appendices. Releasing a cell changes its role in the projection but not its original front-based classification. Targets outside the physical interval $[0,\sum_iV_i]$, beyond the final acceptance tolerance, are rejected.

\subsection{Multiplier solution}
\label{app:indicator_solver}

For a feasible non-empty adjustable set, the projected phase volume is
\begin{equation}
M_I(\lambda_I)=V_{\mathrm{fix},I}
+\sum_{i\in\mathcal{A}_I}
\mathcal{C}_{[L_i,U_i]}\!\left(I_i^{\mathrm{FT}}+\lambda_I\right)V_i.
\label{eq:indicator_volume_function}
\end{equation}
This function is continuous, non-decreasing, and piecewise linear. The bracket
\begin{equation}
\lambda_I^{\mathrm{lower}}=-1-\max_{i\in\mathcal{A}_I}I_i^{\mathrm{FT}},
\qquad
\lambda_I^{\mathrm{upper}}=2-\min_{i\in\mathcal{A}_I}I_i^{\mathrm{FT}}
\label{eq:indicator_lambda_bracket}
\end{equation}
places every adjustable value at its lower or upper bound at the corresponding endpoint. Thus, for a feasible target, the endpoint volumes bracket $M_I^{\mathrm{target}}$. The multiplier is obtained by a safeguarded semismooth Newton iteration using the directional slope of the projected volume. The bracket is updated from the evaluated residual, and a Newton step outside it is replaced by a safeguarded secant step or the bracket midpoint. If the directional slope vanishes, the iteration advances towards the next bound crossing. At most 120 iterations are allowed, and the best evaluated multiplier is retained.

With the domain volume $V_\Omega=\sum_iV_i$, the iterative residual tolerance is $10^{-15}V_\Omega$ and the independent final acceptance tolerance is
\begin{equation}
\tau_{\mathrm{volume}}=10^{-10}V_\Omega.
\label{eq:indicator_volume_tolerance}
\end{equation}
Once the corrected values have been stored, the corrected volume is recomputed globally and must satisfy $|M_I(\lambda_I)-M_I^{\mathrm{target}}|\le\tau_{\mathrm{volume}}$. If $\mathcal{A}_I$ is empty, the fixed field is accepted only when it passes this same check; otherwise, expansion or rejection is required. A target volume below a small cutoff, at most $\tau_{\mathrm{volume}}$, is represented by the zero field.

\subsection{Neighbour guard}
\label{app:indicator_guard}

With a single shift $\lambda_I$, a mixed cell can be driven to zero or unity. If this happens to a cell whose face neighbour is at the opposite value, fixed or driven there by the same shift, the projection has created a jump from zero to unity that the tracked front does not have. After each solve, such cells are detected, and their bounds are tightened towards the value they were heading for: $L_i=\varepsilon_{\mathrm{guard}}$ or $U_i=1-\varepsilon_{\mathrm{guard}}$, with $\varepsilon_{\mathrm{guard}}=10^{-3}$. The same projection is then solved again, so that the volume is still recovered exactly and the field is still the closest admissible one. Since a new shift can drive another cell to a bound, detection and solve are repeated until no new such cell appears, at most $20$ times, and each tightening is limited so that the target volume remains attainable. In the calculations of this paper, $|\lambda_I|\le2\times10^{-4}$, so that the guard can act only where the front-derived indicator jumps from nearly zero to nearly unity within one cell.

\section{Implementation details of the phase-concentration transport}
\label{app:scalar_implementation}

This appendix gives the implementation of the concentration extension and of the mass projection of Section~\ref{sec:scalar_correction}.

\subsection{Extension of the phase-averaged concentration}
\label{app:concentration_extension}

Cells outside the resolved set are first marked as unknown. For a six-faced cell, the six Cartesian directions are identified from the orientation of its face normals: a face is assigned to the axis with which its normal is best aligned, provided that the angle between them is at most $60^\circ$. Cells with a different number of faces have no Cartesian directions. In each layer, every unknown cell evaluates Eq.~\eqref{eq:concentration_extension} from the values of the previous layer. When no direction provides a known neighbour, the cell takes the mean of its known face neighbours, and a cell without known neighbours remains unknown for the next layer. Before each layer, the values of the previous layer are exchanged between the parallel processes, so that the extension near a subdomain boundary uses the same data as in a serial calculation. After $N_\ell$ layers, the remaining unknown cells receive the carrying-phase mean concentration of Eq.~\eqref{eq:phase_mean_concentration}. The extended field is also the state $q^n$ that the second and third Runge--Kutta stages combine with their stage values.

\subsection{Solution of the \texorpdfstring{mass}{mass} projection}
\label{app:scalar_feasibility}
\label{app:scalar_solver}

The mass projection determines a common concentration shift $\lambda_C$ that restores the target mass of Eq.~\eqref{eq:scalar_target_mass} while maintaining non-negativity. The corrected phase support remains fixed throughout this calculation.

The multiplier is obtained by a semismooth Newton iteration on Eq.~\eqref{eq:scalar_projection_multiplier}, starting from $\lambda_C=0$. In each iteration, the slope of $G$ is the carrying-phase volume $\sum_iS_i^{n+1}V_i$ of the cells whose shifted scalar is positive, and the multiplier is updated by the ratio of the mass residual to this slope. The iteration terminates when $|G(\lambda_C)|\le10^{-14}|M_C^{\mathrm{target}}|$, when the slope vanishes, or after at most $60$ iterations; the residual of the final iterate is written to the solver output. Because $G$ is piecewise linear and the shift required in each step is small, a few iterations suffice in practice.

\section{Soluble surfactant: sphere at rest \texorpdfstring{against a reference solution}{against a reference solution}}
\label{app:surfactant_sphere}
\setcounter{figure}{0}

To validate the adsorption and desorption of Section~\ref{sec:surfactant_bubble} against a reference solution, a sphere of diameter $D$ filled with surfactant, $q=q_0$ and $\Gamma=0$ at $t^*=0$, is placed at rest at $(2,2,2)$ in the domain $[0,8]^3$, periodic in all three directions, both fluids having the same properties. Since nothing moves, the problem is scaled by diffusion, with $L=D/2$ and $U=2\mathcal{D}/D$ as in the corresponding test of \citet{Muradoglu2014}, so that the groups of Eq.~\eqref{eq:dimensionless_groups} are $\mathrm{Pe}_c=\mathrm{Pe}_s=1$, $K=0.25$, $\mathrm{Bi}=3.1$ and $\mathrm{Da}=2$. The calculation ends at $t^*=0.415$, with the time steps $\Delta t^*=1.28\times10^{-3}$, $6.4\times10^{-4}$ and $3.2\times10^{-4}$ at $D/h=10$, $20$ and $40$. The problem is spherically symmetric and $\Gamma$ is uniform on the interface, so that the surface diffusion plays no role. The bulk concentration of the radial reference solution, $q^{\mathrm{rad}}(r,t)$, then satisfies
\begin{equation}
\frac{\partial q^{\mathrm{rad}}}{\partial t}=\frac{\mathcal{D}}{r^2}\frac{\partial}{\partial r}\left(r^2\frac{\partial q^{\mathrm{rad}}}{\partial r}\right),\quad 0<r<R,\qquad
\frac{\partial q^{\mathrm{rad}}}{\partial r}\bigg|_{r=0}=0,\qquad
-\mathcal{D}\frac{\partial q^{\mathrm{rad}}}{\partial r}\bigg|_{r=R}=\frac{\mathrm{d}\Gamma^{\mathrm{rad}}}{\mathrm{d}t}=k_a\,c_s^{\mathrm{rad}}\,(\Gamma_\infty-\Gamma^{\mathrm{rad}})-k_b\,\Gamma^{\mathrm{rad}},
\label{eq:sphere_radial}
\end{equation}
with $R=D/2$, $c_s^{\mathrm{rad}}=q^{\mathrm{rad}}(R,t)$, $q^{\mathrm{rad}}(r,0)=q_0$ and $\Gamma^{\mathrm{rad}}(0)=0$. Because of the Langmuir kinetics, this problem has no closed-form solution: for the related problem of a stationary drop that adsorbs surfactant from the surrounding liquid, \citet{Muradoglu2014} found that the similarity solution of \citet{Muradoglu2008} is not exact and compared with a finite-difference solution of the radial equation instead. Likewise, Eq.~\eqref{eq:sphere_radial} is solved here with a finite-volume method on $3000$ radial cells clustered towards the interface, with $c_s^{\mathrm{rad}}$ from a quadratic through the three outermost cells and an implicit backward-differentiation-formula (BDF) time integration with a relative tolerance of $10^{-8}$. With $1500$ cells, $\Gamma^{\mathrm{rad}}(t)$ is the same to within $8\times10^{-8}$, far below the errors of the three-dimensional calculations, and this converged solution serves as the reference. The calculations use the proposed method. Since nothing moves, the transport has no effect, and the error is that of the coupling between the front and the bulk.

\begin{figure}[htbp]
\centering
\includegraphics[width=\textwidth]{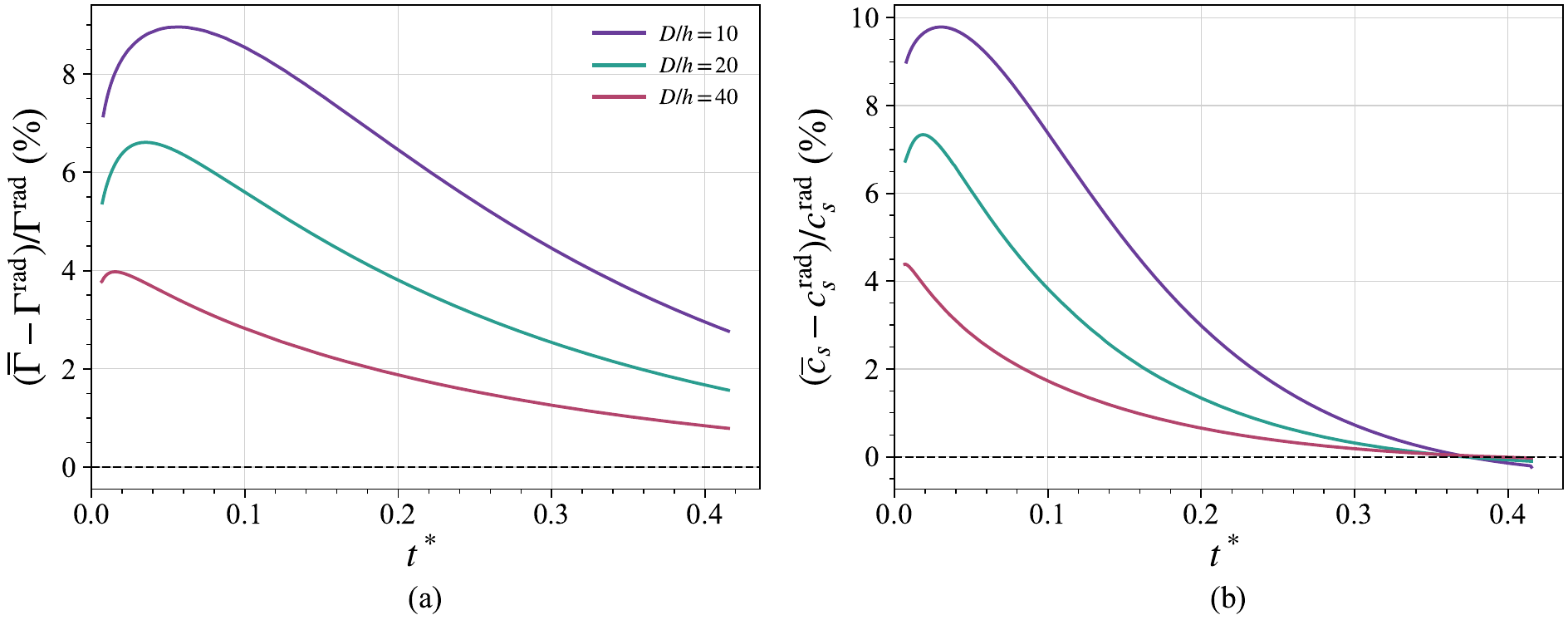}
\caption{Sphere at rest with a soluble surfactant inside, at $D/h=10$, $20$ and $40$: (a) relative error of the mean surface concentration and (b) of the bulk concentration read at the interface, against the reference radial solution, Eq.~\eqref{eq:sphere_radial}, in percent.}
\label{fig:surfactant_sphere}
\end{figure}

Figure~\ref{fig:surfactant_sphere} shows the relative errors of the mean surface concentration (a) and of $c_s$ (b). Both are largest while the surface concentration rises steeply, with peaks of $9.0$, $6.6$ and $4.0\%$ for $\Gamma$ at $D/h=10$, $20$ and $40$, and decrease afterwards. At $t^*=0.415$, the error of $\Gamma$ is $2.8$, $1.6$ and $0.79\%$ and converges at first order, and that of $c_s$ is below $0.3\%$. Translating the sphere along the direction of Section~\ref{sec:source_transport} with the speed $19.3U$, so that the scalar is carried across the grid, changes the final error of $\Gamma$ by at most $0.5$ percentage points, at $D/h=10$, and not at $D/h=40$ ($0.78$ against $0.79\%$).

\section{Localized source with seven advection schemes}
\label{app:scheme_source}
\setcounter{figure}{0}

\begin{figure}[htbp]
\centering
\includegraphics[width=0.8\textwidth]{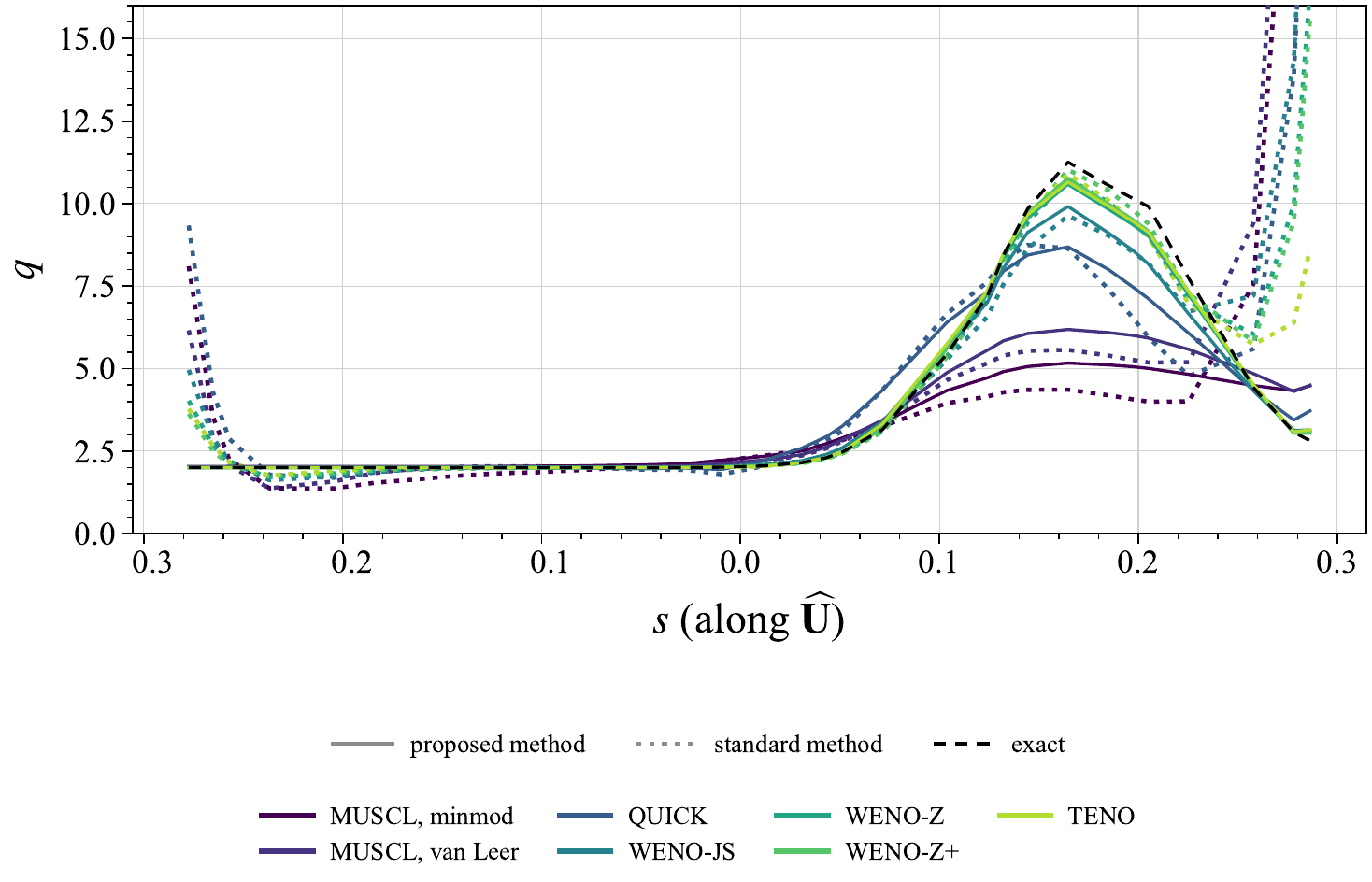}
\caption{Localized source at $D/h=20$ and $t=3.2$ with seven face reconstructions (colours): phase-averaged concentration $q=C/S^{\mathrm{FT}}$ along the translation direction through the source centre, as in Fig.~\ref{fig:source_profiles}, for the proposed method (solid) and the standard method (dotted), with the exact concentration (black dashed). Values above $16$, which the standard method reaches at the leading interface, are cut.}
\label{fig:scheme_source}
\end{figure}

Figure~\ref{fig:scheme_source} shows the localized source of Section~\ref{sec:source_transport} at $D/h=20$ with the seven face reconstructions of Section~\ref{sec:scheme_dependence}; the errors are listed in Table~\ref{tab:scheme_study}.

\bibliographystyle{model1-num-names}

\end{document}